\documentclass[a4paper,11pt]{article}
\pdfoutput=1

\RequirePackage{ifluatex}

\usepackage{jheppub}
\usepackage{physics}
\usepackage{enumerate}
\usepackage{empheq}
\usepackage{booktabs}
\usepackage{amsmath,amssymb,amsbsy,amstext, amsthm, simplewick}
\usepackage{graphicx}
\usepackage{amsfonts}
\usepackage{color}
\usepackage{caption}
\usepackage{subcaption}
\usepackage{wasysym}
\usepackage{wrapfig}
\usepackage{multirow}
\usepackage{array}
\usepackage[normalem]{ulem}
\usepackage{xcolor}

\usepackage{colortbl}
\definecolor{summersky}{cmyk}{0.71,0.33,0,0.5}
\definecolor{flamingo}{cmyk}{0,0.51,0.71,0.5}
\definecolor{rp}{cmyk}{0.2, 1, 0.6, 0}
\definecolor{pacificblue}{cmyk}{0.95,0.3,0, 0.5}
\definecolor{gray60}{cmyk}{0.4,0.4,0,0.8}
\graphicspath{
    {Figures/}
}

\hypersetup{
    pdftoolbar=true,        
    pdfmenubar=true,        
    pdffitwindow=true,     
    pdfstartview={FitH},    
    pdfnewwindow=true,      
    colorlinks=true,       
    linkcolor=pacificblue,          
    citecolor=flamingo,        
    filecolor=magenta,      
    urlcolor=pacificblue           
}

\newcommand{\ex}[1]{\langle #1 \rangle}
\newcommand{\be}{\begin{eqnarray} }
\newcommand{\ee}{\end{eqnarray} }
\newcommand{\bs}{\begin{split} }
\newcommand{\es}{\end{split} }

\newcommand{\la}{\langle}
\newcommand\redsout{\bgroup\markoverwith{\textcolor{red}{\rule[0.5ex]{2pt}{1pt}}}\ULon}

\newcommand{\Mpl}{M_{\text{Pl}}}

\renewcommand{\L}{\mathcal{L}}
\renewcommand{\O}{\mathcal{O}}

\newcommand{\ra}{\rangle}
\newcommand{\then}{\quad \Rightarrow\quad}
\DeclareMathOperator{\sign}{sign}

\newcommand{\JS}[1]{{\color{blue}#1}}

\newcommand{\nc}{\newcommand}

\newcommand{\bvec}[1]{\mathbf{#1}}
\nc{\nn}{\nonumber}
\nc{\eps}{\epsilon}

\newcommand{\D}{\partial}
\renewcommand{\dim}[1]{\text{dim}\left\{#1\right\}}

\newtheorem{theorem}{Theorem}[section]

\title{\centering The Boostless Bootstrap:  \\ \vspace{0.1cm} Amplitudes without Lorentz boosts}

\author[a]{Enrico Pajer,}
\author[a]{David Stefanyszyn,}
\author[a]{Jakub Supe\l{} }

\affiliation[a]{Department of Applied Mathematics and Theoretical Physics, Centre for Mathematical Sciences,
University of Cambridge, Wilberforce Road, Cambridge CB3 0WA, UK}
\emailAdd{ep551@cam.ac.uk}
\emailAdd{d.stefanyszyn@damtp.cam.ac.uk}
\emailAdd{js2154@cam.ac.uk}

\abstract{\noindent  
Poincar\'e invariance is a well-tested symmetry of nature and sits at the core of our description of relativistic particles and gravity. At the same time, in most systems Poincar\'e invariance is not a symmetry of the ground state and is hence broken spontaneously. This phenomenon is ubiquitous in cosmology where Lorentz boosts are spontaneously broken by the existence of a preferred reference frame in which the universe is homogeneous and isotropic. This motivates us to study scattering amplitudes without requiring invariance of the interactions under Lorentz boosts. In particular, using on-shell methods we show that the allowed interactions around Minkowski spacetime are severely constrained by unitarity and locality in the form of consistent factorization. Our analysis assumes massless, relativistic and luminal particles of any spin, and a restricted ansatz for the four-particle amplitude, which can be shown to be equivalent to having Lorentz covariant fields in the Lagrangian description. We find that the existence of an interacting massless spin-2 particle enforces (analytically continued) three-particle amplitudes to be Lorentz invariant, even those that do not involve a graviton, such as cubic scalar couplings. We conjecture this to be true for all $  n $-particle amplitudes. Also, particles of spin $  S>2 $ cannot self-interact nor can be minimally coupled to gravity, while particles of spin $  S>1 $ cannot have electric charge. Given the growing evidence that free gravitons are well described by massless, luminal relativistic particles, our results imply that cubic graviton interactions in Minkowski must be those of general relativity up to a unique Lorentz-invariant higher-derivative correction of mass dimension 9. 
Finally, we point out that consistent factorization for massless particles is \textit{highly IR sensitive} and therefore our powerful flat-space results do not straightforwardly apply to curved spacetime.}

\begin{document}
\maketitle
\flushbottom

\newpage

\section{Introduction and summary}





Symmetry is a physicist's compass and Poincar\'e invariance is perhaps the most precisely tested symmetry in nature \cite{Bluhm:2005uj,Kostelecky:2008ts,Tasson:2014dfa,Will:2014kxa}. Empirically, we observe it everywhere: from electromagnetism to the reign of subatomic particles and the expanse of the cosmos. But just as importantly, Poincar\'e invariance sits at the heart of our description of the laws of nature. On the one hand, it provides us with the organizing principle to model the interactions of subatomic particles through Quantum Field Theory (QFT), and constitutes one of the pillars of the standard model of particle physics. On the other hand, Poincar\'e symmetry is so powerful and rigid that it makes our theoretical description inevitable. We can appreciate this from two complementary points of view. \\

Weinberg argues in \cite{Weinberg:1996kw} that Poincar\'e invariance, combined with quantum mechanics and locality (in the form of cluster decomposition), uniquely selects QFT as the necessary language of nature, at least at low energies. Moreover, from this standpoint, microscopic causality and the analyticity of the S-matrix follow from the above assumptions rather than being invoked as general principles. But fields come at a cost: the spectrum of massless particles cannot fit inside a set of Poincar\'e covariant fields and we are obliged to invoke unobservable ``gauge'' symmetries. Also, the scattering of particles cannot be uniquely mapped into the interactions of fields, as is evident in perturbative field redefinitions. These observations have motivated physicists to look for an alternative description of scattering that does not invoke fields or gauge redundancies. Modern on-shell methods for amplitudes, an intellectual descendant of the S-matrix program of the 60's (see e.g. \cite{Eden:1966dnq}), have made tremendous progress towards precisely this goal (reviews include \cite{Benincasa:2013,Elvang:2013cua,TASI}). It is from this complementary point of view that the rigidity imposed by Poincar\'e invariance becomes once again manifest. All (analytically continued) non-perturbative three-particle amplitudes for massless fields of any spin are uniquely fixed by symmetry, and in theories such as Yang-Mills \cite{BCF,BCFW} and general relativity \cite{Benincasa:2007qj} all higher tree-level amplitudes are uniquely determined in terms of these building blocks. \\

In the discussion so far we have implicitly assumed that Poincar\'e invariance is a symmetry of the ground state of the theory. While this is a good approximation for some particle physics applications, the vast majority of physical systems are not Poincar\'e invariant in their ground state. Indeed, the specific way in which Poincar\'e is thus spontaneously broken determines much of the behavior of a given system. While all possibilities have been classified \cite{Zoology}, a particularly simple and interesting case arises when the ``vacuum'' consists of a static, homogeneous and isotropic medium that permeates spacetime. Observers at rest with respect to this medium are special, as they observe a more symmetric configuration, hence Lorentz boosts are spontaneously broken. This is the case for many condensed matter systems but also for cosmological models as we will discuss in detail shortly. Some even go a step further and speculate about possible explicit breaking of Poincar\'e invariance, perhaps arising in a UV-complete theory of gravity. \\

The above considerations beg the question of what happens to the rigidity of the laws of nature when Poincar\'e invariance is not respected by the ground state, as it is for example the case in our universe at cosmological distances. If the free theory is Poincar\'e invariant, what can we say about interactions? In particular, we will focus on the following formulation of this question: 
\textit{\begin{center} What boost-breaking interactions are allowed for massless, relativistic spinning particles? 
\end{center}}

This question is not just academic. Rather it's motivated by practical considerations. For example, we have recently observed that the free propagation of gravitational waves is extremely well described by the relativistic theory of a (classical) massless spin-2 particle \cite{Monitor:2017mdv}. What does this imply for the interactions that gravitons can have in a consistent theory? More precisely, in this work we will derive all possible on-shell three-particle amplitudes, and the allowed singularities of four-particle amplitudes, for relativistic, massless, luminal particles, while allowing for boost-breaking interactions. Whether Lorentz boosts are broken explicitly, or more likely only spontaneously, will be irrelevant for our discussion (see \cite{Alberte:2020eil} for a recent discussion of Goldstone theorem for boosts). Our assumption that the free theory is Poincar\'{e} invariant leads us to a particular ansatz for four-particle amplitudes, which can be shown to be equivalent to assuming that the underlying Lagrangian is constructed out of Lorentz covariant fields with the breaking of boosts due to the freedom to add time derivatives at will. Although this does not capture the most general set of boost-breaking theories, it provides us with an excellent testing ground and already produces some surprising results. Indeed, we will find that internal consistency severely restricts the allowed set of interactions, especially in the presence of a massless spin-2 particle. We summarise our results in Section \ref{res}.


\subsection{Motivations}\label{res}

Because of the very general methodology that we adopt, our results can be approached and interpreted from a variety of perspectives. In the following, we motivate our analysis from three points of view.

\begin{figure}
\centering
\includegraphics[width=0.6\textwidth]{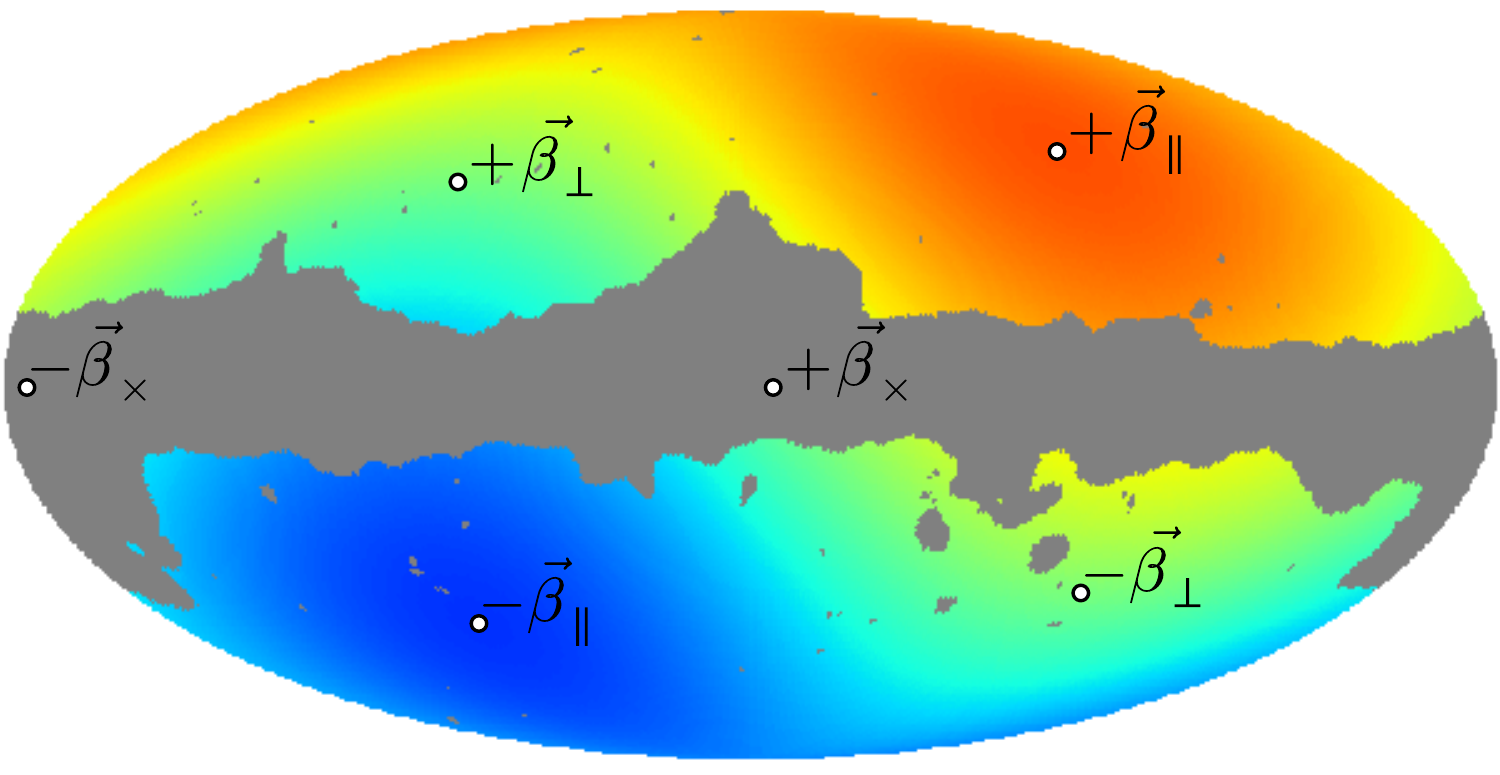}
\caption{\label{fig1} The figure shows CMB dipole at the level of $  3 $ mK align with the $ \pm \beta_{\parallel} $ direction. The two perpendicular directions $  \pm \beta_{\times} $ and $  \pm \beta_{\perp} $ are also shown for reference. This observation highlights the existence of a preferred frame in our universe and hence implies the spontaneous breaking of boost invariance.}
\end{figure}

\paragraph{Cosmology} The expansion of the universe spontaneously breaks time translations and boosts\footnote{Everywhere in this paper we assume invariance under spacetime translations and rotations, but for conciseness we will avoid stating this repeatedly.}. Both breakings are manifest in many cosmological phenomena. For example, the breaking of time translations can be thought of as the root cause of the redshift of light as it travels freely across the cosmos: in the absence of time translation invariance, energy is not conserved and the energy of a free photon can change with time. The breaking of boost invariance is evident in the existence of the Cosmic Microwave Background (CMB) or the cosmic neutrino background. The CMB picks out a preferred reference frame in which the universe looks homogeneous and isotropic. The Earth moves with respect to this preferred frame and so we observe the CMB to be anisotropic to one part in a thousand. Measurements of this CMB dipole by the Planck satellite are shown in Figure \ref{fig1} \cite{Aghanim:2013suk}. \\

A priori, it is impossible to compare the breaking of time translations with that of boosts because the respective parameters have different dimensions\footnote{This is evident in the examples above. In observing the CMB, we see the breaking of boosts in the presence of a dipole, but we can safely neglect the breaking of time translations because observations are conducted over tens of years while the CMB changes in time over $  10^{5} $ years. Conversely, the redshift of photons from distant sources is mostly caused by the breaking of time translations, while the effect of peculiar motion, which is evident in redshift space distortions, is much smaller.}: the breaking of time translations is characterized by a certain time scale $  t_{b} $, while that of boosts by a certain velocity $  v_{b} $. Since in this work we will study the time-translation invariant dynamics of massless particles with broken boosts, it is important to understand under what conditions our results have a chance to be relevant for cosmology. \\

First, we notice that for the scattering of particles at energy $ E  $, the breaking of time translation should be parameterized by $  1/(E t_{b}) $, which is negligible at sufficiently high energies. So in cosmology, where the characteristic time scale is the Hubble parameter, $  t_{b}^{-1}\sim H $, time-translation invariance is often a good approximate symmetry at energies $  E\gg H $. Conversely, for the scattering of massless luminal particles, which are the focus of our study, the typical center of mass velocity is always of order the speed of light. Hence, in cosmology, where the speed of light is often the characteristic speed $  v_{b}\sim c $, the breaking of boosts can be a large effect. \\

Second, in many models of the very early universe and of dark energy, additional symmetries are invoked to suppress the breaking of time translations. The archetypal example is that of a so-called superfluid or $ P  $-of-$ X  $ theory, namely a shift-symmetric scalar field whose evolution is assumed to be approximately linear in time\footnote{In general, the existence of a shift symmetry is not sufficient to ensure time-translation invariance. Rather, its general consequences are new cosmological soft theorems \cite{Finelli:2017fml} and recursive relations for the time-dependence of the  low-energy coupling constants \cite{Finelli:2018upr}. It is only when one further assumes a linear evolution for the shift-symmetric scalar that a diagonal symmetry emerges, which plays the role of time-translation invariance, a general mechanism that  goes under the name of spontaneous symmetry probing \cite{Nicolis:2011pv}. See \cite{GJS} for a recent discussion on using a constant shift symmetry, and other symmetries, to realise a diagonal form of unbroken translations in the presence of additional non-linearly realised symmetries.}. In this case, while time-translations, which are generated by $  T^{0\mu} $, and shifts, which are generated by $  j^{\mu} $, are separately broken spontaneously, an (approximate) unbroken diagonal linear combination $  t^{\mu} $ exists
\begin{align}
t^{\mu}=T^{0\mu}+j^{\mu} \then \nabla_{\mu}t^{\mu}=0\,.
\end{align}
In inflationary models this unbroken diagonal symmetry is eventually responsible for the (approximate) scale invariance of primordial perturbations that we have observed in the data. One might ask whether a similar mechanism can be developed to suppress or eliminate the breaking of boosts. As pointed out recently in \cite{Green:2020ebl} (see also \cite{Baumann:2019ghk}), this is problematic because one would need to invoke a higher-spin symmetry, which in flat space is forbidden by the Coleman-Mandula theorem \cite{Coleman:1967ad}. Indeed, it was proven in \cite{Green:2020ebl} that if one insists on having unbroken boost invariance for cosmological correlators in single-clock inflation, all interactions are forbidden and the theory must be free. Thus, the breaking of boosts cannot be eliminated and in principle it could always affect the interactions. \\

The discussion above highlights the importance for cosmology of time-translation invariant theories that (spontaneously) break boosts. In this work we study some of these theories in the context of scattering amplitudes. It will turn out that the application of our results to cosmology shows an unexpected and very interesting twist. We will discuss this in Section \ref{ssec:pofx}.

\paragraph{Cosmological correlators} The calculation of primordial initial conditions from models of the early universe provides a major motivation for the study of boost-breaking amplitudes. The key observation is that the correlators of $  n $ fields of momenta $  \vec k_{a} $ with $  a=1,\dots,n $ in an expanding universe encode the information of $ n  $-particle scattering amplitudes in Minkowski in the residue of the highest $  k_{T} $ pole (see \cite{Maldacena:2011nz,Raju:2012zr}), where $  k_{T}=\sum |\vec k_{a}| $ is sometimes called the ``total energy''. Schematically, the relation takes the form\footnote{There are many exceptions to this result. For example, when the amplitude vanishes, this relation should be modified since the leading pole disappears. This is what happens in the DBI theory, due to the increased symmetry in the flat space-limit, as recently noticed in \cite{GJS}.}
\begin{align}\label{BtoA}
\lim_{k_{T}\to 0} \ex{\prod_{a=1}^{n}\phi_{a}}' \sim \frac{\Re A_{n}}{\left(  \prod_{a=1}^{n}k_{a}\right)^{2}k_{T}^{p}}+\dots
\end{align}
where the dots represent subleading terms in $  k_{T}\to 0 $, $  \phi_{a} $ are fields (not necessary scalars), $  A_{n} $ is the flat space amplitude for the scattering of the particles created by the $  \phi_{a} $'s, and a prime denotes that we are dropping the momentum conserving delta function. The value of the positive exponent $p$ depends on the interactions included in the theory, with larger $p$'s corresponding to the inclusion of operators of higher and higher dimension \cite{THC}. This relation gives us a handle to leverage our knowledge of amplitudes to better understand cosmological correlators. \\

The idea to constrain cosmological correlators from symmetries has been pursued from various angles over the years. In \cite{Maldacena:2011nz} it was shown that the graviton bispectrum is completely fixed non-perturbatively by the isometries of de Sitter to be a linear combination of only two shapes, one corresponding to the Einstein-Hilbert term and the other to a higher-derivative term. In \cite{Creminelli:2011mw}, de Sitter isometries were used to fixed the bispectrum of a spectator scalar. In \cite{Mata:2012bx}, it was shown how an approximate version of de Sitter isometries constrains the leading-order scalar-scalar-tensor bispectrum. In \cite{Ghosh:2014kba,Kundu:2014gxa,Kundu:2015xta} the study was extended to the scalar bispectrum and trispectrum. In \cite{Pajer:2016ieg}, it was shown that the $ \zeta $ bispectrum in the de Sitter-invariant limit of single-field inflation is fully fixed by approximate de Sitter isometries. More recently, in \cite{Arkani-Hamed:2018kmz,Baumann:2019oyu,Sleight:2019mgd,Sleight:2019hfp,Baumann:2020dch} an ambitious program has been proposed to systematically use not only symmetries but also general principles such as unitarity and locality to ``bootstrap'' correlators, in analogy with the on-shell methods for amplitudes. In the current incarnation of this \textit{cosmological bootstrap}, the isometries of de Sitter spacetime still play an essential role, analogously to the role Poincar\'e invariance plays for amplitudes. On the one hand, it is clear from the above literature that de Sitter isometries are so constraining that many correlators are uniquely specified by them. On the other hand, we know that most observationally interesting correlators, such as for example equilateral and orthogonal non-Gaussianity, are not de Sitter invariant, and so cannot be studied directly with these methods. More generally, in \cite{Green:2020ebl} it was proven that in single-field inflation, the only theory whose $  \zeta $ correlators are invariant under de Sitter isometries is the free theory. It is therefore very important to extend the cosmological bootstrap to less symmetric cases. In particular, it is the invariance under de Sitter boosts that should be relaxed, as this has not been observed in the data and indeed is not present in many models, for example those with a reduced speed of sound, $  c_{s}<1 $. Much insight can already be gained by perturbative calculations \cite{Arkani-Hamed:2015bza,Arkani-Hamed:2017fdk,Arkani-Hamed:2018bjr,Benincasa:2018ssx,Benincasa:2019vqr,Hillman:2019wgh}. \\ 

The amplitudes that emerge on the total energy pole in \eqref{BtoA} when de Sitter boosts are broken are not Lorentz invariant, rather they break Lorentz boosts. So one crucial step to extend the cosmological bootstrap to correlators with broken de Sitter boosts is to understand boost-breaking amplitudes. This is one of our primary motivations for this work.

\paragraph{Gravitational waves} The recent detection of gravitational waves has ushered a new era in astronomy. But the detection of this 100 year old prediction of general relativity (GR) has implications well beyond the study of binary compact objects. It provides strong constraints on modified gravity (see e.g. \cite{MG1,MG2,MG3,MG4}) and on the properties of the graviton. In particular, the concurrent observation of GW170817 \cite{TheLIGOScientific:2017qsa} and the gamma-ray burst GRB170817A \cite{Goldstein:2017mmi} has put extremely strong constraints on the difference $  \Delta v $ between the speed of gravity and the speed of light \cite{Monitor:2017mdv}
\begin{align}
-3 \times 10^{-15}< \Delta v/c < 7 \times 10^{-16}\,.
\end{align}  
More general Lorentz-breaking modifications of the graviton dispersion relation were classified and severely constrained in \cite{Kostelecky:2015dpa} using gravitational \v Cerenkov radiation by cosmic rays, and the constraints are even stronger when the GW170817 and GRB170817A data is included \cite{Monitor:2017mdv}. In particular, Lorentz-breaking deviations from a relativistic dispersion relation $  E^2 = c^2 \bvec{p}^2 $ have to be smaller than a part in $  10^{-13} $, and some specific modifications must be as small as a part in $  10^{-45} $. The mass of the graviton is also strongly constrained by a variety of measurements. Largely model-independent bounds on the graviton mass $  m_{g} $ can be as strong as $ m_{g} < 10^{-22} $ eV from observations such as Yukawa-like corrections to Newton's law \cite{Talmadge:1988qz} or gravitational waves from binary mergers \cite{TheLIGOScientific:2016src} (see \cite{deRham:2016nuf} for a recent summary and more details). More model-dependent bounds can be as strong as $m_{g} < 10^{-32}$ eV from observations of gravitational lensing \cite{Choudhury:2002pu} or of the earth-moon precession \cite{Dvali:2002vf}. All of these bounds strengthen our confidence that GR provides a good description of free gravitons.\footnote{Finally, from a more theoretical perspective, \cite{Hertzberg:2017nzl} argues that the special relativistic energy-momentum relation is a consequence of locality and of the existence of massless gravitons mediating long range forces.} \\

It is then natural to ask: \textit{what gravitational interactions are compatible with the observation that the graviton is a relativistic, massless spin two particle?} Any theoretical guidance in answering this question is of particular relevance also because it is much harder to directly probe the non-linear dynamics of gravitons, due to the weakness of gravity. It has been known for half a century that Lorentz invariance forces the self-interaction of a massless spin-2 particle, as well as the interactions with any other particle, to be universal in the infra-red around Minkowski spacetime and to correspond to the interactions of GR \cite{Weinberg:1965nx,Weinberg:1965rz}. More generally, from a purely on-shell perspective, there are only three possible cubic (analytically continued) amplitudes for three gravitons, which reduce to two if one assumes parity \cite{Benincasa:2007xk}. These are the interactions of GR, coming from the Ricci scalar $  R $, and higher derivative interactions from the (dimension 9) Riemann cubed terms, which are highly suppressed at low energies. Self-interactions with broken Lorentz boosts have received less attention. In \cite{Kostelecky:2003fs}, it is argued that the explicit breaking of Lorentz symmetry is inconsistent with dynamical gravity, while this obstruction may be absent if the breaking is spontaneous. In \cite{Khoury:2013oqa}, the authors show that assuming only spatial covariance, the leading order couplings of the graviton must display Lorentz invariance, which from this perspective appears as an emergent symmetry. \\

In this work we will take a complementary approach. We will only discuss physical on-shell (massless) particles, thus avoiding any mention of gauge symmetries such as general covariance. General principles such as unitarity and locality will then enforce Lorentz invariance and agreement with GR, within the assumptions that we make about the form of our four-particle amplitudes. Our results are summarized below in Section \ref{res}.


\subsection{Summary of the main results}\label{res}

The main body of the paper consists of a detailed derivation of our results. We attempted to make our derivation pedagogical and the presentation self-contained, so that this paper can be approached without much familiarity with on-shell methods and the spinor helicity formalism. While many of our derivations are technical in nature, our final results can be stated in simple terms. For the reader who is not interested in the details, we therefore outline our main findings here. All the statements below are valid under the following assumptions:
\begin{itemize}
\item The spacetime is Minkowski.
\item All particles are relativistic, massless and luminal, i.e. they all propagate at \textit{the same speed}, which we set to one and call the ``speed of light'', even when no photons are present in the spectrum.
\item All interactions respect spacetime translations and rotations, but we allow for interactions that are not invariant under Lorentz boosts. Whether Lorentz boosts are non-linearly realized or explicitly broken plays no role in our analysis. 
\item While our results for three-particle amplitudes are non-perturbative in nature, our factorization constraints on the four-particle amplitudes ignore loop contributions.
\item The helicity scaling of four-particle amplitudes is fixed in terms of ``angle" and ``square" spinor helicity brackets only. This assumption amounts to assuming that the underlying Lagrangian is a function of Lorentz covariant fields with the breaking of Lorentz boost induced by time derivatives, which can appear at will. This assumption means that our results do not apply to theories that are written in terms of $SO(3)$ covariant fields such as the Framid and the Solid of \cite{Zoology}. We will explain in Section \ref{sec:4p} why the amplitudes of these theories are not captured by our ansatz.
\end{itemize}    
From these assumptions and demanding unitarity and locality through the consistent factorizations of four-particle amplitudes, we are able to show that the set of consistent interactions is severely restricted. In more detail:

\begin{itemize}
\item We derive all possible boost-breaking cubic amplitudes for relativistic massless particles of any spin. Unlike in the Lorentz-invariant case, there are always infinitely many possibilities, which are characterized by a generic function of the particles' energies (see \eqref{eq:3pFinalResult}). This result is completely non-perturbative.
\item If interactions with a massless spin-2 particle are allowed, three-particle amplitudes must be Lorentz invariant, even those that do not involve a graviton (see Section \ref{sec:GravitonCouplings}). For example, amplitudes corresponding to boost-breaking cubic scalar interactions such as $\dot \phi^{3} $, $  \dot\phi (\partial\phi)^{2} $ and all other higher-derivative ones are forbidden. We conjecture this to be true for all other higher-particle amplitudes. This is strong evidence that Lorentz invariance follows from having consistent interactions involving a massless spin-2 particle, at least as long as the Lagrangian is written in terms of covariant fields as we stated in the above assumptions.
\item The cubic graviton amplitudes must be those of GR at low energies (corresponding to dimension-5 operators). As for the Lorentz-invariant case, the only other graviton amplitudes correspond to the two possible Riemann$  ^{3} $ couplings (dimension-9 operators). 
\item Particles with spin $  S>1 $ cannot have an electric charge (see Section \ref{sec:PhotonCouplings}). Particles with spin $S > 2$ cannot have cubic self-interactions of dimensionality lower than $3S$. They also cannot interact gravitationally via the GR vertex (see Section \ref{sec:4pIdenticalSpinS} and \ref{sec:GravitonCouplings}). Lower spin particles ($S < 2$) can indeed be minimally coupled to the graviton and these couplings are fixed by the coupling of the GR vertex. This is the on-shell manifestation of the equivalence principle.
\item  Unlike for the Lorentz-invariant case, cubic self-interactions of a single massless spin-1 particle do exist (dimension-6 operators) when boosts are broken (see Section \ref{sec:photon}). All lower dimension operators are forbidden.
\item We find large classes of self-consistent, boost-breaking interactions among scalars, photons and spin-$1/2$ fermions, already at leading order in spatial derivatives. In other words, QED, scalar QED and scalar theories allow for the breaking of boosts at the cubic level (see Section \ref{sec:PhotonCouplings}).
\item We point out that the four-particle test for massless particles is \textit{highly IR sensitive} (see Section \ref{ssec:pofx}). As a consequence, the results that follow from it cannot be straightforwardly applied to cosmology, where the Hubble parameter that characterizes the curvature of spacetime constitutes an IR modification of Minkowski spacetime. Conversely, all those results that are exclusively based on symmetries, such as for example the form of the three-particle amplitude (see Section \ref{sec:3p}) are robust and do apply to curved spacetime as well.
\end{itemize}

\paragraph{Notation and conventions} Since we will be dealing with boost-breaking theories, for dimensional analysis we will have to separate units of length from units of time. Working with $  \hbar =1=c $, we will indicate by ``$  \dim{\dots} $'' the scaling of an object with \textit{spatial} momentum, which has units of inverse length, \textit{excluding the dimension of all coupling constants}. For Lorentz-invariant theories this gives to the standard energy/mass dimension, as e.g. in \cite{TASI}. For example, 
\begin{align}
\dim{(ii)}&=0\,,& \dim{[ij]}&=\dim{\ex{ij}}=1\,.
\end{align}
We will work with the mostly minus metric signature $\eta_{\mu\nu} = \text{diag}(+1,-1,-1,-1)$ and follow \cite{Dreiner} for spinor conventions. We use the beginning of the Greek alphabet for $SU(2)$ indices $(\alpha,\beta,\gamma,\ldots)$, and the middle of the alphabet for $SO(1,3)$ indices $(\mu,\nu,\rho,\sigma,\ldots)$. Our basis for the Pauli matrices $\sigma^{\mu}_{\alpha \dot{\alpha}}$ and $(\bar{\sigma}^{\mu})^{\dot{\alpha} \alpha}$ is
\begin{align}
(\sigma^{0})_{\alpha\dot \alpha} = (\bar{\sigma}^{0})^{\alpha\dot \alpha} &= \left( {\begin{array}{*{20}c}
   1 & \qquad & 0 \\
   0 & \qquad & 1 \\    
 \end{array} } \right), & (\sigma^{1})_{\alpha\dot \alpha} = -(\bar{\sigma}^{1})^{\alpha\dot \alpha} &= \left( {\begin{array}{*{20}c}
   0 & \qquad & 1 \\
   1 & \qquad & 0 \\    
 \end{array} } \right), \\
 (\sigma^{2})_{\alpha\dot \alpha} = -(\bar{\sigma}^{2})^{\alpha\dot \alpha} &= \left( {\begin{array}{*{20}c}
   0 & \qquad & -i \\
   i & \qquad & 0 \\    
 \end{array} } \right), & (\sigma^{3})_{\alpha\dot \alpha} = -(\bar{\sigma}^{3})^{\alpha\dot \alpha} &= \left( {\begin{array}{*{20}c}
   1 & \qquad & 0 \\
   0 & \qquad & -1 \\    
 \end{array} } \right),
\end{align}
and amongst the many useful identities these matrices satisfy 
\begin{align}
\sigma^{\mu}_{\alpha \dot{\alpha}}\bar{\sigma}_{\mu}^{\dot{\beta} \beta} &= 2 \delta_{\alpha}{}^{\beta}\delta^{\dot{\beta}}{}_{\dot{\alpha}}\,, \\
\sigma^{\mu}_{\alpha \dot{\alpha}}(\sigma_{\mu})_{\beta \dot{\beta}} &= 2 \epsilon_{\alpha \beta} \epsilon_{\dot{\alpha} \dot{\beta}}\,, \\
(\bar{\sigma}^{\mu})^{\dot{\alpha}\alpha} \bar{\sigma}_{\mu}^{\dot{\beta}\beta} &= 2 \epsilon^{\alpha \beta}\epsilon^{\dot{\alpha}\dot{\beta}}\,,
\end{align}
where the components of the epsilon and delta tensors are
\begin{align}
\epsilon^{12} &= - \epsilon^{21} = \epsilon_{21} = -\epsilon_{12} = 1\,, &  \delta_{\alpha}{}^{\beta}&=\left( {\begin{array}{*{20}c}
   1 & \qquad & 0 \\
   0 & \qquad & 1 \\    
 \end{array} } \right)\,.
\end{align}
We use these epsilon tensors to raise and lower the dotted and undotted $SU(2)$ indices as
\begin{align}
\psi_{\alpha} = \epsilon_{\alpha \beta}\psi^{\beta}, \qquad \psi^{\alpha} = \epsilon^{\alpha \beta}\psi_{\beta}, \qquad \bar{\psi}_{\dot{\alpha}} = \epsilon_{\dot{\alpha} \dot{\beta}}\bar{\psi}^{\bar{\beta}}, \qquad \bar{\psi}^{\dot{\alpha}} = \epsilon^{\dot{\alpha}\dot{\beta}}\bar{\psi}_{\dot{\beta}}.
\end{align}

\paragraph{Note added} During the completion of this work a paper appeared \cite{Hertzberg:2020yzl} that argues that the consistent description of a massless spin-2 particle requires certain tree-exchange diagram to be Lorentz invariant. One of our main results in this work is in complete agreement with this finding, while other results for gravitons are new. In a similar vein, \cite{Hertzberg:2020ird} recovers the central tenets of electromagnetism, such as charge conservation, without imposing boost invariance. Our point of view and methodology are complementary to that in \cite{Hertzberg:2020yzl}, \cite{Hertzberg:2020ird} since we only use on-shell methods and make no use of the field theory apparatus.

\section{On-shell methods: symmetries and bootstrap techniques} \label{sec:Methods}

The aim of the S-matrix bootstrap program is to construct, directly at the level of the S-matrix, consistent scattering amplitudes exhibiting a given set of (linearly realised) symmetries. This on-shell technique bypasses the usual Lagrangian formalism of effective field theories, thereby avoiding redundancies such as field redefinitions and gauge transformations. In this section we introduce the basic principles of this bootstrap program. 


\subsection{Symmetries and on-shell conditions for free particles}

We begin by discussing the symmetries we are assuming so that we can clearly compare and contrast our results with those in the literature \cite{Benincasa:2007xk,Elvang:2013cua,TASI,McGadyRodina,Schuster:2008nh,NAH}. Up to now, on-shell methods and the four-particle test of \cite{Benincasa:2007xk} have been applied to theories for which the vacuum is assumed to be invariant under the full Poincar\'{e} group $ISO(1,3)$, consisting of spacetime translations, spatial rotations and Lorentz boosts. In this work we relax the assumption that Lorentz boosts leave the vacuum unchanged, while assuming that spacetime translations and spatial rotations remain good linearly realised symmetries. We will be agnostic about whether boosts are explicitly broken or spontaneously broken and non-linearly realized. In four spacetime dimensions our symmetry group is therefore $\mathbb{R}^{4} \rtimes SO(3)$. Throughout our paper, we will use the following terminology:
\begin{align}
&\textit{Boost-invariant theories:} ~~~ \text{unbroken $ISO(1,3)$} \\
&\textit{Boost-breaking theories:} ~~~ \text{unbroken $\mathbb{R}^{4} \rtimes \text{SO}(3)$}. 
\end{align}
In the bootstrap program one has to provide the on-shell data which includes the \textit{on-shell conditions} relating the energy and spatial momentum of each free particle. In boost-invariant theories massless particles satisfy the usual on-shell condition $E^2 - \bvec{p}^2 = 0$, while in boost-breaking theories many other on-shell conditions are allowed due the reduced symmetry. Below we classify these possibilities:
\begin{itemize}

\item \textit{Relativistic}: each free particle satisfies $E^2 - c_s^2 \bvec{p}^2 = 0$ with the speed of sound $c_s$ being \textit{the same} for each particle. Without loss of generality, in this case we can choose to work in units such that $c_{s} = c = 1$ and we will do this in the rest of the paper. 

\item \textit{Linear}: each free particle satisfies $E^2 - c_s^2 \bvec{p}^2 = 0$, where at least two particles have a different $c_s$.

\item \textit{General}: the on-shell condition for each particle is $S(E, p) = 0$ and is not captured by the two cases above. 
\end{itemize}

In this paper we consider the relativistic case where each particle has a Lorentz invariant propagator and leave generalisations to other on-shell conditions for future work. So, we focus on theories where all boosts are broken at the level of the interactions only which will lead us to a natural ansatz for four-particle amplitudes. We therefore combine the energy and spatial momentum into the usual $4$-vector $p_{\mu}$ satisfying $p^{\mu}p_{\mu} = 0$ for each particle.


\subsection{Little group scaling and the spinor helicity formalism}
 
Let us now emphasise that the usual classification of massless particles in terms of helicity remains valid for boost-breaking theories. In this subsection we also present the spinor helicity formalism, which for boost-invariant theories has been reviewed in many cases e.g. \cite{Benincasa:2013,TASI,Elvang:2013cua,Schwartz:2013pla,NAH}, and for boost-breaking theories was introduced in \cite{Maldacena:2011nz} (see also Appendix C of \cite{Baumann:2020dch}).\\

Spacetime translation symmetry alone entails that there exists a basis of one particle states $|\mathbf{p}, E \rangle$, which are the eigenstates of the momentum and energy operators:
\begin{equation}
\hat{p}_i | \mathbf{p}, E \rangle = p_i | \mathbf{p}, E \rangle, \qquad \hat{E} | \mathbf{p}, E \rangle = E | \mathbf{p}, E \rangle.
\end{equation}
States with the same $\bvec{p}$ and $E$ may be degenerate and additional quantum numbers are collectively indicated by an index $\sigma$ i.e. $|\mathbf{p}, E; \sigma \rangle$. An important subgroup of the full Lorentz group is the \textit{little group} which is the group of transformations that leave the $4$-momentum $p_{\mu}$ invariant. Such transformations map
\begin{align}
|\mathbf{p}, E; \sigma \rangle \mapsto D_{ \sigma}^{\ \sigma'} |\mathbf{p}, E; \sigma' \rangle. 
\end{align}

Single particle states can then be further classified according to their eigenvalues under the little group. In both boost-invariant and boost-breaking theories, this is the projective $SO(2)$\footnote{In the boost-invariant case, the little group for massless particles is $ISO(2)$, but we recover $SO(2)$ if we make the reasonable assumption that the fields transform trivially under the noncompact subgroup representing the translations in $ISO(2)$. (See \cite{Weinberg:1995}, Chapter 2 for more details.) Once boosts are broken, the little group becomes $SO(2)$ straight away.}, and the states $|\mathbf{p}, E \rangle$ carry a label corresponding to helicity $h = 0, \pm \frac{1}{2}, \pm 1, \ldots$. Clearly the relevant symmetry here is spatial rotations, rather than Lorentz boosts. The helicity of a particle is the same in all frames related by a rotation and changes sign under a spatial reflection. For that reason, we may consider the allowed helicity states for a massless particle of spin $S > 0$ to be $+S$ and $-S$. \\

Throughout this work we will make use of spinor helicity formalism as a powerful tool to present amplitudes in a compact form. This formalism, introduced below, provides a compact way of expressing amplitudes and its simplicity is beautifully captured by the Parke-Taylor formula for gluon scattering \cite{ParkeTaylor}. Here we extend these methods along the lines of \cite{Maldacena:2011nz} for application in boost-breaking theories.\\

We start by using the Pauli matrices (we follow the conventions of \cite{Dreiner}) to map the momentum $4$-vector $p_{\mu}$ into a $2 \times 2$ matrix\footnote{Since $\sigma^{\mu}_{\alpha \dot{\alpha}} \bar{\sigma}_{\mu}^{\dot{\beta} \beta} = 2 \delta_{\alpha}{}^{\beta} \delta^{\dot{\beta}}{}_{\dot{\alpha}}$ we have $p_{\mu} = \frac{1}{2}\bar{\sigma}_{\mu}^{\dot{\alpha}\alpha}p_{\alpha \dot{\alpha}}$}
\begin{align}
p_{\alpha \dot{\alpha}} = \sigma^{\mu}_{\alpha \dot{\alpha}}p_{\mu} = \left( {\begin{array}{*{20}c}
   p_{0}+p_{3} & p_{1}-ip_{2}  \\
   p_{1} + ip_{2} & p_{0}-p_{3}  \\    
 \end{array} } \right),
\end{align}
where $\sigma^{\mu} = (1,\sigma^{i})$. The dotted and undotted indices transform in the fundamental and anti-fundamental representation of $SL(2, \mathbb{C})$\footnote{In $4$ dimensions, the group of proper Lorentz transformations is $SO(1,3) \simeq SL(2, \mathbb{C})/\mathbb{Z}_2$. Thus, projective representations of the Lorentz group can be identified with representations of $SL(2, \mathbb{C})$.} respectively, such that $p_{\alpha \dot{\alpha}}$ transforms in the $(1/2,1/2)$ representation. The dotted and undotted indices run over two values, e.g. $\alpha = 1,2$, and in a boost-invariant theory dotted and undotted indices are contracted with the epsilon tensors $\epsilon^{\dot{\alpha} \dot{\beta}}, \epsilon^{\alpha \beta}$. Using $p_{\alpha \dot{\alpha}}$ alone, the only Lorentz invariant quantity we can construct is $p^{\alpha \dot{\alpha}}p_{\alpha \dot{\alpha}} = 2\det(p) = 2 p^{\mu}p_{\mu} = 0$. It follows that $p_{\alpha \dot{\alpha}}$ is at most rank one thereby allowing us to write
\begin{align}
p_{\alpha \dot{\alpha}} = \lambda_{\alpha} \tilde{\lambda}_{\dot{\alpha}},
\end{align}
where $\lambda$ and $\tilde{\lambda}$ are two-component spinors. Note that 
these objects \textit{are not} Grassmanian, rather they are complex numbers satisfying $\lambda_{\alpha} \tilde{\lambda}_{\dot{\alpha}} = \tilde{\lambda}_{\dot{\alpha}} \lambda_{\alpha}$. We also note that these spinors are not unique and are only defined up to a little group, or helicity, transformation. Indeed the transformation
\begin{equation}
( \lambda_{\alpha} , \tilde{\lambda}_{\dot{\alpha}} ) \ \mapsto \  ( t^{-1} \lambda_{\alpha} , t \tilde{\lambda}_{\dot{\alpha}}),
\label{eq:Hel}
\end{equation}
where $t$ is a nonzero complex number, leaves $p_{\alpha \dot{\alpha}}$ invariant. For physical processes, the external momenta are always real and therefore the spinors can be chosen to satisfy the reality condition $\tilde{\lambda}_{\dot{\alpha}} = \pm (\lambda^{*})_{\dot{\alpha}}$ and we can restrict the transformation parameter $t$ to a phase. However, to study the analytic structure of the S-matrix we must keep the momenta complex, and therefore the spinors are in general independent. \\

What scalar quantities can we construct from these spinors? In boost-invariant theories we have the following two inner products
\begin{eqnarray}
\la i  j \ra  =  \epsilon^{\alpha \beta} \lambda^{(i)}_{\alpha} \lambda^{(j)}_{\beta}, \qquad \ [ i  j ]  =  \epsilon^{\dot{\alpha} \dot{\beta}} \tilde{\lambda}^{(i)}_{\dot{\alpha}} \tilde{\lambda}^{(j)}_{\dot{\beta}},
\end{eqnarray}
defined for two particles $i$ and $j$. We refer to these products as angle and square brackets, respectively. Since the epsilon tensors are anti-symmetric and the spinors are not Grassmanian, these brackets are anti-symmetric i.e. $\la i  j \ra = -\la j  i \ra$ and $[ i  j ] = - [ j  i ]$, which of course implies $\la i  i \ra = [i i] = 0$. From these brackets we can construct the familiar Mandelstam variables for four-particle scattering amplitudes. Taking all particles as \textit{incoming}, we have 
\begin{align}
s = (p_{1} + p_{2})^{2} = (p_{3} + p_{4})^2 = \langle 12 \rangle [12] = \langle 34 \rangle [34], \\
t = (p_{1} + p_{3})^{2} = (p_{2} + p_{4})^2 = \langle 13 \rangle [13] = \langle 24 \rangle [24], \\
u = (p_{1} + p_{4})^{2} = (p_{2} + p_{3})^2 = \langle 14 \rangle [14] = \langle 23 \rangle [23].
\end{align}

For our interests, however, we have a reduced set of symmetries and therefore additional scalar quantities are allowed. Indeed, in boost-breaking theories we can mix the dotted and undotted indices by contracting the spinors with $(\bar \sigma^{0})^{\alpha \dot{\alpha}}$. We therefore have an additional inner product which we denote as
\begin{equation}
(i  j) = (\bar \sigma^{0})^{\alpha \dot{\alpha}} \lambda^{(i)}_{\alpha} \tilde{\lambda}^{(j)}_{\dot{\alpha}},
\end{equation}
and refer to as round brackets. As will be explained in section \ref{sec:3p}, for three-particle kinematics only the diagonal components of this new bracket i.e. $(ii)$ are independent objects, while for four-particle kinematics one of the off-diagonal brackets is independent. For the relativistic on-shell condition, the $0$-component of the momentum $4$-vector for each particle is the energy of the particle, which we denote by $E$. The diagonal round brackets pick out precisely this component: $(ii) = 2 E_i$. \\

For spinning particles there is a key piece of on-shell data which we haven't yet discussed: the polarisation tensors. These form non-trivial representations of the little group and therefore encode the helicity of the particle in question. For a spin-$S$ particle we write the rank-$S$ polarisation tensor as a product of $S$ polarisation vectors which in the spinor helicity variables take the form
\begin{align}\label{PolarisationVector}
e^{+}_{\alpha \dot{\alpha}} = \frac{\eta_{\alpha}\tilde{\lambda}_{\dot{\alpha}}}{\langle \eta \lambda \rangle}, \qquad e^{-}_{\alpha \dot{\alpha}} = \frac{\lambda_{\alpha}\tilde{\eta}_{\dot{\alpha}}}{[ \tilde{\eta} \tilde{\lambda} ]},
\end{align}
for $+1$ and $-1$ helicity respectively. The form of the polarisation vectors follows from the fact that they should be orthogonal to the corresponding momentum. Indeed, 
\begin{align}
p^{\alpha \dot{\alpha}}e^{+}_{\alpha \dot{\alpha}} = [\tilde{\lambda} \tilde{\lambda}] = 0 =  p^{\alpha \dot{\alpha}}e^{-}_{\alpha \dot{\alpha}} = \langle \lambda \lambda \rangle .
\end{align}
For each particle, the reference spinors $\eta$ and $\tilde{\eta}$ are linearly independent from $\lambda$ and $\tilde{\lambda}$ respectively, but are otherwise arbitrary. Different choices for the reference spinors can alter the polarisation vectors, but only by a gauge transformation, which leaves the amplitude unchanged. We have seen above that for boost-breaking theories we can mix dotted and undotted indices using $(\bar \sigma^{0})^{\alpha \dot{\alpha}}$. This allows us to make choices for the reference spinors for which the zero-component of the polarisation vectors vanishes \cite{Maldacena:2011nz}. In a gauge invariant theory this choice is as good as any other, but if the underlying Lagrangian is constructed out of $SO(3)$ covariant fields only, then this choice is forced upon us since the fields do not have time components. In this paper we are assuming that the fields are Lorentz covariant and so we are not restricted to this choice for the reference spinors. \\

For an $n$-particle scattering amplitude, we have $n$ distinct momenta and therefore $n$ distinct helicity transformation generators $\hat{\mathcal{H}}_i$, corresponding to rotations of a particle around its momentum vector. If we treat all particles as incoming and represent the initial state as $|p; h \ra = | p_1; h_1 \ra \otimes \ldots \otimes | p_n; h_n \ra$, then the $i^{\text{th}}$ helicity generator is represented on the space of initial states as $\hat{H}_i = id \otimes id \otimes \ldots \otimes \hat{\mathcal{H}}_i \otimes \ldots id$, and we have $\hat{H}_i | p; h \ra = h_i |p; h \ra$. The amplitude itself must transform under $\hat{H}_i$ in the same way the initial state does, i.e.
\begin{equation}
\hat{H}_i \mathcal{A}_n(p;h) = h_i \mathcal{A}_n(p;h),
\end{equation}
which in turn implies that under $\{ \lambda^{(i)}, \tilde{\lambda}^{(i)}  \} \mapsto \{ t_i^{-1} \lambda^{(i)}, t_i \tilde{\lambda}^{(i)} \}$ the amplitude transforms as
\begin{equation}
\mathcal{A}_n(\{ \lambda^{(i)}, \tilde{\lambda}^{(i)}; h_i  \}) \mapsto \mathcal{A}_n(\{ t_i^{-1}\lambda^{(i)},  t_i \tilde{\lambda}^{(i)}; h_i  \}) = \prod t_i^{2h_i}\mathcal{A}_n(\{ \lambda^{(i)}, \tilde{\lambda}^{(i)}; h_i  \}) .
\label{eq:HelicityScaling}
\end{equation}
This little group scaling of the amplitude can very powerfully constrain the allowed structure of the amplitude, see e.g. \cite{Benincasa:2013, TASI}. For boost-invariant theories it completely fixes the non-perturbative form of the three-particle amplitudes, while in boost-breaking theories it completely fixes the amplitude up to an arbitrary function of the energies of the three particles, as we shall see in section \ref{sec:3p}.

\subsection{Unitarity, analyticity and the four-particle test}

Analytic properties of the S-matrix have been extensively studied in boost-invariant theories. Analyticity, the singularity structure and crossing symmetry of amplitudes are very important aspects of the S-matrix bootstrap. In this paper we rely on the possibility of extending these essential S-matrix properties to a more general setting and so here we outline why these properties do not require the theory to be invariant under the full Poincar\'{e} group. \\

Let us start with \textit{analyticity} of the S-matrix. By \textit{analyticity}, we mean that once the S-matrix is stripped of the momentum conserving delta function, the remaining factor, when continued into the complex space, is an analytic function of the kinematic variables, except for a finite number of singularities and (possibly) branch cuts. In this paper we will be considering tree level exchange for four-particle amplitudes and so will not encounter any branch cuts. Our three-particle amplitudes are however non-perturbative and are almost completely fixed by symmetry. An argument for analyticity (away from singularities, which are going to be discussed shortly), which does not rely on the invariance of physics under boosts was presented in \cite{White:Overview} and so we will take it for granted that scattering amplitudes are (locally) analytic functions of the kinematic variables discussed above. Our amplitudes will also be crossing symmetric. Crossing symmetry \cite{PeskinSch} is a symmetry of the S-matrix under the following transformation: for a given particle of momentum $p_{\mu}$ in the final state, consider instead its own antiparticle with momentum $-p_{\mu}$ in the initial state. The S-matrix, understood as an analytic function of the complex energies and momenta, must not change under such a transformation. Thus, without loss of generality, we will consider all particles participating in a given process as \textit{incoming} (an incoming particle with negative energy is to be interpreted as an outgoing antiparticle). \\

The most powerful constraint on effective theories and their interactions will come from the singularity structure of the S-matrix. The factorisation theorem, following from locality and unitarity, states that

\begin{figure}
\centering
\includegraphics[width = 6 cm]{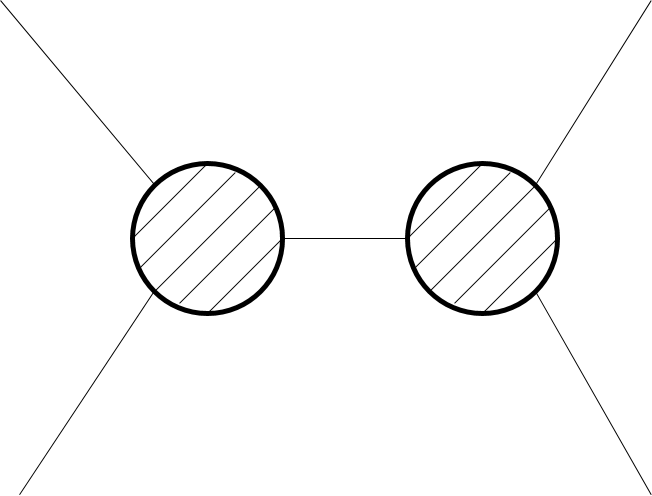}
\caption{Exchange diagram. Circles represent non-perturbative, exact 3-particle amplitudes.}
\label{fig:FD}
\end{figure}

\begin{theorem} (Factorization Theorem) \label{FactorizationThm}
Singularities of codimension $1$ in $4$-particle amplitudes may appear at vanishing energies ($E_i = 0$) or else are at most simple poles in the momenta. Each singularity of the latter type is in one-to-one correspondence with an exchange diagram (Fig. \ref{fig:FD}), in the limit when the exchanged particle $I$ goes \textbf{on-shell}. The residue of each pole factorises into a product of three-particle amplitudes:
\begin{equation}
\lim_{s=0} \left( s \mathcal{A}_4 \right) = \mathcal{A}_3(1,2, -I) \times \mathcal{A}_3(3,4, I)
\end{equation}
where $s$ is the propagator of the intermediate particle, and $s \to 0$ corresponds to the intermediate particle going on-shell.
\end{theorem}

While the above result is almost trivial in perturbation theory and its intuitive physical meaning is not hard to grasp, it can also be demonstrated with mathematical rigour. Starting from the Weak Causality Postulate (\textit{If initial state consists of wave packets colliding at time $t_1$ and the final state consists of wave packets colliding at time $t_2$, and $t_1 - t_2$ is much larger than the typical spatial width of the wave packets, then the scattering amplitude should be small\footnote{More rigorously \cite{Chandler:1969}: scattering amplitude should decay faster than any power of $\Delta t = t_1 - t_2$ as $\Delta t \to \infty$.}}) and by considering wave packets sharply localized in momentum space, Peres \cite{Peres} has shown that the existence of an interacting particle of mass $M \neq 0$ leads to a contribution $A_1 A_2 / (E_I^2 - p_I^2 - M^2 + i \epsilon )$, which is to be identified with processes that involve two collisions of the wave packets (with amplitudes $A_1$ and $A_2$ respectively) separated by a macroscopic time interval. Conversely, if the amplitude in the vicinity of a pole takes the form $A_1 A_2  / (E_I^2 - p_I^2 - M^2 + i \epsilon ) + $ \textit{regular terms}, then the first term represents the amplitude for scattering of wave packets through two or more subsequent collisions, which will be non-negligible provided that the $4$-vector connecting the collisions is approximately parallel to the $4$-momentum $(E_I, p_I)$. This is then interpreted as a propagating particle of mass $M$. The argument of \cite{Peres} does not rely on invariance under boosts\footnote{Although the author does fix Lorentz frame to the center of mass frame, this convenient trick serves illustrative and pedagogical purposes only and can be eliminated altogether.} and can be easily generalized to on-shell conditions of the form $E^2 - \omega^2(p) = 0$, provided there is a mass gap. Other derivations of factorisation, which do not rely on invariance under Lorentz boosts and emphasise the important role of unitarity, can be found in \cite{OliveDiagrams} and Section 10.2 of \cite{Weinberg:1995}. See also \cite{Schwartz:2013pla} for further discussions\footnote{While, strictly speaking, there is no rigorous proof of the Factorization Theorem for massless particles, Feynman rules entail that tree-level diagrams in perturbation theory retain the stipulated property. Moreover, there is no known counterexample to the Factorization Theorem for massless particles. With this in mind, we will follow the many papers we have mentioned previously in the context of this theorem and assume that the theorem holds for massless theories.}. \\

None of the above proofs can on its own exclude the possibility that the poles corresponding to an intermediate particle going on-shell have order higher than $1$. For this we need an additional argument: consider an exchange channel which, according to the Factorization Theorem, leads to a contribution $A_1 A_2  / (E_I^2 - p_I^2 - M^2 + i \epsilon ) + $ \textit{regular terms} to the amplitude. We want to show that the first term contains only first order pole in $(E_I^2 - p_I^2 - M^2 + i \epsilon )$. The essential observation is that if it contained a higher order pole, then one of the three-particle amplitudes, $A_1$ or $A_2$, would have to be singular on some large subset of the $s=0$ hypersurface. But $A_1$ and $A_2$ are three-particle amplitudes in a physical configuration (because the original amplitude could be taken to be in the physical configuration and the intermediate particle is on-shell), so they cannot be singular anywhere. This last statement is also confirmed by an explicit calculation starting from (\ref{eq:3pFinalResult}) - this quantity is finite in a generic configuration.\\

Let us now comment on S-matrix singularities at $E_i =0$. These do not appear in Lorentz invariant theories, as they would clearly violate Lorentz invariance. More generally, such singularities \textit{cannot} appear if the Lagrangian is local and can be written solely in terms of $X_{\mu_1 \mu_2 \ldots}, \eta_{\mu \nu}, \eps_{\mu \nu \sigma \rho}, \partial_{\mu}$ and $\partial_t$ (where $X_{\mu_1 \mu_2 \ldots}$ collectively denotes Lorentz covariant fields). This is because the factor $1/E_i$ is generated only when the some of the tensor field indices are spatial indices. In that case the associated polarization tensor $e^{\pm S}$ has a vanishing temporal component, so it must have a predetermined reference spinor as we eluded to above:
\be
e^{+S}_{\alpha_i \dot{\alpha}_i}(\bvec{k}) =  \prod\limits_{i=1}^{S} \frac{(\eps. \tilde{\lambda})_{\alpha_i} \tilde{\lambda}_{\dot{\alpha}_i}}{2 k}, \quad e^{-S}_{\alpha_i \dot{\alpha}_i}(\bvec{k}) =  \prod\limits_{i=1}^{S} \frac{\tilde{\lambda}_{\alpha_i} (\eps. \lambda)_{\dot{\alpha}_i}}{2k}.
\ee
We see that $e^{\pm s}(\bvec{k})$ has a singularity at $E_k \equiv k = 0$, which might therefore appear also in the helicity amplitude by virtue of the relation
\be
\mathcal{A}_4 = e^{h_1, \mu_1} e^{h_2, \mu_2} e^{h_3, \mu_3} e^{h_4, \mu_4} A_{4, \mu_1 \mu_2 \mu_3 \mu_4},
\ee
where $A_{4, \mu_1 \mu_2 \mu_3 \mu_4}$ is the \textit{covariant amplitude}, which only has singularities when an exchanged particle goes on-shell. As we have explained above, we will be assuming that the Lagrangian is written in terms of Lorentz covariant fields so we don't expect such inverse powers of the energies to arise, but in many cases we see that allowing for these inverse powers does not affect our results.\\ 

Summarizing, four-particle scattering amplitudes in boost-invariant or boost-violating theories have the following singularity structure:
\begin{itemize}
\item The amplitude has only simple poles in the Mandelstam variables $s,t$ and $u$, as well as poles in the individual energies $E_i$.
\item On the $s,t$ and $u$ poles the amplitude factorises into a product of three-particle amplitudes. 
\end{itemize}

These properties form the basis of the \textit{four-particle test} \cite{Benincasa:2007xk}. This test requires the singularity structure of four-particle amplitudes to satisfy these two conditions, and for each pole in $s,t$ or $u$ to be interpreted as the propagation of a physical particle. Ensuring consistency in all three channels ($s$,$t$ and $u$) is highly non-trivial and rules out almost all interactions for massless particles in boost-invariant theories, see \cite{Benincasa:2007xk,TASI,McGadyRodina,Schuster:2008nh,NAH,Benincasa:2011pg}\footnote{The test was originally formulated using BCFW momentum shifts \cite{BCFW}. Indeed, the authors of \cite{Benincasa:2007xk} demanded that two different BCFW shifts gave rise to the same answer for the four-particle amplitudes. As discussed in \cite{TASI,Schuster:2008nh}, the test can actually be formulated as above where only complex factorisation is required.}. The reason why the test is non-trivial is that the residue on say the $s$-channel pole can contain inverse powers of $t$ and $u$, as we shall see. In this paper we will see that the four-particle test is also very constraining when we allow for boost-breaking interactions. \\

We will use the factorization theorem to constrain the constructible part of the tree-level four-particle amplitudes. For this application, it will be sufficient that the tree-level propagator corresponds to a relativistic on-shell condition. If one made the stronger assumption that this is the case also for the full non-perturbative propagator, then one might be able to use our results to derive some constraints on non-perturbative four-particle amplitudes. \\

It should be noted that for massless particles, the $  s\to 0 $ limit of the amplitude makes perfect sense in Minkowski spacetime but this is not the case in curved spacetime. For example, in an FLRW spacetime this limit always takes us outside the validity of the flat-space approximation. Hence, the constraints imposed by Theorem 2.1 apply to flat spacetime but care is required when considering cosmological spacetimes. We discuss this in detail in Section \ref{ssec:pofx}.\\

\section{Three-particle amplitudes} \label{sec:3p}

In this section we construct general on-shell three-particle amplitudes using the spinor helicity techniques outlined in Section \ref{sec:Methods}. Then, as an example, we discuss the cases where all three particles are identical.

\subsection{Non-perturbative structure for all spins} 

We assume that every particle is massless, has a definite helicity, and satisfies the relativistic on-shell condition $p^{\mu}p_{\mu}$ = 0. We take all particles as incoming and therefore by momentum conservation we have 
\begin{align}\label{cons}
p^{\mu}_{1}+p^{\mu}_{2}+p^{\mu}_{3} = 0,
\end{align}
where $1,2,3$ label the external particles. The amplitudes only depend on the observable quantities that can be defined on the asymptotic states and these in turn can be fully recovered from the spinors and helicities $h_i$. The amplitudes are then only a function of $\lambda^{(i)}, \tilde{\lambda}^{(i)}$ and $h_i$. Indeed, written in terms of the spinor helicity variables, \eqref{cons} becomes
\begin{align} \label{MomentumConservation}
\lambda^{(1)}_{\alpha} \tilde{\lambda}^{(1)}_{\dot{\alpha}} + \lambda^{(2)}_{\alpha} \tilde{\lambda}^{(2)}_{\dot{\alpha}} + \lambda^{(3)}_{\alpha} \tilde{\lambda}^{(3)}_{\dot{\alpha}} = 0.
\end{align}
The simple form of this equation is the main reason why computations are considerably simpler when dealing with relativistic on-shell conditions. For any other on-shell condition, such as linear or general, \eqref{MomentumConservation} does not hold and the following analysis needs to be modified. \\

As explained in Section \ref{sec:Methods}, the quantities from which we should construct amplitudes are the three inner products: $\la i j \ra,  [i j],  (i j)$. However, momentum conservation and the fact that each particle is on-shell ensures that any contraction of two distinct momenta is zero. Indeed, 
\begin{align}
(p_{1}+p_{2})^{2} = 2p_{1} \cdot p_{2} = p_{3}^{2} = 0, \\
(p_{2}+p_{3})^{2} = 2p_{2} \cdot p_{3} = p_{1}^{2} = 0, \\
(p_{1}+p_{3})^{2} = 2p_{1} \cdot p_{3} = p_{2}^{2} = 0. 
\end{align}
In the spinor helicity variables this translates into
\begin{align}
\la 12 \ra [12] = \la 13 \ra [13] = \la 23 \ra [23] = 0.  
\end{align}
It follows that if $\la 12 \ra \neq 0$, we have $[12] = 0$ but by momentum conservation we have
\begin{align}
\la 12 \ra [23] = - \la 11 \ra [13] - \la 13 \ra [33] = 0,
\end{align}
and therefore $[23] = 0$ too. We also have $\la 12 \ra [13] = 0$ which requires $[13] = 0$. So having one angle bracket non-zero requires the three square brackets to vanish and vice versa. This tells us that three-particles amplitudes split up into holomorphic and anti-holomorphic configurations:
\begin{align}
\text{Holomorphic kinematics}: ~~~ &[12] = [13] = [23] = 0, \\
\text{Anti-holomorphic kinematics}: ~~~ &\la 12 \ra  = \la 13 \ra = \la 23 \ra = 0.
\end{align}
Furthermore, the off-diagonal components of $(ij)$ are degenerate with other brackets. Indeed for $i \neq j$ we can write
\begin{align}
(ij)\la jk \ra = - (ii)\la ik \ra, \qquad (ij)[ik] = - (jj)[jk],
\end{align}
which allows us to solve for the off-diagonal components of $(ij)$ for both the holomorphic and anti-holomorphic configurations. The brackets we can use to construct amplitudes are therefore $\la ij \ra, [ij]$ for $  i\neq j $ and $(ii)$. Recalling that for the relativistic on-shell condition $(ii) = 2 E_i$, we therefore write the amplitudes as a sum of holomorphic and anti-holomorphic pieces as
\begin{equation}
\mathcal{A}_3(\{ \lambda^{(i)}, \tilde{\lambda}^{(i)}; h_i  \}) = M_H(\la ij \ra, E_i ; h_i)+ M_{AH}([ij], E_i ; h_i).
\end{equation}

We are now in a position to constrain the amplitude by demanding it scales in the correct way under a helicity transformation $(\lambda^{(i)}, \tilde{\lambda}^{(i)}) \mapsto  (t_i^{-1}\lambda^{(i)}, t_{i} \tilde{\lambda}^{(i)})$. As explained in Section \ref{sec:Methods}, under this transformation the amplitude scales as
\begin{equation}
\mathcal{A}_3(\{ t_i^{-1}\lambda^{(i)},  t_i \tilde{\lambda}^{(i)}; h_i  \}) = \prod_{j=1}^{3} t_j^{2h_j}\mathcal{A}_3(\{ \lambda^{(i)}, \tilde{\lambda}^{(i)}; h_i  \}),
\end{equation} 
which constrains the dependence of the angle and square brackets. Note that the diagonal round brackets, or the energies, are invariant under this helicity transformation and so this symmetry does not constrain how they enter the amplitude. First consider $M_H$, which we can write as
\begin{equation} 
M_H(\la ij \ra, E_i ; h_i) = 
\la 1 2 \ra^{d_3} \la 2 3 \ra^{d_1} \la 3 1 \ra^{d_2} F^H_{h_1, h_2, h_3}(E_1, E_2, E_3).
\end{equation}
Demanding the correct scaling of the amplitudes fixes
\begin{align}
d_{1} = h_1 - h_2 - h_3, \\
d_{2} = h_{2} - h_{3} - h_{1}, \\
d_{3} = h_{3}-h_{1}-h_{2}. 
\end{align}
Likewise, for $M_{AH}$ we have
\begin{equation} 
M_{AH}([ij], E_i ; h_i)  = 
[ 1 2 ]^{-d_3} [ 2 3 ]^{-d_1} [ 3 1 ]^{-d_2} F^{AH}_{h_1, h_2, h_3}(E_1, E_2, E_3).
\end{equation} 

Now consider the three cases $h >0$, $h<0$ and $h = 0$ where $h = h_1 + h_2 + h_3$ is the sum of the three helicities. If $h> 0$, we have $d_1 + d_2 + d_3 < 0$ meaning that the $M_{H}$ part of the amplitude would become singular in the entire region defined by $\la ij \ra = 0$ (as long as $F^H \neq 0$ in that region). Three-particle amplitudes cannot have such singularities, so we require $F^H = 0$ whenever $\la ij \ra = 0$. But $F^H$ is just a function of energies, not of the $\la i j \ra$ brackets, and it is impossible to generate these brackets from the energies alone. So in fact when $h> 0$ we require $F^H =M^H = 0$ everywhere. A similar analysis for $h <0$ shows that we require $F^{AH} = M^{AH} = 0$ everywhere. For the third possibility, $h=0$, both contributions to the amplitude can be non-zero. \\

We can also argue this by locality of the interactions. Let us define the mass dimension of an object $A$ by $\dim{A}$ where \textit{we do not include the functions of energy in the mass dimension}. Now since each angle and square bracket has mass dimension $1$, we have $\dim{M_{H}} = -h $ and $\dim{M_{AH}} = h$. The helicity part of the amplitudes cannot have a negative mass dimension as that would require inverse powers of Lorentzian derivatives in the interactions which cannot occur in a local theory. We therefore require $h \leq 0$ for the holomorphic configuration and $h \geq0$ for the anti-holomorphic one. \\

In conclusion, three-particle amplitudes for boost-breaking theories take the general form
\begin{equation}
\mathcal{A}_3(\{ \lambda^{(i)}, \tilde{\lambda}^{(i)}; h_i  \}) = \begin{cases}
\la 1 2 \ra^{ h_{3}-h_{1}-h_{2}} \la 2 3 \ra^{h_{1}-h_{2}-h_{3}} \la 3 1 \ra^{h_{2}-h_{3}-h_{1}}F^H_{h_1, h_2, h_3}(E_1, E_2, E_3),  &h \leq  0,  \\
[ 1 2 ]^{ h_{1}+h_{2} - h_{3}} [ 2 3 ]^{h_{2}+h_{3}-h_{1}} [ 3 1 ]^{h_{3}+h_{1}-h_{2}}F^{AH}_{h_1, h_2, h_3}(E_1, E_2, E_3) , & h  \geq 0.
\end{cases}
\label{eq:3pFinalResult}
\end{equation}
Note that in our convention particles are arranged cyclically in the order $123$, and energy conservation $\sum E_i = 0$ ensures that $F^H$ and $F^{AH}$ can be reduced to functions of two variables only. Thus we will sometimes write 
\begin{align}
F(E_1, E_2)\equiv F(E_1, E_2, E_3 = -E_1 - E_2)\,.
\end{align}
We will also drop the $H/AH$ index unless it is necessary. Qualitatively, therefore, the only difference between the boost-invariant (see \cite{Benincasa:2013,TASI}) and boost-breaking amplitudes is an arbitrary function of the energies that we can add to the latter thanks to the reduced set of symmetries. Our task in Section \ref{sec:4p} will be to constrain these functions using the four-particle test. 
To recover the boost-invariant amplitudes one can simply set $  F^{H,AH} $ to a constant. \\

Before going on to discuss some examples, we first show that the functions $F^{H}$ and $F^{AH}$ are not independent. They are related by a parity transformation (space inversion) $P$, which does not belong to the connected component of the identity of the Lorentz group. The amplitude can either stay the same (scalar) or inherit a minus sign (pseudoscalar) under $P$. The transformation of all the $4$-momenta $(E, \bvec{p}) \mapsto (E, -\bvec{p})$ can be represented in spinor-helicity formalism by transforming the spinors according to\footnote{The presence of a factor of $i$ is due to the requirement that the (+) polarization tensor should be transformed exactly into the (-) polarization tensor under spatial reflection.}
\begin{align}
\lambda_{\alpha}  \mapsto \lambda'_{\alpha} &= (- i \tilde{\lambda}_2, i \tilde{\lambda}_1)\,, &\tilde{\lambda}_{\dot{\alpha}}  \mapsto \tilde{\lambda}'_{\dot{\alpha}} &= (i \lambda_2, - i \lambda_1)\,,
\end{align}
which leads to $[ij] \mapsto - \la i j \ra$ and $\la i j \ra \mapsto - [i j]$. The helicities also change sign under $P$ and so the helicity dependent part of the amplitude transforms as
\begin{align}
[ 1 2 ]^{-d_3} [ 2 3 ]^{-d_1} [ 3 1 ]^{-d_2} \mapsto (-1)^{d} \la 1 2 \ra^{d_3} \la 2 3 \ra^{d_1} \la 3 1 \ra^{d_2},
\end{align}
where $d = d_{1}+d_{2}+d_{3} = -h$, and vice versa. Therefore requiring the amplitude to transform as scalar or psuedoscalar under $P$ fixes 
\begin{equation} \label{eq:ParityTransformation}
F^H_{h_1, h_2, h_3}(E_1, E_2, E_3) = \pm (-1)^{h} F^{AH}_{-h_1, -h_2,  -h_3}(E_1, E_2, E_3),
\end{equation}
with $+$ for a scalar transformation and $-$ for the pseudoscalar. We will therefore often quote results for $F^{H}$ or $F^{AH}$ only.\\

Let us finally emphasise that we have not assumed anything here other than the symmetries of the theory and locality. These amplitudes hold completely non-perturbatively and for any external particles, both bosonic and fermionic\footnote{Fermions always come in pairs and so the exponents are always integers.}.


\subsection{Identical particles: symmetric and alternating polynomials}

As an example, in this subsection we discuss the three-particle amplitudes for identical spin-$S$ particles. Note that the spin-statistic theorem implies that $S$ must be an integer in this case i.e. the particles are bosons. This is clear from \eqref{eq:3pFinalResult} since for fermions each of the brackets has a fractional exponent and therefore when we exchange two fermions the amplitude does not transform into minus itself as it should by Fermi statistics. At the Lagrangian level there is no way to contract the $SU(2)$ indices of three fermions to create a scalar quantity. This is the case for both boost-invariant and boost-breaking theories. \\

There are two fundamentally distinct helicity configurations with either two or three identical helicities. The corresponding amplitudes have mass dimension $S$ and $3S$ respectively and so come from different operators. We can read off the amplitudes from (\ref{eq:3pFinalResult}). First consider the lowest dimension amplitudes $(\pm S,\pm S,\mp S)$ which take the form
\begin{align} \label{LowestDimIden}
\mathcal{A}_3(1^{+S} 2^{+S} 3^{-S}) &= \left( \frac{[12]^3}{[2 3][3 1]} \right)^S F^{AH}_{+S, +S, -S}(E_1, E_2), \\
\mathcal{A}_3(1^{-S} 2^{-S} 3^{+S}) &= \left( \frac{\la12 \ra^3}{\la 2 3\ra \la 3 1\ra} \right)^S F^{H}_{-S, -S, +S}(E_1, E_2),
\end{align}
where we have eliminated $E_{3}$ by energy conservation. Now, since particles $1$ and $2$ have the same helicity and they are bosons, the amplitudes must be invariant under their exchange. The spinor helicity part of these amplitudes inherits a factor of $(-1)^S$ under this transformation and so the functions of energy must be symmetric if the particles have even spin and anti-symmetric if they have odd spin:
\begin{align}
F^{AH}_{+S, +S, -S}(E_1, E_2) &= (-1)^S F^{AH}_{+S, +S, -S}(E_2, E_1), \\
F^{H}_{-S, -S, +S}(E_1, E_2) &= (-1)^S F^{H}_{-S, -S, +S}(E_2, E_1).
\end{align}
To make further progress, we will assume that the functions $F$ are polynomials divided by powers of $E_1, E_2$ and $E_1+E_2$\footnote{The factorisation constraints we derive in Section \ref{sec:4p} will actually hold for more general functions of the energies too. In local theories with covariant fields we would expect no inverse powers of the energies but our results do indeed apply to more general scenarios.}:
\begin{equation} \label{eq:RationalFn}
F(E_1,E_2) = \frac{f(E_1,E_2)}{E_1^{a} E_2^{b} (E_1+E_2)^{c}},
\end{equation}
 It is easy to see that symmetry implies $a = b$ for any spin\footnote{We can naturally assume that $a,b$ and $c$ are minimal. If $a>b$, then we would have $E_2^{a-b} f(E_1, E_2) =\pm E_1^{a-b} f(E_2, E_1) $ and thus $f(E_1, E_2)$ would be divisible by $E_1$, contradicting the assumption that $n$ was minimal.}. \\
 
Now let us restrict to the case of even $S$ where the functions $f$ are required to be symmetric polynomials. By the fundamental theorem of symmetric polynomials, $f$ can be written purely in terms of \textit{elementary symmetric polynomials}. For $n$ variables, there is a single elementary symmetric polynomial of degree $m$ for all non-negative integers $m \leq n$. If we label the $n$ variables as $x_{1} \ldots x_{n}$ then the degree-$m$ elementary symmetric polynomial is
\begin{align}
e_{m}(x_{1}, \ldots x_{n}) = \sum_{1 \leq j_{1} < j_{2}< \ldots <j_{m} \leq n}x_{j_{1}} \ldots x_{j_{m}}.
\end{align}
For example, for $n=2$ we have
\begin{align}
\{1,  x_{1} + x_{2},  x_{1}x_{2}\}.
\end{align} 
On the other hand, if $S$ is odd, the functions of energy in the numerators should be \textit{alternating} polynomials. An alternating polynomial\footnote{Notice that the only object that is anti-symmetric under all possible permutations is zero. That's why anti-symmetric polynomials don't exist. The non-trivial objects are alternating polynomials, which are symmetric or anti-symmetric depending on the sign of the permutation.} is defined by the property
\begin{align}
\text{Poly}(x_{\sigma(1)},\dots,x_{\sigma(n)})=\sign(\sigma) \text{Poly}(x_{1},\dots,x_{n})\,,
\end{align}
for any permutation $  \sigma $ of the $  n $ variables. All alternating polynomials can be written as the Vandermonde polynomial $  v_{n} $ multiplied by sums and products of any number of elementary symmetric polynomials and numerical coefficients (it's an ideal on the ring of polynomials). The Vandermonde polynomial is defined as
\begin{align}
V_{n}(x_{1},\dots,x_{n})\equiv \prod_{1\leq i< j \leq n} (x_{j}-x_{i})\,,
\end{align}
and it is an alternating polynomial of order $  n(n-1)/2 $. In the case at hand the functions are of two variables ($n=2$) and therefore the relevant Vandermonde polynomial is $V_{2} = E_{1} - E_{2}$. For the above amplitudes we therefore have
\begin{align}\label{eq:N1vector}
f_{+S,+S,-S} = \left\{ 
\begin{array}{ll}  \text{Poly}(E_{1}+E_{2},E_{1}E_{2}) &\quad \text{for $  S $ even,} \\
(E_{1}-E_{2}) \text{Poly}(E_{1}+E_{2},E_{1}E_{2}) & \quad \text{for $  S $ odd,}  
 \end{array}
 \right.
\end{align}
and similarly for $f_{-S,-S,+S}$. \\

The remaining two three-particle amplitudes have mass dimension $3S$ and take the form
\begin{align} \label{HighestDim}
\mathcal{A}_3(1^{+S} 2^{+S} 3^{+S}) &= \left([12][23][31]\right)^S F^{AH}_{+S, +S, +S}(E_1, E_2,E_{3}), \\
\mathcal{A}_3(1^{-S} 2^{-S} 3^{-S}) &= \left( \la12 \ra \la 2 3\ra \la 3 1\ra \right)^S F^{H}_{-S, -S, -S}(E_1, E_2,E_{3}).
\end{align}
Now the amplitudes need to be invariant under the exchange of any two external particles as they all have the same helicity. Thus, in \ref{eq:RationalFn} we require $a=b=c$. For even $S$ the functions $ f $ must be symmetric polynomials, meaning that they are constructed out of the elementary symmetric polynomials with $n=3$, namely
\begin{align}
\{1,  x_{1} + x_{2}+x_{3},  x_{1}x_{2}+x_{2}x_{3}+x_{1}x_{3}, x_{1}x_{2}x_{3}\}.
\end{align} 
For odd $S$ the functions are constructed from these elementary symmetric polynomials multiplied by the order $3$ alternating polynomial $  V_{3} $. We therefore have
\begin{align}\label{N1vector}
f_{+S,+S,+S} = \left\{ 
\begin{array}{ll}  \text{Poly}(E_{1}E_{2}+E_{1}E_{3}+E_{2}E_{3},E_{1}E_{2}E_{3}) &\quad \text{for $S$ even,} \\
 V_{3}\left( E_{1},E_{2},E_{3} \right) \text{Poly}(E_{1}E_{2}+E_{1}E_{3}+E_{2}E_{3},E_{1}E_{2}E_{3}) & \quad \text{for $S$ odd,}  
 \end{array}
 \right.
\end{align}
and similarly for $f_{-S,-S,-S}$. Note that for $n=3$ we have $ E_{1}+E_{2}+E_{3} = 0$ since we are constructing on-shell amplitudes. So there are only two non-trivial elementary symmetric polynomials. Here we did not eliminate $E_{3}$ using energy conservation, so as to ensure that the permutation invariance of $F_{+S,+S,+S}$ remains manifest.


\subsubsection*{Scalar} 

If the identical particles are three scalars, i.e. $S=0$, then the amplitude is simply a function of the energies:
\begin{align}
\mathcal{A}_3(1^{0} 2^{0} 3^{0}) &= F_{0,0,0}(E_1, E_2,E_{3}).
\end{align} 
The helicity part of the amplitude disappears because scalars transform in a trivial way. In the boost-invariant case the amplitude is just a constant $  F_{0,0,0}= $ const. 

\subsubsection*{Photon}
For identical $S=1$ particles, each of the four amplitudes presented above requires the functions of energy $  F_{\pm 1,\pm 1,\mp 1} $ and $  F_{\pm 1, \pm 1, \pm1} $ to be alternating polynomials, possibly divided by powers of $E_1 E_2$ and $(E_1 + E_2)$. This rules out the possibility of three-particle amplitudes for a photon in a boost-invariant theory, since a constant polynomial cannot be alternating. More generally any odd number of photons cannot self-interact. This well-known fact can be understood at the level of a Lagrangian where three-particle interactions for a single massless vector should be invariant under the $U(1)$ gauge symmetry $A_{\mu} \rightarrow A_{\mu} + \partial_{\mu} \Lambda(x)$. The building block of invariant Lagrangians is the field strength $F_{\mu\nu} = \partial_{\mu}A_{\nu} - \partial_{\nu}A_{\mu}$ with the indices contracted with $\eta^{\mu\nu}$ or $\epsilon^{\mu\nu\rho\sigma}$ to produce a Lorentz scalar. Three-particle vertices therefore contain at least three derivatives and so the mass dimension of the three-particle amplitudes is $\dim{A_{3}} \geq 3$. This is the Lagrangian reason why the $(\pm1,\pm1,\mp1)$ amplitudes vanish since they have mass dimension $1$. For the $(\pm1,\pm1,\pm1)$ amplitudes we can try to contract three powers of the field strength. However, all Lorentz scalars cubic in the fields, e.g. $F^{\mu}{}_{\nu}F^{\nu}{}_{\rho}F^{\rho}{}_{\mu}$, $\epsilon^{\mu\nu\rho\sigma}F_{\mu\nu}F_{\rho \kappa}F_{\sigma}{}^{\kappa}$, vanish by symmetry\footnote{We can write down non-zero gauge invariant operators at quartic or higher order in the field strength, which describe the interaction of an \textit{even} number of photons. Such terms appear in the Euler-Heisenberg Lagrangian, an effective description of QED below the mass of the electron.}. This Lagrangian interpretation requires us to jump through a few hoops, most notably the introduction of a gauge redundancy to remove the additional degrees of freedom required to write down a manifestly Lorentz invariant and local Lagrangian. The on-shell approach where such redundancies are not required is clearly more efficient and elegant. \\

In a boost-breaking theory, we can use alternating polynomials in energies to ensure that each of the four three-particle amplitudes have the correct Bose symmetry. It is interesting that we can write down an amplitude of this form even though it has no boost-invariant counterpart. But one must first check if these amplitudes pass the four-particle test before declaring that such a theory is consistent (within our assumptions). \\

\subsubsection*{Graviton and higher spins}

For identical particles with $S \geq 2$ and $  S $ even, we can write down three-particle amplitudes in both boost-invariant and boost-breaking theories, while for particles with $  S $ odd we can only write down such amplitudes in a boost-breaking theory, just like for $S=1$. Note that the graviton helicity amplitudes are literally the square of the photon amplitudes. When we allow for multiple spin-$1$ particles, where Bose symmetry in boost-invariant theories is satisfied thanks to the anti-symmetric couplings (the structure constants), the structure of the amplitude is unchanged up to the addition of some colour indices. This simple observation is one of the reasons for the symbolic expression ``$\text{GR} = \text{YM}^{2}$'' \cite{DoubleCopy}. \\


\section{Four-particle amplitudes and the four-particle test} \label{sec:4p}

Having constructed general, non-perturbative three-particle amplitudes, we are now in the position to constrain the almost arbitrary functions of energy using the four-particle test. As explained in Section \ref{sec:Methods}, tree-level four-particle amplitudes contain poles and regular pieces. The latter correspond to contact diagrams while the former come from particle exchange illustrated in Figure \ref{fig:FD}. When the exchanged particle is taken on-shell, the amplitude approaches a singularity whose residue should factorise into a product of three-particle amplitudes. We use this feature to bootstrap consistent four-particle amplitudes due to exchange diagrams in boost-breaking theories. This bootstrap does not constrain the regular parts of the four-particle amplitude; we are constraining the singularity structure of four-particle amplitudes and therefore the cubic couplings in the process.\\

\begin{figure}[h]
    \centering
    \begin{subfigure}[t]{0.36\textwidth}
        \centering
        \vspace{-5cm}
        \includegraphics[width = 6 cm]{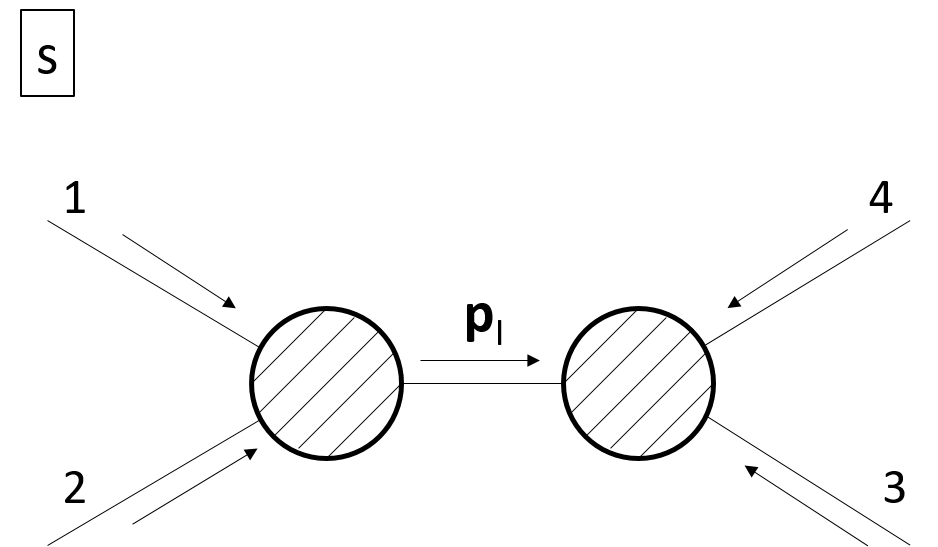}
\small
\label{fig:SCH}
    \end{subfigure}%
~
\begin{subfigure}[t]{0.3\textwidth}
        \centering
        \includegraphics[width=5 cm]{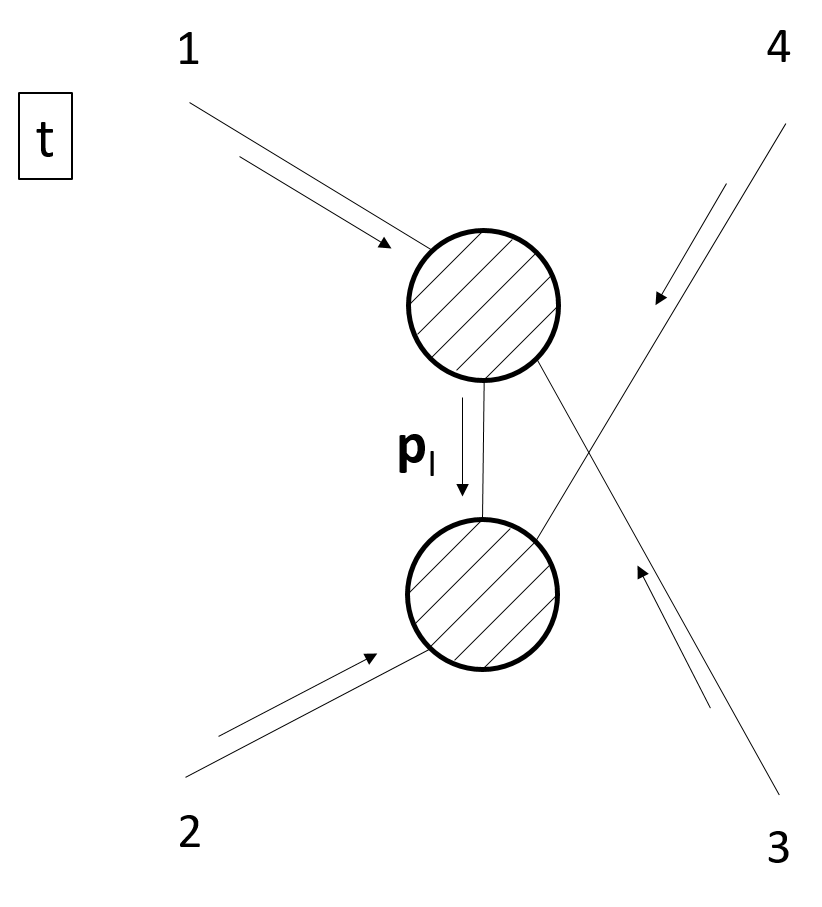}
\label{fig:TCH}
\normalsize
    \end{subfigure}
    ~
\begin{subfigure}[t]{0.3\textwidth}
        \centering
        \includegraphics[width=5 cm]{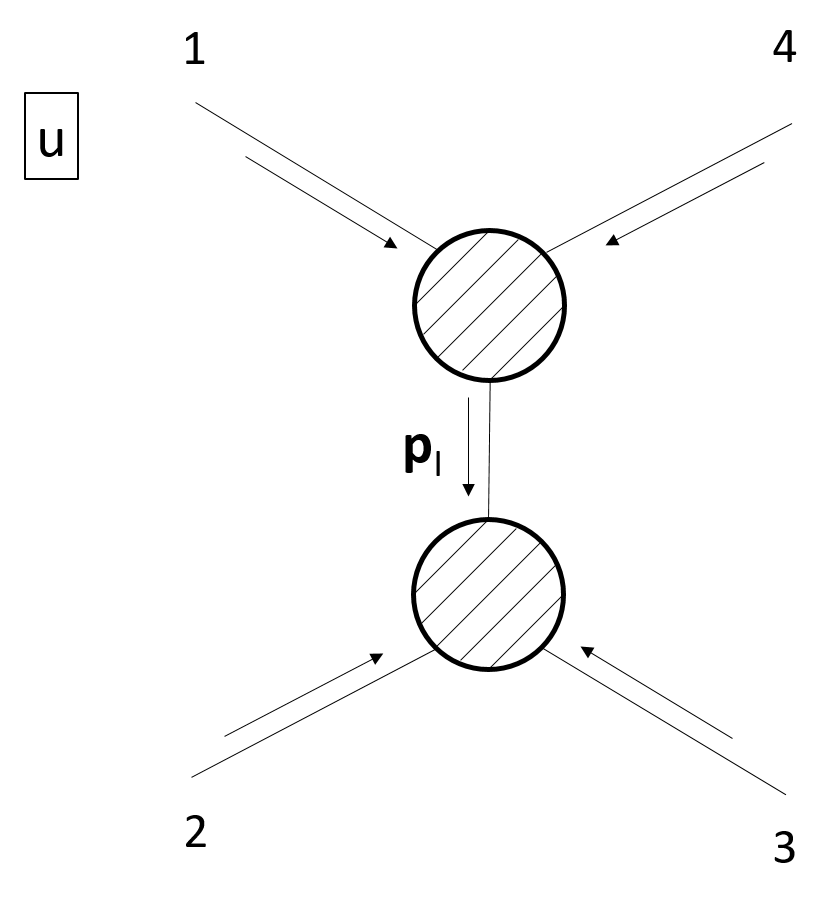}
\label{fig:UCH}
\normalsize
    \end{subfigure}
    \caption{$s$, $t$ and $u$-channel exchange diagrams, respectively.}
    \label{fig:STU-CH}
\end{figure}

To illustrate the idea behind this approach, we may first consider a naive attempt at writing down a four-particle amplitude that factorises into three-particle amplitudes. We have three channels, shown in figure \ref{fig:STU-CH}, and so one could initially allow for three separate terms with an order one pole in $s$, $t$ or $u$ as follows\footnote{We remind the reader that we are working with relativistic dispersion relations for each particle meaning that we only encounter poles in the usual boost-invariant Mandelstam variables.}
\begin{equation} \label{eq:NaiveFactorization}
\mathcal{A}_4 \stackrel{?}{=}  \frac{ \mathcal{A}_3(1,2, -I) \times \mathcal{A}_3(3,4, I) }{s} + 
\frac{ \mathcal{A}_3(1,3, -I) \times \mathcal{A}_3(2,4, I) }{t} +
\frac{\mathcal{A}_3(1,4, -I) \times \mathcal{A}_3(2,3,I) }{u} 
\end{equation}
where $I$ and $-I$ label the exchanged particle outgoing from the vertex involving particle 1, or incoming into that vertex respectively\footnote{Throughout our analysis in the spinor helicity variables we send $p_{I} \rightarrow -p_{I}$ by $\lambda^{(I)} \rightarrow \lambda^{(I)}$, $\tilde{\lambda}^{(I)} \rightarrow - \tilde{\lambda}^{(I)}$. See Appendix \ref{AppendixSigns} for a justification of this method.}. All external particles are incoming. If more than one intermediate particle is allowed, we need to sum over all the species of $I$. Now it would appear that this amplitude has the residues required by Theorem \ref{FactorizationThm}. However, it is possible that $\mathcal{A}_3(1,2, -I) \times \mathcal{A}_3(3,4, I)$, when analytically continued beyond the loci of $s=0$, has a pole at $t=0$ or $u = 0$. In this case, the first term contributes to the $t = 0$ or $u = 0$ residue and the formula \ref{eq:NaiveFactorization} could give an incorrect residue at $t = 0$. Finding a four-particle amplitude with the correct residues in \textbf{all} three channels is therefore a non-trivial matter. This is known as the four-particle test \cite{Benincasa:2007xk,Schuster:2008nh}, and as we shall see, it allows us to constrain, or altogether eliminate, certain types of cubic interactions in boost-breaking theories. \\


Before we begin, we must identify a set of $SO(3)$-invariant variables that are sufficient to fully determine the on-shell data for the scattering of four particles. In addition to the four external helicities, we must use some of the brackets $\la i j \ra$, $[i j]$ and $(ij)$, which constitute a complete list of invariants of mass dimension $1$. However, not all of these are independent: all but one of the off-diagonal $(ij)$ brackets can be determined in terms of the other brackets and the energies by using momentum conservation.\footnote{We verified this via algebraic manipulation in Mathematica.} Therefore, any $SO(3)$ invariant can be written in terms of $\la i j \ra$, $[i j]$, $E_i$ and just one of the off-diagonal $(ij)$. These variables are still not all independent, but this won't present a problem for us. On the other hand, it must be emphasized that without at least one $(ij)$ bracket we would be unable to fully determine the kinematic data in the general case. This means that in boost-breaking theories, four-particle amplitudes could depend on one of the $(ij)$'s and this dependence cannot be eliminated by application of bracket identities. \\

There is a special class of Lagrangians for which four-particle amplitudes are functions of $\la i j \ra$, $[i j]$ and $E_i$ only. These Lagrangians take the form
\be \label{eq:SemiCovLagr}
\mathcal{L} = \mathcal{L}\left[ X_{\mu_1 \mu_2 \ldots}, \eta_{\mu \nu}, \eps_{\mu \nu \sigma \rho}, \partial_{\mu}, \partial_t \right],
\ee
where $X_{\mu_1 \mu_2 \ldots}$ collectively denotes Lorentz covariant fields. If a physical four-particle amplitude can be written solely in terms of $\la i j \ra$, $[i j]$ and $E_i$, then there exists a Lagrangian of the form (\ref{eq:SemiCovLagr}) which generates this amplitude. Such a Lagrangian can be constructed as follows: first, write down a Lorentz-invariant Lagrangian that generates the four-particle amplitude with the energy dependence stripped off, and then insert time derivatives acting on appropriate fields to reinstate the desired energy dependence of the amplitude. Suppose, on the other hand, that a four-particle amplitude in some theory cannot be written without at least one round bracket $(ij)$ (which, as we remarked, cannot be determined solely in terms of the $\la i j \ra$, $[i j]$ and $E_i$). Then the corresponding Lagrangian must depend on some objects other than the ones listed in (\ref{eq:SemiCovLagr}). For example, the Larangian could be constructed out of $SO(3)$ covariant fields rather than Lorentz covariant ones.   \\


As an example of the latter kind of theory, let us consider the Framid EFT \cite{Zoology} which arises from the spontaneous breaking of Poincar\'{e} symmetry to an unbroken subgroup of translations and rotations. Indeed, the Framid degrees of freedom are the Goldstone modes of broken Lorentz boosts. With respect to the unbroken $SO(3)$ symmetry, the Framid consistents of three degrees of freedom: a massless tranverse vector and a massless longitudinal scalar with speeds $c_{T}$ and $c_{L}$.  Taking $c_L = c_T$, in which the scalar and vector modes have identical propagation speeds as we have been assuming in this work, the Framid Lagrangian up to cubic order in fields takes the form \cite{Zoology}
\be
\mathcal{L} = \frac{M_{1}^{2}}{2}\left(\dot{\eta}_{i}^2-c_{L}^{2}\partial_{i}\eta_{j} \partial_{i}\eta_{j}  \right) + M_{1}^{2}\left( c_{L}^{2}-1 \right) \eta_{i}  \partial_{i}\eta_{j}  \dot{\eta}_{j} + \mathcal{O}\left( \eta^4 \right)\, . \label{eq:FramidLagrangian0}
\ee
After defining rescaled fields $\chi_i = c_L M_1 \eta_i$ and replacing $t$ with the rescaled time coordinate $t' = t/c_L$, we obtain
\be
\mathcal{L} = \frac{1}{2}\left(  \dot{\vec \chi}^{2} - \partial_{i}\chi_{j} \partial_{i}\chi_{j}  \right) + \left( \frac{c_{L}^{2}-1}{c_L^2 M_1} \right) \chi_{i}  \partial_{i}\chi_{j}  \dot{\chi}_{j} + \mathcal{O}\left( \chi^4 \right)\, . \label{eq:FramidLagrangian}
\ee
Using the above Lagrangian (and rescaled coordinates), we computed the four-particle amplitude $\mathcal{A}_4(1^0 2^+ 3^0 4^-)$ from tree-level exchange to verify and illustrate that it has an explicit dependence on one of the off-diagonal $(ij)$, which cannot be eliminated. The result is the simplest, albeit still quite lengthy, if we allow for the dependence on $(42)$, in which case the amplitude reads as follows:
\be  \nn
\mathcal{A}_4(1^0 2^+ 3^0 4^-) & = & \frac{1}{4 e_4} \left( \frac{c_{L}^{2}-1}{c_L^2 M_1} \right)^2 \times \\ \nn
& \quad & \times \Big\{ \frac{1}{s} \left[ 
F_{(1,a)}(E_1, E_2, E_3, E_4; s, t) (42)^2 \right. \\ \nn
& \quad &  \quad  \quad  \quad + F_{(1,b)}(E_1, E_2, E_3, E_4; s, t) [23] \la 3 4 \ra (42) \\ \nn
& \quad & \left.  \quad  \quad  \quad + F_{(1,c)}(E_1, E_2, E_3, E_4; s, t) [23]^2 \la 3 4 \ra^2
\right]  \\ \nn
& \quad & +  \frac{1}{t} \left[ 
F_{(2,a)}(E_1, E_2, E_3, E_4; s,t) (42)^2 \right. \\ \nn
& \quad & \left.  \quad  \quad  \quad  \quad
+ F_{(2,b)}(E_1, E_2, E_3, E_4; s, t) [23] \la 3 4 \ra (42) \right] \\ \nn
& \quad & + \frac{1}{u} \left[ 
F_{(1,a)}(E_3, E_2, E_1, E_4; u, t) (42)^2 \right. \\ \nn
& \quad &  \quad  \quad  \quad  \quad - F_{(1,b)}(E_3, E_2, E_1, E_4; u, t) [23] \la 3 4 \ra (42)  \\
& \quad & \left. \quad  \quad  \quad   \quad + F_{(1,c)}(E_3, E_2, E_1, E_4; u, t) [23]^2 \la 3 4 \ra^2
\right] \Big\}, \label{eq:Framid4pt}
\ee
where functions $F_{(i, x)}$ are defined in Appendix \ref{AppFramid} and $e_4 \equiv E_1 E_2 E_3 E_4$.  \\

Since an ansatz that depends on round brackets would be too general to be constrained effectively, to make progress we will assume that four-particle amplitudes take the form
\be
\mathcal{A}_4 = \mathcal{A}_4 \left(\la i j \ra, [i j], s,t,u, E_i \right),
\ee
meaning that the underlying Lagrangians take the form of \eqref{eq:SemiCovLagr}. For more general Lagrangians, we would have to allow for the presence of $(42)$ (or some other off-diagonal $(ij)$) in the four-particle amplitude. We plan to come back to this in the future.

\subsection{Single spin-$S$ particle} \label{sec:4pIdenticalSpinS}

We begin by constraining the lowest dimension three-particle amplitudes for identical spin-$S$ bosons presented in \eqref{LowestDimIden}, namely the $(\pm S,\pm S,\mp S)$ amplitudes. Consider the four-particle amplitude $\mathcal{A}_{4}(1^{-S} 2^{+S} 3^{-S} 4^{+S})$ due to exchange of the spin-$S$ particle. By little group scaling we can fix the helicity part of the amplitude leaving only the dependence on the little group invariants $(s,t,u,E_{i})$ left to fix by the four-particle test. The amplitude takes the general form 
\begin{align}
\mathcal{A}_{4}(1^{-S} 2^{+S} 3^{-S} 4^{+S}) = \la 13 \ra^{2S} [24]^{2S} \mathcal{G}(s,t,u,E_{i}),
\end{align}
and its mass dimension (recall that we don't count the explicit energy dependence in the mass dimension) is
\begin{align} \label{MassDim1}
\dim{\mathcal{A}_{4}} = 4S + \dim{\mathcal{G}}.
\end{align}
Now for exchanges in the $s$ and $u$ channels both constituent three-particle amplitudes have mass dimension $S$ and this can also be achieved in the $t$ channel for one of the two possible helicity configurations of the exchanged particle. Since factorisation requires $\lim_{s \rightarrow 0} (sA_{4}) = A_{3} \times A_{3}$, for the case at hand the mass dimension of the four-particle amplitude is 
\begin{align}  \label{MassDim2}
\dim{\mathcal{A}_{4}} = 2S - 2.
\end{align}
By equating \eqref{MassDim1} and \eqref{MassDim2} we find that the mass dimension of $\mathcal{G}$ satisfies
\begin{align}
\dim{\mathcal{G}} = -2S-2.
\end{align}
However, locality dictates that the amplitude can only contain simple poles in $s,t$ and $u$ and so we require $\dim{\mathcal{G}} \geq -6$ yielding the constraint
\begin{align}
S \leq 2.
\end{align}
This tell us that the above four-particle amplitude is inconsistent for bosonic particles with $S \geq 3$, even in boost-breaking theories. We require the corresponding $(\pm S,\pm S,\mp S)$ amplitudes to vanish, so we set $F_{-S,-S,+S} = F_{+S,+S,-S} = 0$ for $S \geq 3$. This very simple argument leads to a profound result: massless, higher spinning particles cannot have low-energy cubic self-interactions (under the assumption that the underlying Lagrangian is written in terms of covariant fields). \\

\begin{figure}[h!]
\centering
\includegraphics[width = 8 cm]{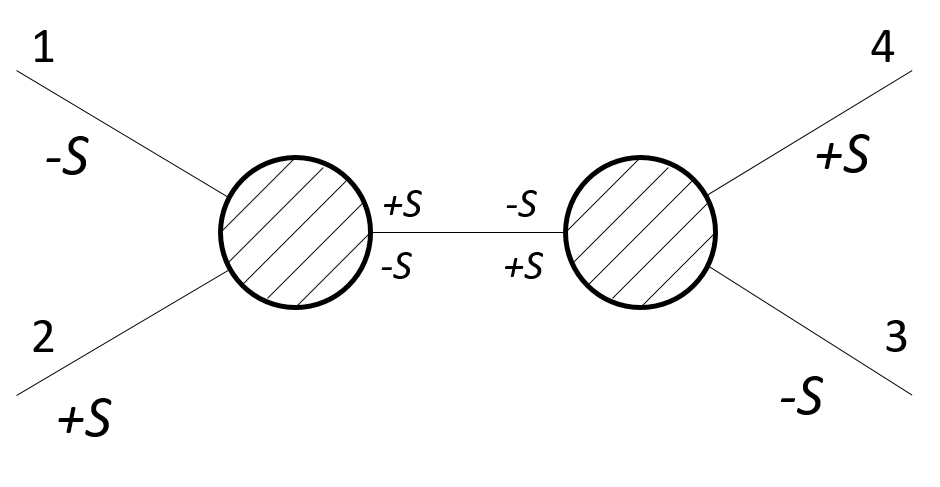}
\caption{Two choices for the helicity configuration of the exchanged particle.}
\label{fig:SCHExchange}
\end{figure}

Let us consider this amplitude in more detail for $S=0,1,2$ where dimensional analysis did not exclude the possibility of consistent factorization. In the $s$ and $u$ channels there are two distinct diagrams since we have two choices for the helicity configuration of the exchanged particle (see Figure \ref{fig:SCHExchange} for the two $s$-channel possibilities). In the $t$ channel there is only one diagram. We therefore have two residues to compute in the $s$ and $u$ channels and we label these as $R_{s}^{-+}, R_{s}^{+-}$ and $R_{u}^{-+}, R_{u}^{+-}$.
Using the three-particle amplitudes \eqref{LowestDimIden} the residue on the $s=0$ pole is
\begin{eqnarray} 
R_s =&& R_s^{-+} + R_s^{+-} \label{eq:Spole}  \\ =  &&   \left( \frac{\la I 1 \ra^3}{\la 1 2 \ra \la 2 I \ra} \right)^S \left(\frac{[4 I]^3}{[I 3] [3 4]} \right)^S F_{-S,-S,+S}(-E_1 - E_2, E_1) F_{+S,+S,-S}(E_4, -E_3 - E_4) \nonumber  \\ 
+&&
\left( \frac{[ 2 I ]^3}{[ I 1 ] [ 1 2 ] } \right)^S \left(\frac{\la I 3 \ra^3}{ \la 3 4 \ra \la 4 I \ra} \right)^S 
 F_{+S,+S,-S}(E_2, -E_1 - E_2) F_{-S,-S,+S}(-E_3 - E_4, E_3),\nonumber
\end{eqnarray}
where we have used energy conservation to eliminate $E_{I}$. Now in the spinor helicity variables there is not a unique way to approach $s=0$. We have $s = \la 1 2 \ra [12] = \la 34 \ra [34] = 0$ and this has two main solutions. If $[12] = 0$, then by momentum conservation we have $0 = [12] \la 23 \ra = [14] \la 34 \ra$ and so to avoid imposing additional constraints on the kinematics we have to choose $\la 3 4 \ra = 0$. Similarly, if $\la 12 \ra = 0 $, then $[34] = 0$ too.\\

For $[12] = \la 3 4 \ra = 0$, the second term in \eqref{eq:Spole} vanishes\footnote{Once we eliminate $I$ from all brackets, one sees that the numerator vanishes faster than the denominator.} leaving
\begin{align} 
R_s = R_s^{-+}  = &  \left( \frac{\la I 1 \ra^3}{\la 1 2 \ra \la 2 I \ra} \right)^S \left(\frac{[4 I]^3}{[I 3] [3 4]} \right)^S F_{-S,-S,+S}(-E_1 - E_2, E_1) F_{+S,+S,-S}(E_4, -E_3 - E_4) \nonumber \\
 = & \frac{(\la 1 3 \ra^2 [2 4]^2 )^S}{t^{S}} F_{-S,-S,+S}(-E_1 - E_2, E_1) F_{+S,+S,-S}(E_4, -E_3 - E_4),
\end{align}
where using conservation of momentum at each vertex we eliminated all factors of $I$, for example, $\la 1 I \ra [I 4] = \la 12 \ra [24]$. For $\la 1 2 \ra = [3 4] =  0$ the first term vanishes leaving 
\begin{align} 
R_s = R_s^{+-}  = &  \left( \frac{[ 2 I ]^3}{[ I 1 ] [ 1 2 ] } \right)^S \left(\frac{\la I 3 \ra^3}{ \la 3 4 \ra \la 4 I \ra} \right)^S 
 F_{+S,+S,-S}(E_2, -E_1 - E_2) F_{-S,-S,+S}(-E_3 - E_4, E_3) \nonumber \\ 
 = &  \frac{(\la 1 3 \ra^2 [2 4]^2 )^S}{t^{S}}  F_{+S,+S,-S}(E_2, -E_1 - E_2) F_{-S,-S,+S}(-E_3 - E_4, E_3).
\end{align}
Again we see how $S \geq 3$ amplitudes are ruled out: for $S \geq 3$, the $s$-channel residue contains higher order poles when $t=0$ and so the corresponding amplitude is inconsistent. One may also think that $S=2$ is problematic since the denominator is quadratic in $t$. However, when $s=0$ we can write $t^{2} = -tu$.  Before moving on to the other channels, we note that the residue in the $s$-channel should not differ if we approach the pole in two different ways and so we match the two different expressions for $R_{s}$ yielding our first constraint on the three-particle amplitudes\footnote{Here is a brief justification. Near $s = 0$, the schematic form of the amplitude is $A \sim s^{-1} (f_1(\lambda) F_1(E) + f_2(\lambda) F_2(E))$, where $f_i$ are functions of the Lorentz invariants and $F_i$ are functions of the energies only. The amplitude has the same dependence on the Lorentz invariants in the two limits, which can then differ only by a function of energies. Hence, we can write $A \sim s^{-1} f(\lambda) F(E)$. Since we can take either of the limits $\la 1 2 \ra \to 0$ or $[1 2] \to 0$ while keeping the energies fixed, we must get the same $F(E)$, which is to be identified with the energy-dependent functions in the main text.}:
\begin{align} \label{MatchingResidues}
F_{-S,-S,+S}(-E_1 - E_2, E_1) F_{+S,+S,-S}(E_4, -E_3 - E_4) \nonumber \\ = F_{+S,+S,-S}(E_2, -E_1 - E_2) F_{-S,-S,+S}(-E_3 - E_4, E_3).
\end{align}
In the boost-invariant limit the two residues are trivially the same. \\

The $u$-channel also contains two diagrams and the corresponding residues can easily be obtained from the $s$-channel ones by interchanging particles $2$ and $4$. With \eqref{MatchingResidues} imposed the two residues are equivalent. We have, for example,
\begin{equation} 
R_u = R_u^{-+} = \frac{(\la 1 3 \ra^2 [2 4]^2 )^S}{t^{S}} F_{+S,+S,-S}(E_4, -E_1 - E_4) F_{-S,-S,+S}(-E_3 - E_2, E_3) .
\end{equation}
Finally, the $t$-channel is qualitatively different since it involves two particles of the same helicity on each side of the diagram. There is therefore only a single choice for the exchange particle's helicity if this contribution to the amplitude is to have the same mass dimension as the other channels. The residue is
\begin{eqnarray} \nonumber
R_t & = &  \left( \frac{\la 1 3 \ra^3}{\la 3 I \ra \la I 1 \ra} \right)^S \left(\frac{[2 4]^3}{[4 I] [I  2]} \right)^S F_{-S,-S,+S}(E_1, E_3) F_{+S,+S,-S}(E_2, E_4) \\
&  =  & \frac{(\la 1 3 \ra^2 [2 4]^2 )^S}{s^{S}} F_{-S,-S,+S}(E_1, E_3) F_{+S,+S,-S}(E_2, E_4) .
\end{eqnarray}

In summary, the residues are 
\begin{eqnarray} \label{eq:SpinSs}
R_s & = & \frac{(\la 1 3 \ra^2 [2 4]^2 )^S}{t^{S}} F_{-S,-S,+S}(-E_1 - E_2, E_1) F_{+S,+S,-S}(E_4, -E_3 - E_4), \\ \label{eq:SpinSt}
R_t & = & \frac{(\la 1 3 \ra^2 [2 4]^2 )^S}{s^{S}} F_{-S,-S,+S}(E_1, E_3) F_{+S,+S,-S}(E_2, E_4), \\  \label{eq:SpinSu}
R_u & = & \frac{(\la 1 3 \ra^2 [2 4]^2 )^S}{t^{S}} F_{+S,+S,-S}(E_4, -E_1 - E_4) F_{-S,-S,+S}(-E_3 - E_2, E_3),
\end{eqnarray}
and are subject to \eqref{MatchingResidues}. Let us now zoom in on the three different allowed values for $S$. \\


\subsubsection*{Scalar} \label{sec:scalar}
For a single scalar, $S=0$, consistent factorisation is trivial. Indeed, each residue is simply a function of the energies and does not contain spurious poles. The consistent four-particle amplitude is
\begin{align}
\mathcal{A}_4(1^{0},2^{0},3^{0},4^{0}) &= \frac{F(-E_1 - E_2, E_1) F(E_4, -E_3 - E_4)}{s} \nonumber \\ &+ \frac{F(E_1, E_3) F(E_2, E_4)}{t} \nonumber \\ &+ \frac{F(E_4, -E_1 - E_4) F(-E_3 - E_2, E_3)}{u},
\end{align}
where $F \equiv F_{0,0,0}$. The only constraint we have on the function of energy is that it should be a symmetric function as explained in Section \ref{sec:Methods}. \\

We can understand this result from a Lagrangian point of view. In the boost-invariant case the three-particle amplitude is a constant with consistent factorisation of the four-particle amplitude for scalar scattering. One may wonder about cubic vertices with derivatives. It is easy to contract the indices in a Lorentz invariant way but these vertices always involve, up to integration by parts, the $\Box = \partial^{\mu}\partial_{\mu}$ operator acting on at least one of the fields and therefore it vanishes on-shell and can be removed by a field redefinition in favour of four-point vertices which only contribute to the regular part of the four-particle amplitude.\\

In the boost-breaking case we write operators using the usual Lorentzian derivative $\partial_{\mu}$, but also have the freedom to add extra time derivatives. Because any terms with Lorentzian derivatives can be removed by a field redefinition, the only non-trivial three scalar vertices have zero derivatives, corresponding to a constant amplitude, or contain time derivatives only giving rise to functions of energy in the amplitude. A well-known example is the $\dot{\phi}^3$ vertex appearing in the flat space, decoupling limit of the EFT of single-field inflation. Generalisations with more derivatives are easy to write down. \\


\subsubsection*{Photon} \label{sec:photon}
For a photon, $S=1$, consistent factorisation becomes a nontrivial problem: $R_{s}$ has a pole when $t = 0$, $R_{t}$ has a pole when $u = 0$, and $R_{u}$ has a pole when $s=0$. Therefore the full amplitude must take the form
\begin{equation}
\mathcal{A}_{4}(1^{-1} 2^{+1} 3^{-1} 4^{+1}) = \la 1 3 \ra^2 [2 4]^2  \left( \frac{A}{st} + \frac{B}{tu} + \frac{C}{us} \right),
\end{equation}
where $A,B$ and $C$ are constrained by
\begin{eqnarray}
R_s & = &  \la 1 3 \ra^2 [2 4]^2  \left( \frac{C-A}{u} \right), \\
R_t & = &  \la 1 3 \ra^2 [2 4]^2  \left( \frac{A-B}{s} \right), \\
R_u & = &  \la 1 3 \ra^2 [2 4]^2  \left( \frac{B-C}{t} \right),
\end{eqnarray}
where again we have used $s+t+u=0$. As explained in Section \ref{sec:3p}, $F_{-1,-1,+1}$ and $F_{+1,+1,-1}$ are proportional with the proportionality factor $\pm$ for parity odd and even theories respectively. Since only their product appears in each residue the following analysis is the same in both cases, so without loss of generality let us take $F = F_{-1,-1,+1} = F_{+1,+1,-1}$. Matching our two expressions for the residues yields 
\begin{eqnarray}
C - A & = & -F(E_2, -E_1 - E_2) F(-E_3 - E_4, E_3), \label{VectorCondition1} \\
A - B & = & F(E_1, E_3) F(E_2, E_4), \label{VectorCondition2} \\
B - C & = & F(E_4, -E_1 - E_4) F(-E_2-E_3, E_3), \label{VectorCondition3}
\end{eqnarray}
with 
\begin{align} \label{eq:SingleVector1}
F(-E_1 - E_2, E_1) F(E_4, -E_3 - E_4)  = F(E_2, -E_1 - E_2) F(-E_3 - E_4, E_3),
\end{align}
such that the residues in the $s$ and $u$ channels are the same regardless of how we approach the pole. Taking the sum of \eqref{VectorCondition1}, \eqref{VectorCondition2} and \eqref{VectorCondition3} yields the main $S=1$ factorisation constraint
\begin{eqnarray} \label{eq:SingleVector2}
&&F(E_2, -E_1 - E_2) F(-E_3 - E_4, E_3) \nonumber \\ &-& F(E_1, E_3) F(E_2, E_4) \nonumber \\
&-& F(E_4, -E_1 - E_4) F(-E_2-E_3, E_3) = 0,
\end{eqnarray}
which must be satisfied for all $E_{i}$ subject to $E_{1}+E_{2}+E_{3}+E_{4} = 0$. \\

Recall from Section \ref{sec:3p} that $F$ must be an alternating polynomial (possibly divided by some powers of energies) such that the three-particle amplitudes have the correct Bose symmetry. Since $F$ is an alternating function of two variables, we can write
\begin{equation}
F(x,y) = \frac{(x - y) P[x + y, xy]}{x^m y^m (x+y)^k},
\end{equation}
with $xy \nmid P[x+y, xy]$ if $m > 0$ and $(x+y) \nmid P[x+y, xy]$ if $k > 0$ ($\nmid$ means ``does not divide''). Writing the factorisation constraint \eqref{eq:SingleVector2} in terms of $P$, we can prove that it requires $P \equiv 0$. The reason for this is that $P$, as we show in Appendix \ref{AppendixA}, has to satisfy infinitely many distinct constraints of the form $P[x, a_k x^2] = 0 ~ \forall x$ and thus we need $(a_k x^2 - y)$ to divide $P[x,y]$ for all the $a_k$, which is impossible if $P$ is a nonzero polynomial. We therefore conclude that the four-particle test requires the $(\pm 1,\pm 1,\mp 1)$ three-particle amplitudes for a single photon in a boost-breaking theory (formulated in terms of covariant fields) to vanish: even when boosts are broken there are no consistent three-point vertices for a single photon giving rise to these lowest dimension amplitudes. Note that this result did not require us to impose the additional constraint \eqref{eq:SingleVector1} from matching the residues. One may wonder if consistent amplitudes are possible if we include additional particles, but we will show in Section \ref{sec:PhotonCouplings} that additional exchanges do not change this result.  \\

In a theory with only a single photon the four-particle test cannot constrain the other three-particle amplitudes, namely those with $(\pm 1,\pm 1,\pm 1)$ helicities since these amplitudes do not contain inverse powers of brackets and therefore residues constructed out of these amplitudes cannot contain poles. These three-particle amplitudes are therefore only constrained by Bose symmetry which for $S=1$ tells us that $F_{-1,-1,-1}$ and $F_{+1,+1,+1}$ are alternating functions in the three energies. Amplitudes of the lowest possible dimension are\footnote{We acknowledge Maria Alegria Gutierrez's findings on the possible structures of $F_{\pm 1, \pm 1, \pm 1}$.}
\begin{align} 
\mathcal{A}_3(1^{-1} 2^{-1} 3^{-1}) &= g \la12 \ra \la 2 3\ra \la 3 1\ra \frac{(E_{1}-E_{2})(E_{2}-E_{3})(E_{1}-E_{3})}{E_1 E_2 E_3}, \label{PhotonAmplitudes1} \\
\mathcal{A}_3(1^{+1} 2^{+1} 3^{+1}) &= \pm g [12][23][31] \frac{(E_{1}-E_{2})(E_{2}-E_{3})(E_{1}-E_{3})}{E_1 E_2 E_3},\label{PhotonAmplitudes2}
\end{align}
while the first amplitudes arising from a $U(1)$ gauge invariant theory are
\begin{align} 
\mathcal{A}_3(1^{-1} 2^{-1} 3^{-1}) &= g' \la12 \ra \la 2 3\ra \la 3 1\ra (E_{1}-E_{2})(E_{2}-E_{3})(E_{1}-E_{3}), \label{PhotonAmplitudesSub1} \\
\mathcal{A}_3(1^{+1} 2^{+1} 3^{+1}) &= \pm g' [12][23][31] (E_{1}-E_{2})(E_{2}-E_{3})(E_{1}-E_{3}),\label{PhotonAmplitudesSub2}
\end{align}
where we allow for parity-even and parity odd possibilities and $g, g'$ are coupling constants. All of these amplitudes are consistent since the four-particle test for photon scattering does not impose any conditions on $(+1,+1,+1)$ and $(-1,-1,-1)$ interactions. \\

Let us briefly comment on the Lagrangian approach to all-plus (and all-minus) amplitudes. Despite the fact that (\ref{PhotonAmplitudes1}) - (\ref{PhotonAmplitudes2}) are allowed by symmetry and the 4p test, they cannot arise from a gauge invariant cubic term. This is because gauge invariance requires us to construct interactions out of the field strength $F_{\mu \nu}$, which already contains three derivatives, but $F^{\mu}_{\ \nu} F^{\nu}_{\ \rho} F^{\rho}_{\ \mu}$ vanishes identically. In contrast, for (\ref{PhotonAmplitudesSub1}) - (\ref{PhotonAmplitudesSub2}) there exists an underlying local Lagrangian which is gauge invariant. By taking boost-invariant interactions and adding time derivatives we find both a parity-even and parity-odd possibility given by \begin{align} \label{VectorConsistentVertex}
\ddot{F}^{\mu}{}_{\nu}\dot{F}^{\nu}{}_{\rho}F^{\rho}{}_{\mu}, \qquad \epsilon^{\mu\nu\rho\sigma}\ddot{F}_{\mu\nu}\dot{F}_{\rho \kappa}F_{\sigma}{}^{\kappa}.
\end{align} 
In Appendix \ref{AppendixB} we show that the latter interaction does indeed give rise to the purported amplitudes (\ref{PhotonAmplitudesSub1}) - (\ref{PhotonAmplitudesSub2}). The calculation for the first interaction is similar. \\

In conclusion, boost-breaking theories of a single photon do exist but any gauge invariant cubic interactions require at least 6 derivatives meaning that its low energy consequences are heavily suppressed. In addition, in Section \ref{sec:GravitonCouplings} we will show that in the presence of gravity these interactions do not pass the four-particle test! \\


\subsubsection*{Graviton} \label{sec:graviton}
The graviton, $S=2$, is the final case to consider. Here we see that each residue contains a pole in the other two Mandelstam variables and so consistent factorisation is non-trivial. This tells us that a four-particle amplitude with consistent factorisation must take the form
\begin{equation}
\mathcal{A}_{4}(1^{-2} 2^{+2} 3^{-2} 4^{+2}) = \la 1 3 \ra^4 [2 4]^4 \frac{A}{stu},
\end{equation}
with the function $A$ constrained by matching to each residue. Our $S=2$ factorisation conditions are 
\begin{eqnarray} \label{eq:SingleGraviton}
-A & = & F(-E_{1}-E_{2},E_{1}) F(E_{4},-E_{3}-E_{4}) \\& = & F(E_1, E_3) F(E_2, E_4), \\ 
& = & F(E_4, -E_1 - E_4) F(-E_3 - E_2,E_{3}),
\end{eqnarray}
where again we have dropped the subscripts denoting the helicities, and cover both parity even and parity odd cases. We also need to satisfy \eqref{MatchingResidues}. \\

In Appendix \ref{AppendixA} we show that the only solution to this set of equations, given that $F$ is now a symmetric polynomial multiplied by inverse energies, is $F = \text{const}$. This reduces the $(\pm2,\pm2,\mp2)$ three-particle amplitudes, and the four-particle amplitude due to these vertices, to the boost-invariant limit. The four-particle amplitude is then what one finds in General Relativity (GR). Indeed, in this boost-invariant limit the three-particle amplitudes have mass dimension $2$ which is due to the two-derivative nature of the Einstein-Hilbert action. Note that the minus sign in the overall amplitude is because gravity is an attractive force. We denote the magnitude of the three-gravity coupling as $\kappa$. \\

As with the photon case, we may have anticipated this result from a Lagrangian point of view. In GR the required gauge redundancy is diffeomorphism invariance under which the spacetime coordinates transform. Furthermore, the quantum effective theory of GR is best understood by expanding the Einstein-Hilbert action around the vacuum solution $g_{\mu\nu} = \eta_{\mu\nu} + h_{\mu\nu}$. One finds a tower of two-derivative terms with each coupling fixed by diffeomorphisms relating operators at different orders in $h_{\mu\nu}$. Given that in this work the two-derivative kinetic term is assumed to be of the boost-invariant form, adding time derivatives to the cubic vertex would break the (linearised) diffeomorphsim symmetry and one would therefore expect issues to arise. However, let us again emphasise that although this Lagrangian interpretation can yield some intuition, the on-shell analysis presented here is preferable given that it is independent of gauge redundancies and field redefinitions. As we shall see in Section \ref{sec:GravitonCouplings}, the analysis is also robust against adding additional particles.  \\

Now in contrast to the photon case, here we can constrain the other three-particle amplitudes $(\pm2,\pm2,\pm2)$ thanks to the non-vanishing GR amplitudes. The dimension $6$ amplitudes are 
\begin{align} \label{HighestDim}
\mathcal{A}_3(1^{-2} 2^{-2} 3^{-2}) &= \left( \la12 \ra \la 2 3\ra \la 3 1\ra \right)^2 F_{-2,-2,-2}(E_1, E_2,E_3), \\
\mathcal{A}_3(1^{+2} 2^{+2} 3^{+2}) &= \left([12][23][31]\right)^2 F_{+2,+2,+2}(E_1, E_2,E_{3}) ,
\end{align}
where $F_{-2,-2,-2}$ and $F_{+2,+2,+2}$ are symmetric polynomials. Now consider the four-particle amplitude $\mathcal{A}_{4}(1^{+2},2^{+2},3^{+2},4^{-2})$. We can arrange the helicities of the exchanged particle such that each residue has mass dimension $8$ and going through an analysis mirroring those above we see that the amplitude takes the form
\begin{align}
\mathcal{A}_{4}(1^{+2},2^{+2},3^{+2},4^{-2}) = [12]^4 [23]^4 \la 24 \ra^4 \frac{B}{stu},
\end{align} 
and consistent factorisation requires
\begin{align}
-B = &\kappa F_{+2,+2,+2}(E_{1},E_{2}, -E_{1}-E_{2}) \\
 = &\kappa F_{+2,+2,+2}(E_{1},E_{3}, -E_{1}-E_{3}) \\
  = &\kappa F_{+2,+2,+2}(E_{2},E_{3}, -E_{2}-E_{3}).
\end{align}
It is clear that the only solution to this system, for generic energies, is $F_{+2,+2,+2} = \text{const}$. We therefore also have $F_{-2,-2,-2} = \text{const}$ by parity and so the amplitudes are reduced to their boost-invariant limits. \\

At the Lagrangian level, these mass dimension $6$ three-particle amplitudes are due to terms cubic in the Riemann tensor. Note that there are no three-particle amplitudes with mass dimension $4$. One may expect terms quadratic in curvature, $R^{2}$, $R_{\mu\nu}^2$ and $R_{\mu\nu\rho\sigma}^2$, to give rise to mass dimension $4$ amplitudes. However, in $4D$ the Riemann squared term is degenerate with the other two up to the Gauss-Bonnet total derivative and both of these can be removed by a field redefinition since they are proportional to $R_{\mu\nu}$ which vanishes on-shell. One may also wonder about terms with four or more powers of curvature, but these do not contribute to three-particle amplitudes since at cubic order in fluctuations at least one curvature would need to be evaluated on the flat background where it vanishes. \\

\subsubsection*{Brief Summary} 
Let us briefly summarise our results for a single spin-$S$ particle:
\begin{itemize}
\item For $S=0$ factorisation is trivial with each residue a function of the external energies.
\item For $S=1$ the four-particle test forces the leading order three-particle amplitudes to vanish. This result assumes that the functions of energies are polynomials divided by some powers of the energies, but does not rely on any specific truncation of such polynomials. The highest dimension three-particle amplitudes are unconstrained by the four-particle test and at the level of a Lagrangian, the leading order gauge invariant vertices are \eqref{VectorConsistentVertex}. 
\item For $S=2$ all three-particle amplitudes are forced to their boost-invariant limit. These are the amplitudes in GR with the addition of a term cubic in curvature. Again we assume that the functions of energies are polynomials divided by some powers of energies and our result does not rely on a truncation of the numerator. Lorentz violation in graviton cubic vertices is therefore impossible for a relativistic on-shell condition, in contrast to the photon.
\item For $S \geq 3$ the four-particle test cannot be passed and there cannot be any cubic self-interactions for these particles, at least to leading order in derivatives. This is potentially tricky to understand at the level of a Lagrangian, but here simple dimensional analysis and the four-particle test ruled out these vertices. 
\end{itemize}
In the following sections we will see that these results are robust against including additional massless particles. 

\subsection{Couplings to a photon: Compton scattering and beyond} \label{sec:PhotonCouplings}

We now move to couplings between spin-$S$ particles and a photon. We take $S \neq 1$ as we will consider multiple spin-$1$ particles in Section \ref{MultipleParticles}. Apart from this restriction, we allow for both bosonic and fermionic particles. We initially consider Compton scattering $\mathcal{A}_{4}(1_{a}^{-S}, 2^{+1}, 3_{b}^{+S}, 4^{-1})$ to constrain the $(+S,-S,\pm1)$ amplitudes, allowing for multiple spin-$S$ particles since in the boost-invariant limit a single copy cannot have a $U(1)$ charge. These amplitudes have mass dimension $1$ and so correspond to the familiar cubic couplings of a charged particle. We then present a complete analysis, i.e. we constrain all amplitudes that can be constrained, for a theory of a single scalar coupled to a photon. Couplings to a graviton are studied in Section \ref{sec:GravitonCouplings}. \\


\subsubsection*{Compton scattering}
Consider the amplitude $\mathcal{A}_{4}(1_{a}^{-S}, 2^{+1}, 3_{b}^{+S}, 4^{-1})$ with $\dim{\mathcal{A}_{4}} = 0$. Each residue must have mass dimension $2$ which in turn must come from two mass dimension $1$ three-particle amplitudes\footnote{It is not possible to exchange a particle such that one three-particle amplitude is dimensionless and the other has mass dimension $2$.}. First consider the $s$-channel where there are two possibilities for the spin of the exchanged particle. We can exchange a spin-$S$ particle or a spin-$|S-2|$ particle. However, we find that the latter case yields spurious poles for \textit{all} $S$ and so consistency demands that the $(\mp S,\pm (S-2),\pm1)$ amplitudes vanish. For the former case we use the three-particle amplitudes 
\begin{align} 
A_{3}(1_{a}^{-S}, 2_{b}^{+S}, 3^{-1}) &=  \la 12 \ra^{-1} \la 23 \ra^{1-2S} \la 31  \ra^{2S+1} F_{ab}^{H}(E_{1},E_{2}), \label{Compton1} \\
A_{3}(1_{a}^{-S}, 2_{b}^{+S}, 3^{+1}) &=  [12]^{-1} [23]^{2S+1} [31]^{1-2S} F_{ab}^{AH}(E_{1},E_{2}), \label{Compton2}
\end{align}
where we have dropped the helicity subscripts on the $F$'s in favour of the internal indices $(a,b)$ labelling the external spin-$S$ particles, and have used energy conservation to eliminate $E_{3}$. Computing the $s$-channel residue we find 
\begin{align}
(R_{s})_{ab} = \frac{(\la 14 \ra [23])^{2S} (\la 34 \ra [23])^{2-2S} }{u} \sum_{e}F^{AH}_{ae}(E_{1}, -E_{1}-E_{2})F^{H}_{eb}(-E_{3}-E_{4}, E_{3}),
\end{align}
where we have summed over the possible spin-$S$ exchanged particles.\\

Moving to the $t$-channel, we see that we must exchange a photon to realise the desired mass dimension. A non-zero residue then requires non-zero three-photon amplitudes $(-1,+1,\pm1)$. In Section \ref{sec:photon} we showed that in the absence of other particles these amplitudes must vanish but since we have now included additional particles, we have to check if this result still holds. Going back to the amplitude $\mathcal{A}_{4}(1^{-1},2^{+1},3^{-1},4^{+1})$, we see that in the $s$ and $u$ channels only photon exchange can yield a dimensionless amplitude while in the $t$-channel we can exchange a photon, as we considered in Section \ref{sec:photon}, but can also exchange a spin-$3$ particle. The required three-particle amplitudes are $(\pm1,\pm1,\mp3)$ but we find that such a residue induces spurious poles in $t$ and therefore consistency requires these three-particle amplitudes to vanish. So our result in Section \ref{sec:photon} on the absence of a consistent mass dimension $1$ three-particle amplitude for photons is unchanged when we allow for additional exchanges. It follows that there is no $t$-channel contribution for Compton scattering.   \\

Finally, for $u$-channel exchange we again find two possibilities for the exchanged particle: we can exchange a spin-$S$ particle or a spin-$(S+2)$ particle. As in the $s$-channel we find that the latter choice yields spurious poles for all $S$ and so the $(\mp S,\pm (S+2),\mp1)$ amplitudes must vanish. For the former case we find that the residue is 
\begin{align}
(R_{u})_{ab} =  \frac{(\la 14 \ra [23])^{2S} (\la 34 \ra [23])^{2-2S}}{s} \sum_{e}F^{H}_{ae}(E_{1},-E_{1}-E_{4})F^{AH}_{eb}(-E_{3}-E_{2},E_{3}),
\end{align}
where again we have summed over the possible spin-$S$ exchanged particles. Now we see a fundamental difference between the two cases $S < 1$ and $S > 1$. For $S > 1$, each residue contains a spurious pole in $(\la 34 \ra [23])$ meaning that no consistent four-particle amplitude is possible. The four-particle test therefore requires the $(+S,-S,\pm1)$ three-particle amplitudes to vanish for $S > 1$, implying that such a particle cannot have a $U(1)$ charge. This result is known in the boost-invariant limit and here we see that it is unchanged when we allow for the breaking of Lorentz boosts. Compton scattering is therefore only possible for low spins with $S=0,1/2$. The test is still non-trivial in these cases, since consistent factorisation yields the constraints
\begin{align}\label{ComptonFactorisation}
& \sum_{e}F^{AH}_{ae}(E_{1}, -E_{1}-E_{2})F^{H}_{eb}(-E_{3}-E_{4}, E_{3}) \nonumber \\
 = &\sum_{e}F^{H}_{ae}(E_{1},-E_{1}-E_{4})F^{AH}_{eb}(-E_{3}-E_{2},E_{3}),
\end{align}
which needs to be satisfied for all $E_{i}$ subject to $E_{1}+E_{2}+E_{3}+E_{4} = 0$. Again these constraints are the same for parity even and parity odd amplitudes so we will drop the $H/AH$ labels in the following. These factorisation constraints are solved by $F_{ab} = f_{ab}F(E_{1} + E_{2})$ where $f_{ab}$ is a constant matrix, and $F$ is an arbitrary function of the sum $E_{1} + E_{2}$\footnote{We haven't shown that there are no other solutions.}. For bosons, $f_{ab}$ needs to be anti-symmetric by Bose symmetry (given the form of \eqref{Compton1} and \eqref{Compton2}), and therefore consistent factorisation is not possible for a single scalar which in the boost-invariant limit is the well known fact that a single scalar cannot have a $U(1)$ charge. For two scalars, a consistent boost-breaking amplitude is possible with $F_{ab} = \epsilon_{ab}F(E_{1} + E_{2})$, and similarly a consistent amplitude exists for a charged $S=1/2$ particle. In Appendix \ref{AppendixC} we provide a Lagrangian description of these boost-breaking versions of massless QED with unbroken $U(1)$ gauge symmetry.


\subsubsection*{Scalar-photon couplings} \label{PhotonScalar}
We now provide a full analysis for a theory of a single scalar coupled to a photon. Many of the possible three-particle amplitudes have already been constrained and our goal in this part is to constrain the others where possible. There are five three-particle amplitudes arising from couplings between the scalar and the photon: $(\pm1,\pm1,0)$, $(-1,+1,0)$ and $(\pm1,0,0$). However, we have already considered the $(\pm1,0,0)$ amplitude above and we find that there are no solutions to \eqref{ComptonFactorisation} for a single scalar and therefore this amplitude must vanish. In addition, there are two amplitudes involving only the photon: $(\pm 1,\pm 1,\pm 1)$. Finally, there is a single amplitude involving only the scalar: $(0,0,0)$. \\

Lets start by constraining the $(-1,+1,0)$ amplitude. Consider the four-particle amplitude $\mathcal{A}_{4}(1^{-1} 2^{+1} 3^{-1} 4^{+1})$ between four photons. By little group scaling this amplitude takes the general form 
\begin{equation}
\mathcal{A}_4(1^{-1} 2^{+1} 3^{-1} 4^{+1}) = \la 1 3 \ra^{2} [2 4]^{2} \mathcal{G}(s,t,u,E_{i}).
\end{equation}
Now in the $s$-channel we can exchange a scalar particle, meaning that this residue will have a vanishing mass dimension. This can also be arranged for in the $u$-channel by exchanging a scalar. If these residues are dimensionless, the four-particle amplitude has $\dim{\mathcal{A}_{4}} =-2$ which in turn requires $\dim{\mathcal{G}} = -6$ and so the amplitude must take the form 
\begin{equation} \label{eq:Ampl-+-+}
\mathcal{A}_4(1^{-1} 2^{+1} 3^{-1} 4^{+1}) = \la 1 3 \ra^{2} [2 4]^{2} \frac{\mathcal{F}(E_{i})}{stu},
\end{equation}
meaning that we require exchanges in \textit{all} channels. In the $t$-channel we would need to exchange a graviton to realise the same mass dimension for the amplitude. However, even in the presence of a graviton the test cannot be passed, since the necessary $(\pm 1, \pm 1, \mp 2)$ amplitudes are forced to vanish by a different test, as we will show in section \ref{sec:GravitonCouplings}. Thus, the $(-1,+1,0)$ three-particle amplitude must vanish.\\

\begin{table}[h!]
\begin{center}
\begin{tabular}{|p{3cm}|p{3cm}|p{4cm}|}
 \hline
Helicities & Amplitude $\mathcal{A}_{3}$& Constraint \\
 \hline
$(-1,-1,+1)$ &$ \la 12 \ra^{3}/(\la 23 \ra \la 31 \ra) F$& $F = 0$ \\ 
$(-1,-1,-1)$ & $\la 12 \ra \la 23 \ra \la 31 \ra F$  & alternating $F$ in $(1,2,3)$ \\
$(-1,-1,0)$ & $\la 12 \ra^{2} F$  & symmetric $F$ in $(1,2)$ \\
$(-1,+1,0)$ & $\la 13 \ra^{2} / \la 23 \ra^{2} F$ & $F = 0$\\
$(-1,0,0)$ & $ (\la 12 \ra \la 31 \ra) / \la 23 \ra F$ & $F=0$ \\
$(0,0,0)$ & $ F$ & symmetric $F$ in $(1,2,3)$ \\
 \hline
\end{tabular}
\caption{Constrains on the three-particle amplitudes in a theory of a scalar coupled to a photon}
\label{tab:li_efts}
\end{center}
\end{table}

We are therefore left with three distinct three-particle amplitudes and their parity counterparts. The others are forced to vanish. This is summarised in Table \ref{tab:li_efts} and one can see that the non-zero amplitudes do not contain inverse powers of the brackets and therefore cannot give rise to spurious poles in four-particle amplitudes. For a theory of a single scalar coupled to a photon, there are therefore no further constraints from the four-particle test. The symmetry constraints on $F$ tell us the minimum number of time derivatives required to write down a consistent boost-breaking interaction. As we discussed above, for the $(\pm 1,\pm 1,\pm 1)$ amplitudes we need at least three time derivatives. For the $(\pm 1,\pm 1,0)$ and $(0,0,0)$ vertices we need at least one and two respectively. The leading order Lagrangian giving rise to these amplitudes is (assuming parity-even interactions only) \begin{align} \label{ScalarVectorL}
\mathcal{L} = &\frac{1}{2}(\partial \pi)^{2} + \frac{1}{4}F_{\mu\nu}F^{\mu\nu} + (a_{1} \pi^3 + a_{2} \pi^{2} \ddot{\pi} + a_{3} \dot{\pi}^{3} + \ldots) \nonumber \\ &+ (b_{1} \pi + b_{2} \dot{\pi} + b_{3} \ddot{\pi} + \ldots) F_{\mu\nu}F^{\mu\nu} + (c_{1} \ddot{F}^{\mu}{}_{\nu} \dot{F}^{\nu}{}_{\rho} F^{\rho}{}_{\mu} + \ldots),
\end{align}
where $a_{i}$ etc are dimensionful Wilson coefficients. \\

\subsubsection*{Brief summary}
Let us briefly summarise our results for a spin-$S$ particle coupled to a photon:
\begin{itemize}
\item Compton scattering is not possible for $S > 1$, while for $S=0,1/2$ consistent boost-breaking theories of massless scalar and fermionic QED with $U(1)$ gauge symmetry exist. We can write down Lagrangians in each case with generalised boost-breaking gauge symmetries (see Appendix \ref{AppendixC}). Along the way we also showed that the absence of $(-1,+1,\pm1)$ vertices is robust against adding additional particles and that the $(\mp S,\pm (S-2),\pm 1)$ and $(\mp S,\pm (S+2),\mp 1)$ amplitudes must vanish for $S \neq 1$.
\item A consistent boost-breaking theory of a single scalar coupled to a photon does exist. Self-interactions for both particles are possible and so are $\pi \gamma \gamma$ vertices. The leading Lagrangian is presented in \eqref{ScalarVectorL}.
\end{itemize} 


\subsection{Couplings to a graviton: gravitational Compton scattering and beyond}\label{sec:GravitonCouplings}
We now move onto couplings between spin-$S$ particles and gravity. This section contains:
\begin{itemize}
\item constraints on the $(\pm 2, +S, -S)$ vertices due to gravitational Compton scattering
\item a full analysis of all possible three-particle amplitudes in a theory of a single scalar coupled to gravity
\item a full analysis of all possible three-particle amplitudes in a theory of a photon coupled to gravity
\item an analysis for theory of a massless $S=3/2$ particle coupled to gravity a.k.a $\mathcal{N} = 1$ supergravity.
\end{itemize}

\subsubsection*{Gravitational Compton scattering} \label{sec:GravitonCouplings}
We begin by constraining the leading, mass dimension $2$, three-particle amplitudes for spin-$S$ particles coupled to gravity, namely the $(\pm2, +S,-S)$ amplitudes. We take $S \neq 2$. Consider the four-particle amplitude $\mathcal{A}_4(1^{-S}, 2^{+2}, 3^{-2}, 4^{+S})$ with $\dim{\mathcal{A}_{4}}=2$. As with the photon case above, there are two ways to achieve the required dimension of the residues in $s$ and $t$ channels and a unique way in the $u$-channel. In the $s$-channel we can exchange a spin-$S$ particle or a spin-$|S-4|$ particle. In the latter case we find spurious poles in the residue and so we set the $(\mp S,\pm 2,\pm(S-4))$ amplitudes to zero for all $S \neq 2$. For spin-$S$ exchange, we need the following three-particle amplitudes 
\begin{align}
\mathcal{A}_{3}(1^{-2},2^{-S},3^{+S}) &= \frac{\la 12 \ra^{2S+2}\la 31 \ra^{2-2S}}{\la 23 \ra^2} F^{H}_{-2,-S,+S}(E_{1},E_{2}),   \label{GravitationalCompton1} \\
\mathcal{A}_{3}(1^{+2},2^{+S},3^{-S}) &= \frac{[12]^{2S+2}[31]^{2-2S}}{[23]^2} F^{AH}_{+2,+S,-S}(E_{1},E_{2}).  \label{GravitationalCompton2}
\end{align}
Computing the residue we find (for both integer and half-integer $S$)
\begin{align} \nonumber
R_s  &=  - \frac{ \la 13 \ra^{2S} \la 3 4 \ra^{4-2S} [24]^4}{t u}  F^{AH}_{+2, +S, -S}(E_2, -E_1 - E_2) F^{H}_{-2, -S, +S}(E_3, -E_3-E_4).
\end{align}
The ordering of particles is especially important in the fermionic case, where changing the order of two fermions gives rise to a minus sign. Here and in the remaining equations we take particle $1$ to always appear before particle $4$.\\

In the $t$-channel, dimensional analysis allows for exchange of a spin-$S$ particle and a spin-$(S+4)$ particle. However in the latter case spurious poles are unavoidable for all $S$. We therefore require the $(\mp S,\mp 2,\pm (S+4))$ amplitudes to vanish. For spin-$S$ exchange we find the residue (for both integer and half-integer $S$)
\begin{align}
R_{t} &=  - \frac{\la 13 \ra^{2S} \la 3 4 \ra^{4 - 2S}  [ 2 4 ]^4 }{s u}  F^{H}_{-2, -S, +S}(E_3, E_1) F^{AH}_{+2, +S, -S}(E_2, E_4) . 
\end{align}

Finally, for $u$-channel exchange there is only a single choice for the spin of the exchanged particle that yields a residue with the desired mass dimension; that particle must be the graviton. The residue therefore depends on the lowest dimension three-graviton amplitude which in Section \ref{sec:graviton} we concluded must be reduced to the boost-invariant GR amplitude. However, now that we have included additional particles we must check if that result is robust against allowing for additional exchanges. Going back to the $\mathcal{A}_{4}(1^{-2},2^{+2},3^{-2},4^{+2})$ amplitude, we see that if the amplitude has $\dim{\mathcal{A}_{4}}=2$ we can only exchange a graviton in the $s$ and $u$ channels, but in the $t$-channel dimensional analysis allows for $S=2$ and $S=6$ exchange. In the latter case, however, we find a spurious pole in $t$ and so only graviton exchange can yield a consistent amplitude - consistency demands that the $(\pm 2,\pm 2,\mp 6)$ amplitudes are zero. Our result of \ref{sec:graviton}, i.e. the $(+2,-2,\pm2)$ amplitudes must be boost-invariant and correspond to those of GR, is robust against including additional massless particles.\\

We can now go back to gravitational Compton scattering. To compute the $u$-channel residue, we now need the lowest dimension three-graviton amplitudes. As shown above, these take the form
\begin{align}
\mathcal{A}_3(1^{-2}, 2^{-2}, 3^{+2}) &=  \left(\frac{\la 12\ra^3}{\la 23 \ra \la 31 \ra}\right)^2 \kappa,  \\
\mathcal{A}_3(1^{+2}, 2^{+2}, 3^{-2}) &=   \left(\frac{[12]^3}{[23][31]}\right)^2 \kappa, 
\end{align}
where $\kappa$ is related to the Planck mass in GR and we have used the fact that GR is a parity-even theory. Now as we have seen a number of times before, there are two choices for the helicity configuration of the exchanged graviton. The total residue is a sum of the two, $R_u  =  R_u^{+-} + R_u^{-+}$, but one of these always vanishes once we declare how we approach the $u$-channel pole. We first consider the case of bosons, meaning we can swap the order of any two particles without introducing minus signs, but we keep factors of $(-1)^{2S}$ to make the formulae easy to generalise to the fermionic case. When $[14] = \la 2 3 \ra = 0$ we have $R_u^{+-} = 0$ and
\begin{align} 
R_u^{-+} = & - \frac{\la 13  \ra^{2S}   \la 34  \ra^{4 - 2S} [24]^4}{s t } \kappa F^{H}_{-2,-S,+S} (-E_1 - E_4, E_1), 
\end{align}
and when $\la 14 \ra = [ 2 3 ] = 0$ we have $R_u^{-+} = 0$ and (for bosons) 
\begin{align}
R_u^{+-} =  (-1)^{2S+1} \frac{\la 13  \ra^{2S}   \la 34  \ra^{4 - 2S} [24]^4}{s t} \kappa F^{AH}_{+2,+S,-S} (-E_1 - E_4, E_4).
\end{align}
If the spin-$S$ particles are fermions, then the expression for $R^{+-}_u$ inherits an overall minus sign (due to the necessity of swapping the order of particles $1$ and $4$), which conveniently cancels out the $(-1)^{2S}$ factor while $R^{-+}$ is unchanged. The $u$-channel residue for both integer and half-integer $S$ is therefore 
\begin{align} 
R_u = - \frac{\la 13  \ra^{2S}   \la 34  \ra^{4 - 2S} [24]^4}{s t } \kappa F^{H}_{-2,-S,+S} (-E_1 - E_4, E_1), 
\end{align}
subject to 
\begin{align}
\label{eq:BranchMatchingGrav}
F_{-2,-S,+S}^{H} (-E_1 - E_4, E_1) = F_{+2,+S,-S}^{AH} (-E_1 - E_4, E_4),
\end{align}
ensuring that the residue is the same regardless of how we approach the pole. This matching condition ensures that operators generating the amplitudes \eqref{GravitationalCompton1} and \eqref{GravitationalCompton2} are parity-even.\\

Now we see from each residue that when $4-2S <0$, i.e. $S \geq 5/2$, a consistent four-particle amplitude cannot be constructed due to the additional poles in $s$. Hence we conclude that the above three-particle amplitudes for a massless particle with $S \geq 5/2$ coupled to gravity are inconsistent and must vanish. In a boost-invariant theory this is the well-known statement that a massless particle with $S \geq 5/2$ cannot couple to gravity, and we see that this statement is unchanged for boost-breaking theories. This is indeed consistent with some recent study in the light-cone formalism in which the only explicitly constructed cubic coupling of higher-spin particles to gravity is non-unitary \cite{Ponomarev:2016lrm}. \\

\noindent For $S < 5/2$ we can construct a consistent amplitude for gravitational Compton scattering. It takes the form
\begin{align}
\mathcal{A}_4(1^{-S} 2^{+2} 3^{-2} 4^{+S})  =  \la 13 \ra^{2S} \la 3 4 \ra^{4-2S} [24]^4 \frac{A}{stu},
\end{align}
and consistent factorisation requires
\begin{align} \nonumber
-A &=   F^{AH}_{+2, +S, -S}(E_2, -E_1 - E_2) F^{H}_{-2, -S, +S}(E_3, -E_3-E_4) \\ \nonumber
&=  F^{H}_{-2, -S, +S}(E_3, E_1) F^{AH}_{+2, +S, -S}(E_2, E_4) \\ 
 &=  \kappa F^{AH}_{+2,+S,-S} (-E_1 - E_4, E_4). 
 \label{eq:GravitonSpinSCondition}
\end{align}
The $F$-functions are related by (\ref{eq:BranchMatchingGrav}) and therefore both can be written as the same $F$. If $F$ contained any inverse powers of energies, then the singularities of the three expressions wouldn't match, so $F$ must be a polynomial of a degree which we denote as $p$. The above equations then imply that $2 p = 2p = p$, and therefore $p = 0$. So only constant solutions are possible: the four-particle test has reduced the amplitudes to their boost-invariant limits! Furthermore, the coupling constants of the $(\pm2, +S,-S)$ amplitudes are not arbitrary. The equations tell us that they are fixed in terms of the pure gravitational coupling $\kappa$: $F^{H}_{-2, -S, +S} = F^{AH}_{+2, +S, -S} = \kappa$. This is the on-shell derivation of the universality of gravity for elementary massless particles with $S \leq 2$: all particles couple to gravity with the same strength. \\

Compared to photon Compton scattering considered above, we see some important differences for gravity. Here boost-breaking interactions are not permitted whereas for a photon coupled to $S=0,1/2$ particles such a breaking is permitted. Here we also see the emergence of the equivalence principle, and allowed couplings to $S=3/2$ particles. We attribute these differences to the presence of a three-particle amplitude for three gravitons which does not exist for three photons. The case of a $S=3/2$ particle coupled to gravity is particularly interesting. The amplitudes we have considered are those appearing in $\mathcal{N}=1$ supergravity and here we have seen that boost-breaking versions, with relativistic on-shell conditions, do not exist. We refer the reader to \cite{McGadyRodina} for some very nice results using factorisation when a massless $S=3/2$ particle is in the spectrum. These results include: the necessity of gravity, the derivation of super-multiplets, and a proof that having $\mathcal{N} > 8$ requires the presence of a $S=5/2$ particle and therefore the test cannot be passed if there is too much supersymmetry. Most of these results come from pole counting and we would therefore expect them to hold for boost-breaking theories with relativistic on-shell conditions too.\\

\subsubsection*{Scalar-graviton couplings}  \label{sec:ScalarGraviton}
We now turn our attention to the boost-breaking theory of a single scalar coupled to gravity. Here we show that for relativistic on-shell conditions \textit{the four-particle test requires all three-particle amplitudes for a scalar coupled to a graviton to be boost-invariant}. We have already seen that the pure graviton three-particle amplitudes are forced to be boost-invariant and so are the $(\pm 2, 0, 0)$ amplitudes. The remaining amplitudes to be discussed are $(\pm2,\pm2,0)$, $(+2,-2,0)$ and $(0,0,0)$. \\

First consider the $(+2,-2,0)$ amplitude, which we can easily show is inconsistent in both boost-invariant and boost-breaking theories. This vertex can contribute to $s$-channel exchange in the four-particle graviton amplitude $\mathcal{A}_{4}(1^{-2}, 2^{+2}, 3^{-2}, 4^{+2})$. This $s$-channel contribution to the amplitude has mass dimension $-2$ since the residue is dimensionless. However, the scaling of this amplitude under a little group transformation requires it to take the form 
\begin{align}
\mathcal{A}_{4}(1^{-2}, 2^{+2}, 3^{-2}, 4^{+2}) = \la 13 \ra^{4}[24]^{4} \mathcal{G}(s,t,u,E_{i}),
\end{align} 
and so if $\dim{\mathcal{A}_{4}} =-2$ the amplitude cannot be consistent, since simple poles require $\dim{\mathcal{G}} \geq -6$, while  $\text{dim} \{  \la 13 \ra^{4}[24]^{4} \} = 8$. \\
 
We now constrain the $(\pm2,\pm2,0)$ amplitudes using $\mathcal{A}_{4}(1^{+2},2^{+2},3^{-2},4^{0})$ with scalar exchange in the $s$-channel. The contribution to the amplitude from this diagram has mass dimension\footnote{This mass dimension can also be achieved by exchanging a spin-$6$ particle but such a residue contains spurious poles.} $4$. The same mass dimension can be realised in the $t$ and $u$ channels by exchanging a graviton and using the leading (mass dimension $2$) three-graviton amplitudes\footnote{Another possibility is to exchange a spin-$4$ particle but in this case the residues again have spurious poles.}. Given that 
\begin{align}
\mathcal{A}_{3}(1^{+2},2^{+2},3^{0}) = [12]^{4}F^{AH}_{+2,+2,0}(E_{1},E_{2}),
\end{align}   
the three residues are given by 
\begin{align}
R_{s} &= - \frac{[12]^{6} \la 13 \ra^{2} \la 23 \ra^{2}}{tu} \kappa F^{AH}_{+2,+2,0}(E_{1}, E_{2}),\\
R_{t} &= - \frac{[12]^{6} \la 13 \ra^{2} \la 23 \ra^{2}}{su} \kappa F^{AH}_{+2,+2,0}(E_{2}, -E_{2}-E_{4}), \\
R_{u} &= - \frac{[12]^{6} \la 13 \ra^{2} \la 23 \ra^{2}}{st} \kappa F^{AH}_{+2,+2,0}(E_{1}, -E_{1}-E_{4}).
\end{align}
Here we have written $F^{AH}_{+2,+2,0}$ as a function of two energies only and it must be a symmetric function by Bose symmetry. Furthermore, we have used the fact that the $(-2,0,0)$ amplitude is boost-invariant with its coupling identical to the graviton self-coupling $\kappa$. A consistent amplitude must therefore take the form
\begin{align}
\mathcal{A}_{4}(1^{+2},2^{+2},3^{-2},4^{0}) = [12]^{6} \la 13 \ra^{2} \la 23 \ra^{2} \frac{B}{stu},
\end{align}
with
\begin{align}
-B &=  \kappa F^{AH}_{+2,+2,0}(E_{1}, E_{2}) \\
& =   \kappa F^{AH}_{+2,+2,0}(E_{2}, -E_{2}-E_{4}) \\
& =  \kappa F^{AH}_{+2,+2,0}(E_{1}, -E_{1}-E_{4}),
\end{align}
which can only be solved if $F^{AH}_{+2,+2,0} = \text{const}$, thereby reducing the $(\pm2,\pm2,0)$ amplitudes to their boost-invariant limits. Note that the coupling constant for these amplitudes is not fixed in terms of $\kappa$. \\

Finally, we can constrain the pure scalar amplitude $(0,0,0)$ using the four-particle amplitude $\mathcal{A}_{4}(1^{0},2^{0},3^{0},4^{+2})$. If we exchange a scalar in each channel with 
\begin{align}
\mathcal{A}_{3}(1^{0},2^{0},3^{0}) = F_{0,0,0} (E_{1},E_{2}),
\end{align}
the three residues are 
\begin{align}
R_{s} &= - \frac{[34]^{2} [24]^{2} \la 23 \ra^{2}}{tu} \kappa F_{0,0,0}(E_{1}, E_{2}),\\
R_{t} &= - \frac{[34]^{2} [24]^{2} \la 23 \ra^{2}}{su} \kappa F_{0,0,0}(E_{1}, E_{3}),\\
R_{u} &= - \frac{[34]^{2} [24]^{2} \la 23 \ra^{2}}{st} \kappa F_{0,0,0}(E_{2}, E_{3}),
\end{align}
and so the consistent amplitude is
\begin{align}
\mathcal{A}_{4}(1^{0},2^{0},3^{0},4^{+2}) = [34]^{2} [24]^{2} \la 23 \ra^{2} \frac{C}{stu},
\end{align}
with
\begin{align}
-C &=  \kappa F_{0,0,0}(E_{1}, E_{2}) \\
&=  \kappa F_{0,0,0}(E_{1}, E_{3}) \\
 &=  \kappa F_{0,0,0}(E_{2}, E_{3}).
\end{align}
Again, the only solution to these factorisation constraints for generic energies is $F_{0,0,0} = \text{const}$, thereby reducing the three-scalar amplitude to its boost-invariant form, which is simply a constant. \\

We have therefore seen that \textit{all} three-particle amplitudes, and therefore all three-point vertices, in a theory of a graviton coupled to a scalar (if the Lagrangian depends on covariant fields only) must reduce to their boost-invariant limits. Let us discuss the allowed boost-invariant interactions in more detail. We have discussed the pure gravity vertices at the level of a Lagrangian earlier on. The only allowed pure scalar amplitude is a constant and so the cubic vertex is simply $\phi^{3}$. The other two allowed interactions mix the scalar and the graviton and have mass dimension $2$ and $4$. The coupling of the former is the same as the three graviton coupling $\kappa$, while the coupling of the latter is independent of $\kappa$ and is therefore a new Wilson coefficient in the effective action. At the level of a Lagrangian they come from the $(\partial \phi)^{2} = g^{\mu\nu} \partial_{\mu} \phi \partial_{\nu} \phi$ and $\phi R^{\mu\nu\rho\sigma}R_{\mu\nu\rho\sigma}$ terms respectively, expanded around the boost-invariant vacuum $g_{\mu\nu} = \eta_{\mu\nu}$, $\phi = 0$.  Note that there is no $\phi^{2}R$ coupling as this can be removed by a field redefinition going from Jordan to Einstein frame. We can also write down a parity-odd vertex $\phi \epsilon^{\mu\nu\rho\sigma} R_{\mu\nu \kappa \lambda}R^{\kappa \lambda}{}_{\rho \sigma}$. In appendix \ref{AmplitudeCalculation}, we provide further clarifications on why a simple $\dot{\phi}^3$ self-interaction for a scalar coupled to $h_{\mu \nu}$ in Minkowski space is inconsistent. \\

In \cite{Conjecture} it was conjectured that in the flat space, decoupling and slow-roll limit of the EFT of inflation, if the scalar Goldstone has a boost-invariant kinetic term, then the only possible UV completion is a free theory. In this language, the decoupling limit boils down to neglecting all interactions with the metric fluctuations and the slow-roll limit corresponds to neglecting all Lorentz-invariant interactions, such as for example a potential $  V(\phi) $. In other words, the conjecture is that any scalar EFT with $c_{s}=1$ and boost-breaking interactions cannot be UV completed. The relation of this conjecture to our results is tantalizing but not straightforward. On the one hand, we also found that for $  c_{s}=1 $ boost-breaking interactions are forbidden, but we crucially needed to assume (i) that the scalar is coupled to gravity, (ii) the theory is in Minkowski and (iii) assume a restricted form of the four-particle amplitude. Also, we did not use any constraints coming from a putative UV completion. All our analysis is based on the low-energy EFT. This is to be contrasted with the discussion in \cite{Conjecture} where the coupling to gravity does not seem to play a role, while all the constraining power comes from demanding a consistent UV completion. Furthermore, the application of our results to the flat-space limit of FLRW spacetimes clashes with the IR sensitivity of the four-particle test. We will discuss this in Section \ref{ssec:pofx}.


\subsubsection*{Photon-graviton couplings} 
We have seen that when a scalar is coupled to $h_{\mu \nu}$, all three-particle amplitudes and therefore all three-point vertices are required to be boost-invariant by the four-particle test. One may therefore expect the presence of the graviton is forcing boost-invariance upon us when free particles satisfy relativistic on-shell conditions. Here we provide more evidence of this by showing that when a photon is coupled to $h_{\mu\nu}$, all three-point vertices involving this photon have to be boost-invariant. This result can be derived because of the existence of a (boost-invariant) three-point $(++-)$ vertex for gravitons, which is absent for photons. \\


Let us recap the relevant results we have derived so far. We have shown that the pure graviton three-particle amplitudes are boost-invariant. The lowest dimension photon amplitudes are forced to vanish by the test, while boost-breaking possibilities have not yet been ruled out for the $(+++)$ and $(---)$ three photon interaction. Now, for mixed amplitudes, we have four possibilities (plus their parity counterparts) left to consider:
\begin{align}
 (+2,+2,+1), \quad (+2,+2,-1), \quad (+1,+1,+2), \quad (+1,+1,-2).
\end{align}  

First consider the dimensionless choice $(+ 1, + 1, - 2)$. These amplitudes have both holomorphic and anti-holomorphic parts, and contribute to e.g. $u$-channel diagram for the $\mathcal{A}_{4}(1^{+1},2^{-1},3^{+2},4^{-2})$ amplitude via a photon exchange. The dimensionality of this amplitude is
\begin{align}
\text{dim} \{ \mathcal{A}_4 \} = 0 + 0 - 2 = -2.
\end{align}
On the other hand, to achieve correct helicity scalings, we need,
\begin{align}
\mathcal{A}_4 \sim [13]^{2}[23]^{2} \la 24 \ra^{4} \mathcal{G}(s,t,u,E_{i}),
\end{align}
but then $\text{dim} \{ \mathcal{G}(s,t,u,E_{i}) \}  = -10 < -6$, which yields a contradiction. We therefore fail the test, which means these amplitudes must vanish. Note that this is the case for both boost-invariant and boost-breaking theories. In \cite{McGadyRodina} it was argued that all dimensionless amplitudes, other than the pure scalar one, must vanish by virtue of the test. This result is based on pole counting so we expect those general results to be valid in our case too. \\

Now consider pure graviton scattering via the amplitude $\mathcal{A}_4(1^{-2} 2^{-2} 3^{+2} 4^{+2})$ which by the little group scaling takes the form
\begin{align} \label{GravityAmp}
\mathcal{A}_4(1^{-2} 2^{-2} 3^{+2} 4^{+2}) = \la 12 \ra^{4} [34]^{4} \mathcal{G}(s,t,u,E_{i}).
\end{align}
Now if we allow for a photon to be exchanged in the $s$-channel, the residue can have mass dimension $6$ if we use the $(+2,+2,-1)$ amplitudes and their parity counterparts. This contribution to the amplitude therefore has mass dimension $4$ and by comparing to \eqref{GravityAmp} we see that we need a $t$ or $u$ channel exchange to construct a consistently factorising amplitude. However, to achieve the required same mass dimension in either the $t$ or $u$ would require the exchange of a spin $3$ particle with non-zero $(+2,-2,\pm 3)$ amplitudes. But such amplitudes are not permitted\footnote{Indeed, if we allow for graviton exchange in the $s$-channel of the $\mathcal{A}_{4}(1^{-3},2^{+2}, 3^{-2}, 4^{+3})$, we see that the residue contains a $1/t^{3}$ piece and therefore the $(+2,-2,\pm 3)$ amplitudes are forced to vanish.}. It is therefore impossible to achieve mass dimension $6$ residues in the $t$ and $u$ channels of $\mathcal{A}_4(1^{-2} 2^{-2} 3^{+2} 4^{+2})$ and so the $(\pm 2,\pm 2,\mp 1)$ amplitudes must vanish. This is the case for both boost-invariant and boost-breaking theories considered here. \\

Now consider the $\mathcal{A}_{4}(1^{+1},2^{+1},3^{+1},4^{-2})$ amplitude which we can use to constrain the $(+1,+1,+1)$ interactions. The process is very similar to what we have seen a number of times. If we exchange a photon in the $s$-channel, we can construct a residue using the $(+1,+1,+1)$ and $(+1,-1,-2)$ amplitudes. The former has not yet been constrained beyond Bose symmetry, while the latter is required to be boost-invariant. By exchanging a photon in the other channels too we find a non-trivial factorisation constraint which fixes $F_{+1,+1,+1} = 0$\footnote{We could also exchange a $S=4$ particle to find residues with the same mass dimension, but these additional exchanges lead to spurious poles.}. So in the presence of gravity, under the assumptions we made, all three-particle amplitudes involving three photons must vanish: there are no cubic self-interactions for a gravitationally coupled photon in a boost-breaking theory with $h_{\mu \nu}$ and $A_{\mu}$ fields, just as is the case for a boost-invariant one. \\

We have two more sets of amplitudes to constrain: $(+1,+1,+2)$ and $(+2,+2,+1)$ (and their parity counterparts).  We find that both are forced to their boost-invariant limit using the four-particle test applied to $\mathcal{A}_{4}(1^{+1},2^{+1},3^{+2},4^{-2})$ and $\mathcal{A}_{4}(1^{+1},2^{+2},3^{+2},4^{-2})$ respectively. In both cases we include all possible exchanges allowed by dimensional analysis and find that any amplitudes involving higher spin ($S>2$) particles are inconsistent. The coupling of $(+1,+1,+2)$ corresponds to a new Wilson coefficient unrelated to the gravitational coupling $\kappa$. Meanwhile, the $(+2,+2,+1)$ amplitudes are forced to vanish by Bose symmetry. \\

In conclusion, all three-particle amplitudes, in theories formulated in terms of covariant fields, are forced to their boost-invariant limits when we have a photon and a graviton in the spectrum. Pure photon vertices are constrained to vanish. The only allowed amplitudes that mix the photon and the graviton are $(+1,-1,\pm2)$, $(\pm1,\pm1,\pm2)$. At the level of a Lagrangian, the parity even operators are the Maxwell kinetic term $F^{\mu\nu}F_{\mu\nu} = g^{\mu\nu} g^{\rho \sigma}F_{\mu \rho}F_{\nu \sigma}$, and the non-minimal coupling term $F^{\mu\nu}F^{\rho\sigma}R_{\mu\nu\rho\sigma}$ expanded around the vacuum $g_{\mu\nu} = \eta_{\mu\nu}$, $A_{\mu} = 0$. Parity-odd amplitudes come from $\epsilon^{\mu\nu\rho\sigma}F_{\mu\nu}F_{\rho\sigma}$ and $\epsilon^{\mu\nu\lambda \kappa}F_{\lambda \kappa}F^{\rho\sigma}R_{\mu\nu\rho\sigma}$.


\subsubsection*{Brief summary}

\begin{itemize}
\item We have seen that massless particles with $S \geq 5/2$ cannot couple to gravity under our assumptions, while particles with $S < 5/2$ can consistently couple to gravity, in which case the test yields universality of the gravitational couplings. No boost-breaking interactions are permitted. Along the way we also showed that allowing for additional particles does not change the fact that the lowest dimension vertices containing three gravitons must be boost-invariant and given by GR. We also saw that the $(\mp S,\pm(S-4),\pm2)$, $(\mp S,\pm(S+4),\mp2)$ amplitudes must vanish since for all $S \neq 2$ they yield spurious poles in gravitational Compton scattering. 
\item We have perfomed a full analysis for the cases of a graviton coupled to a scalar or a photon. In each case we find that \textit{all} three-point vertices, including the self-interactions of the scalar or photon are forced to their boost-invariant limits. 
\end{itemize}

\subsection{Multiple $S=1$ particles}\label{MultipleParticles}

We now move on to considering multiple particles of the same spin. Consistent factorisation is trivial for multiple scalar particles since the three-particle amplitudes remain only functions of the energies and therefore products of these amplitudes cannot yield singularities. In this section we will focus on multiple $S=1$ particles which we take to come in multiplets and therefore carry an additional colour index, $a = 1, 2, \ldots, N$. Our goal is to constrain the interactions between these particles in a boost-breaking theory formulated in terms of covariant fields. Recall that for a single particle ($N=1$), the $(\pm1,\pm1,\mp1)$ amplitudes are excluded by the four-particle test, whereas boost-breaking $(\pm1,\pm1,\pm1)$ amplitudes are allowed (as long as gravity is decoupled). \\

The lowest mass dimension three-particle amplitudes are 
\begin{align} \label{eq:MultipleSpin1a}
\mathcal{A}_3(1^{+1}_a 2^{+1}_b 3^{-1}_c)  & =  \frac{[12]^3}{[23][31]}F^{AH}_{abc}(E_1, E_2), \\ \label{eq:MultipleSpin1b}
\mathcal{A}_3(1^{-1}_a 2^{-1}_b 3^{+1}_c)  & =  \frac{\la 12 \ra^3}{\la 23 \ra \la 31 \ra} F^{H}_{abc}(E_1, E_2),
\end{align}
where we have eliminated $E_{3}$ by energy conservation and have dropped the helicity subscripts on $F^{H/AH}$ in favour of the colour indices. The relationship between $F^H$ and $F^{AH}$ is\footnote{We assume that the parity transformation commutes with the internal symmetry group, so that particle $a$ is mapped to particle $a$ under $P$.}
\begin{align} \label{eq:Fparity}
F^{H}_{abc}(E_1, E_2) =  \pm F^{AH}_{abc}(E_1, E_2),
\end{align}
with the $-/+$ sign corresponding to parity even/parity odd amplitudes respectively, by (\ref{eq:ParityTransformation}). In addition, Bose symmetry constrains the functions to satisfy
\begin{align} \label{eq:Fsymmetries1}
F^{H}_{abc}(E_1, E_2) & =  - F^{H}_{bac}(E_2, E_1), \\ \label{eq:Fsymmetries2}
F^{AH}_{abc}(E_1, E_2) & =  - F^{AH}_{bac}(E_2, E_1).
\end{align}

Now consider the amplitude $\mathcal{A}_4(1^{-1}_a 2^{+1}_b 3^{-1}_c 4^{+1}_d)$ with $S=1$ exchange in each channel. If the amplitude has mass dimension $2$, then there are two choices for the helicity in the $s$ and $u$ channels, and a unique choice for the $t$-channel. Remembering to take proper care of the ordering of indices and energies, we find the two residues to be
\begin{align} 
R^{+-}_s &= \sum\limits_e \frac{\la 1 3 \ra^2 [24]^2}{t} F^{AH}_{bea}(E_2, -E_1 - E_2) F^H_{ecd} (-E_3-E_4, E_3), \\
R^{-+}_s &= \sum\limits_e \frac{\la 1 3 \ra^2 [24]^2}{t} F^{H}_{eab}(-E_1 - E_2, E_1) F^{AH}_{dec} (E_4, -E_3-E_4),
\end{align}
summing over the exchanged particle colour $e$. Matching these two residues yields our first constraint on the three-particle amplitudes:
\begin{align} \label{MatchingResiduesMultiple}
&\sum\limits_e F^{AH}_{bea}(E_2, -E_1 - E_2) F^H_{ecd} (-E_3-E_4, E_3) \nonumber \\
 = & \sum\limits_e F^{H}_{eab}(-E_1 - E_2, E_1) F^{AH}_{dec} (E_4, -E_3-E_4).
\end{align}
Next consider the $u$-channel. The two residues are
\begin{align} 
R^{+-}_u & = -  \sum\limits_e \frac{\la 1 3 \ra^2 [24]^2}{s} F^{AH}_{dea}(E_4, -E_1 - E_4), F^H_{ecb} (-E_3-E_2, E_3) \\
R^{-+}_u & = -  \sum\limits_e \frac{\la 1 3 \ra^2 [24]^2}{s} F^{H}_{ead}(-E_1 - E_4, E_1) F^{AH}_{bec} (E_2, -E_2-E_3),
\end{align}
and these are equivalent thanks to \eqref{MatchingResiduesMultiple}. Finally, the $t$-channel residue is
\begin{align} 
R_t = - \sum\limits_e \frac{\la 1 3 \ra^2 [24]^2}{u} F^{H}_{ace}(E_1, E_3) F^{AH}_{bde} (E_2, E_4).
\end{align}
The full amplitude must therefore take the form
\begin{align}
A_{4}(1_{a}^{-1},2_{b}^{+1},3_{c}^{-1},4_{d}^{+1}) = \la 13 \ra^{2} [24]^{2} \left(\frac{A_{abcd}}{st} + \frac{B_{abcd}}{su} + \frac{C_{abcd}}{tu} \right),
\end{align}
with consistent factorisation fixing 
\begin{align}
A_{abcd} - B_{abcd} &= \sum\limits_e F^{AH}_{bea}(E_{2},-E_{1}-E_{2})F^{H}_{ecd}(-E_{3}-E_{4},E_{3}), \nonumber \\
C_{abcd} - A_{abcd} &= -  \sum\limits_e F^{H}_{ace}(E_{1},E_{3})F^{AH}_{bde}(E_{2},E_{4}), \nonumber \\
B_{abcd} - C_{abcd} &= - \sum\limits_e F^{AH}_{dea}(E_{4},-E_{1}-E_{4})F^{H}_{ecb}(-E_{3}-E_{2},E_{3}). 
\end{align}
Taking the sum of these equations yields
\begin{align} \nonumber 
&\sum\limits_e F^{AH}_{bea}(E_2, -E_1 - E_2) F^H_{ecd} (-E_3-E_4, E_3) \\
-&\sum\limits_e F^{H}_{ace}(E_1, E_3) F^{AH}_{bde} (E_2, E_4)  \\
-&\sum\limits_e F^{AH}_{dea}(E_4, -E_1 - E_4) F^H_{ecb} (-E_2-E_3, E_3) =0,
\end{align}
which is our main factorisation constraint and must be satisfied with \eqref{MatchingResiduesMultiple} subject to $E_{1}+E_{2}+E_{3}+E_{4} = 0$.\\

Now in the boost-invariant limit we have $F^{H}_{abc} = f_{abc} = \text{const}$, $F^{AH}_{abc} = \mp f_{abc} = \text{const}$. Under the assumption of complete antisymmetry of $f_{abc}$, matching the residues is trivial, but the primary factorisation constraint yields
\begin{equation}
\sum\limits_e f_{abe} f_{ecd}  + \sum\limits_e f_{ace} f_{edb} + \sum\limits_e f_{ade} f_{ebc}   = 0.
\end{equation}
The amplitudes in this case are those of Yang-Mills and we see that consistent factorisation of the four-particle amplitude forces the coupling constants to satisfy the familiar Jacobi identity. Note that we have made no reference to an underlying Lie-algebra; this result follows from the basic physical principles of unitarity and locality. \\

Coming back to the boost-breaking case, the system of equations is very difficult to solve in general. To make progress, we make the assumption that $F^{H}_{abc} = f_{abc}F(E_{1},E_{2})$, $F^{AH}_{abc} = \mp f_{abc}F(E_{1},E_{2})$ with $f_{abc}$ the usual couplings of Yang-Mills theory. Our three-particle amplitudes are therefore of the Yang-Mills form multiplied by a function of the energies. Bose symmetry requires these functions to be \textit{symmetric} in the exchange of their two arguments, since $f_{abc}$ are fully antisymmetric. Our factorisation constraint now becomes 
\begin{align} \nonumber 
&\sum\limits_e f_{bea}f_{ecd}F(E_2, -E_1 - E_2) F(-E_3-E_4, E_3) \nonumber \\
-&\sum\limits_e f_{ace} f_{bde}F(E_1, E_3)  F(E_2, E_4) \nonumber  \\
-&\sum\limits_e f_{dea} f_{ecb} F(E_4, -E_1 - E_4)  F(-E_2-E_3, E_3) =0.
\end{align}
Now if we don't want to impose additional constraints on $f_{abc}$, consistent factorisation requires
\begin{align}
&F(E_1, E_3) F (E_4, E_2) \\ =&F(E_2, -E_1 - E_2) F(-E_3-E_4,E_{3}) \\ = &F(E_4, -E_1 - E_4) F (E_3, -E_2-E_3).
\end{align}
Upon using \eqref{MatchingResiduesMultiple}, we see that this constraint is exactly the same as the constraint on the graviton three-particle amplitude (\ref{eq:SingleGraviton}). As shown in Appendix \ref{AppendixA}, the only solution is $F = \text{const}$ and therefore consistent factorisation requires the three-particle amplitudes to take their boost-invariant, Yang-Mills form. One may have expected the constraints for multiple $S=1$ particles to be equivalent to a single $S=2$ particle due to the kinematic-colour duality relating these amplitudes \cite{DoubleCopy}.

\section{Mind the gap: amplitudes and the flat-space limit of cosmology}\label{ssec:pofx}

In this section, we discuss the connection of our results to cosmology. Instead of considering the most general scenario, for concreteness we focus on theories of a single scalar field minimally coupled to gravity, as they are both simple and relevant for models of inflation and dark energy. For so-called $  P(X) $-theories, to be defined below, we will confirm our findings that in Minkowski all interactions must be Lorentz invariant if we impose that the scalar speed of propagation $  c_{s} $ is the same as that of the graviton, $  c=1 $, and require that the graviton be described in terms of a covariant Lagrangian (at least on the level of the free theory). Then, we consider the case in which the background is an FLRW spacetime with non-vanishing Hubble parameter, $  H\neq 0 $, and we study the sub-Hubble limit, i.e. we imagine performing a scattering experiment in a small laboratory of size $  L\ll H^{-1} $, and describe the results in terms of flat-space amplitudes. Our main observation is that for arbitrarily small but non-vanishing $  H $, it is always possible to find amplitudes that break boosts by any amount, within the validity of the Effective Field Theory (EFT), and no violations of unitarity or locality seem to arise. We argue that, despite the appearance, this observation does not imply any pesky physical discontinuity. Rather, we interpret this finding as the fact that \textit{the constraining power of unitarity and locality through consistent factorization for massless theories is extremely fragile to IR modifications}. An analogous principle has already been established in Lorentz invariant contexts, where many interactions prohibited in flat space have consistent counterparts in AdS, regardless of the AdS radius - see \cite{Bekaert:2010hw} and references therein. Nonetheless, we decided to illuminate this issue further by discussing FLRW backgrounds which are more closely related to cosmology. \\

Sensitivity to IR modifications in cosmological scenarios is to be expected on the following grounds. Factorization happens when $  s $, $ t $ or $  u $ go to zero and that's where all the constraining power of the four-particle test comes from. But this regime cannot be reached within the validity of the sub-Hubble limit. Indeed, for a flat-space approximation of FLRW spacetime to make sense, we need to require that the quantum uncertainty $  \Delta x $ on the spacetime position of the scattering particles is well within a Hubble volume $  \Delta x \ll H^{-1} $. But then by the uncertainty principle 
\begin{align}
\Delta p \geq \frac{1}{2\Delta x} \gg H \then \Delta s,\Delta t,\Delta u \gg H^{2}\,,
\end{align}
and therefore we always have an uncertainty in the Mandelstam variables of order $ H^{2} $. In FLRW spacetime, we cannot meaningfully distinguish, say, a pole at $  s=0 $ from one at $  s=H^{2} $. In more physical terms, as long as $  H\neq 0 $, we cannot experimentally reach the poles corresponding to massless on-shell particles while neglecting the expansion of the universe. Our finding that in the presence of an interacting spin-2 particle boost-breaking interactions cannot satisfy consistent factorization on $  s,t,u=0 $, respectively, does not seem to matter in FLRW spacetime where this kinematic regime cannot be reached in the flat-space limit.\\ 

The suspicious reader might complain that our results suggest the presence of an unphysical discontinuity as $ H \to 0 $, but this is not the case. In the deep IR of the theory, a background with $  H \neq 0$ is always very different from one with $ H=0 $ because of the presence of a Hubble ``horizon''. So it is to be expected that any IR property of the theory for $  H\to 0 $ might be different from the corresponding one at $  H=0 $. In other words, one cannot engineer a continuous series of physical thought experiments that give a discontinuous set of results and so there is no problem with our claims in this section. \\

Before proceeding, let's stress that there might be other obstructions to Lorentz breaking interactions when $  c_{s} =1 $, which we don't capture in our analysis. For example, \cite{Conjecture} conjectured that for the theory to have a local and unitary Lorentz invariant UV-completion, all Lorentz-breaking interactions for a single scalar with non-linear boosts must vanish as $  c_{s}\to 1 $. Also, recently \cite{TanguyScott} found some related obstructions considering perturbative unitarity in the sub-Hubble limit, where they showed that the window of validity of an EFT description for amplitudes shrinks to zero when $  c_{s}\to 1 $ in the presence of $  \dot\phi^{3} $ interactions.

 
\subsection{The absence of boost-breaking interactions in Minkowski}\label{ssec:}
For concreteness, consider so-called $  P(X) $ theories minimally coupled to gravity with action
\begin{align}
S&=-\int d^{4}x \sqrt{-g}  \left[ \frac{\Mpl^{2}}{2}R+P(X) \right]\,,& X\equiv \frac{1}{2}g^{\mu\nu}\partial_{\mu}\phi\partial_{\nu}\phi\,,
\end{align} 
which is a good toy model to study the spontaneous breaking of boosts while preserving time translations. The homogeneous equations of motion for the  background $  \phi(t)  $ and the scale factor $  a(t) $ are
\begin{align}
 3\Mpl^{2}H^{2} - 2X P_X + P=0,\quad -\Mpl^{2}\dot H=XP_{X}, \quad \ddot \phi \left(  P_{X}+2XP_{XX}\right)+3H\dot \phi P_{X} =0\,.
\end{align}
The Lagrangian for perturbations $   \varphi(t,\vec x) $ is
\begin{align}\label{l2}
\L=\frac{1}{2} (P_{X}+2XP_{XX})\dot\varphi^{2}-\frac{1}{2}P_{X}\partial_{i}\varphi\partial^{i}\varphi+ \frac{1}{6} P_{XXX} \dot{\phi}^{3} \dot \varphi^{3} +\dots\,,
\end{align}
where the dots stand for higher derivatives of $  P(X) $ with respect of $  X $, which will not be relevant for this discussion (they could be chosen to vanish if desired). The speed of sound is found to be
\begin{align}
c_{s}^{2}=\frac{P_{X}}{P_{X}+2XP_{XX}}\,.
\end{align}
In this class of theories, it is only possible to have a well-defined solution in Minkowski spacetime with $  c_{s}=1 $ if $  X=0 $, in which case all interactions are Lorentz invariant. To see why, note that the following three assumptions cannot all be satisfied at the same time:
\begin{itemize}
\item Spontaneously broken boosts: This implies $  X \neq 0 $. From the equations of motion, setting $  H=0 $ and $  P_{X}=0 $ as appropriate for Minkowski, we get
\begin{align}
 \ddot \phi \left(  P_{X}+2XP_{XX}\right)=0 \then \ddot \phi=0 \text{ or } P_{X}+2XP_{XX}=0\,.
\end{align}
The second option is the cuscuton \cite{Afshordi:2006ad}, which is non-dynamical and so not relevant for the present discussion. From $\ddot\phi=0 $ we deduce that $  X $ is constant, and so if it is non-vanishing it remains so for all times.
\item Luminal propagation: This implies $  c_{s}=1 $ and so
\begin{align}
c_{s}^{2}=\frac{P_{X}}{P_{X}+2XP_{XX}}\overset{!}{=}1 \then  P_{X}\neq 0 \, \& \,( P_{XX}=0\text{ or }X=0)\,.
\end{align}
%
\item Minkowski spacetime with \textit{dynamical} gravity: This implies $  g_{\mu\nu}=\eta_{\mu\nu} $ and so 
\begin{align}
\left\{ \begin{array}{ll}  3\Mpl^{2}H^{2}=2XP_{X}-P\overset{!}{=}0 \\ -\Mpl^{2}\dot H=XP_{X} \overset{!}{=}0\end{array}\right. \then P=0 \,\&\, (P_{X}=0 \text{ or } X=0)\,.
\end{align}
\end{itemize}  
Combining the above requirements we arrive at a contradiction: if we insist that $  X\neq 0 $, so that a Lorentz violation is in principle possible, then the luminality and Minkowski requirements are incompatible because the former leads to $  P_{X}\neq 0 $, while the latter entails $  P_{X}=0 $. While we don't discuss it here in detail, the above result also applies to theories with higher derivatives. Intuitively, this stems from the fact that the higher derivative terms vanish when evaluated on the linearly time-dependent background we considered above. \\

This discussion confirms and complements our result that coupling to gravity (i.e. the covariant $h_{\mu \nu}$) in Minkowski enforces Lorentz invariance. On the one hand, our amplitude discussion is more general as it does not assume a $P(X)$ Lagrangian. On the other hand, the above discussion generalized our findings in that it shows, for $P(X)$ theories, that all $n$-particle amplitudes must be Lorentz invariant if the scalar propagates at the same speed as the graviton. In appendix \ref{AmplitudeCalculation}, we provide further clarifications on why a simple $\dot{\phi}^3$ theory coupled to gravity is inconsistent in Minkowski space.

 
\subsection{Boost-breaking interactions in the sub-Hubble limit}\label{sec:}

The attentive reader will have noticed that when $  P_{X}=0=P_{XX} $, the speed of sound is ill defined, $  c_{s}\overset{?}{=}0/0 $. In particular, the order of taking the limits matters: if we first impose Minkowski by setting $ P_{X}  =0 $, then $ c_{s} =0 $ for any finite $  P_{XX} $; while if we first impose $  c_{s}=1 $ by setting $  P_{XX}=0 $, then we can take the Minkowski limit of FLRW, $  P_{X}\to 0 $, without changing the value of $  c_{s} $. In this section, we discuss in detail this second possibility and find that in this case, Lorentz-breaking interactions are allowed within the regime of validity of the EFT. Let us now study how the Minkowski and $  c_{s}^{2}=1 $ solutions are \textit{approached} from an FLRW solution. \\

Let us first assume the value $  \bar X $ of $  X(t)  $ at some time is such that
\begin{align}
P_{XX}(\bar X)=0 \quad \text{but} \quad  P_{X}(\bar X)\neq 0\,.
\end{align}
Expanding around it, we find
\begin{align}
P_{XX}(X)&=P_{XX}(\bar X)+(X-\bar X)P_{XXX}(\bar X)+\O((X-\bar X)^{2})\,,\\
&=(X-\bar X)P_{XXX}(\bar X)+\O((X-\bar X)^{2})\,.
\end{align}
The background equations of motion to zeroth order in $  X-\bar X $ are
\begin{align}
P_{X}(\bar X)\left(  \ddot \phi +3H \dot \phi\right)+\O\left( X-\bar X \right)&=0\,,\\
\dot \phi \left(  \ddot \phi +3H \dot \phi\right)&\simeq \O\left( X-\bar X \right)\,, \\
\dot X +6H  X &\simeq \O\left( X-\bar X \right)\,,
\end{align}
and so are solved by $  X\propto a^{-6} $. More usefully, for a small time interval $  \Delta t \ll H^{-1} $, we can write
\begin{align}
X &= \bar X + \dot{\bar X} \Delta t + \O ((X-\bar X)^{2}) \\
\then \frac{X-\bar X}{\bar X}&\simeq -6 H \Delta t + \O\left( \left( X-\bar X \right)^{2} \right)\,. \label{Xevol}
\end{align}
So we find that, unlike in Minkowski where a constant $  X $ is always a solution, in FLRW we have to take into account that $  X $ evolves with time at some rate set by $  H $. \\

Consider now the theory of perturbations in \eqref{l2}. 
Since $  X $ depends on time and we don't want to assume $  P(X) $ is just linear in $  X $, which corresponds to the free theory, we cannot set $  c_{s}^{2}=1 $ at all times, but only at the time corresponding to $  X=\bar X $ where $  P_{XX} $ happens to vanish. We can Taylor expand around $  c_{s}-1\to 0 $ and re-write $  c_{s} $ as
\begin{align}
c_{s}^{2}&=\frac{P_{X}}{P_{X}+2XP_{XX}}\\
&=1-\frac{2X P_{XX}}{P_{X}}+\mathcal{O}\left(  \left(  \frac{2X P_{XX}}{P_{X}}\right)^{2}\right)\\
&=1-\frac{2 \bar X (X-\bar X) P_{XXX}(\bar X)}{P_{X}(\bar X)}+\mathcal{O}\left(  \left(  X-\bar X\right)^{2}\right)\,.
\end{align}
Using \eqref{Xevol} for the time evolution of $ X  $, this becomes
\begin{align}
1-c_{s}^{2}&=-\frac{12 H \Delta t \bar X^{2} P_{XXX}(\bar X)}{P_{X}(\bar X)}+\mathcal{O}\left(  \left(  X-\bar X\right)^{2}\right)\,.
\end{align}
Now we want to ask whether we can keep $  1-c_{s}^{2} $ arbitrary small while performing a subHubble scattering experiment in which some $  \varphi $ particles interact via the (spontaneously) boost-breaking coupling $  \dot \varphi^{3} $ in the Lagrangian \eqref{l2}. We canonically normalize $  \varphi $ to $  \varphi_{c} $ and extract the cutoff scale $  \Lambda $ of the $ \dot  \varphi_{c}^{3} $ operator
\begin{align}
\L_{2}&=\frac{1}{2}\left[  \frac{1}{c_{s}^{2}}\dot\varphi_{c}^{2}-\frac{1}{2} \partial_{i}\varphi_{c}\partial^{i}\varphi_{c} \right]+\frac{\sqrt{2}}{3} \frac{ X^{3} P_{XXX}}{ (X P_{X})^{3/2}}   \dot \varphi_{c}^{3}\\
&\equiv \frac{1}{2}\left[  \frac{1}{c_{s}^{2}}\dot\varphi_{c}^{2}-\frac{1}{2} \partial_{i}\varphi_{c}\partial^{i}\varphi_{c} \right]+\frac{\dot \varphi_{c}^{3}}{\Lambda^{2}} \,.
\end{align}
Since we rescaled by $  P_{X} $, which is time dependent, we also pick up additional terms proportional to $  \partial_{t} P_{X} $, such as a mass term. We have neglected writing these terms because, around $  X=\bar X $,
\begin{align}
\partial_{t} P_{X}(X)&=P_{XX}(X)\dot X  \\
&\simeq - 6 H \Delta t \bar X(X-\bar X)  P_{XXX}(\bar X) +\dots \,, \\
&\simeq 36 \left( H \Delta t \right)^{2} \bar X^{2}  P_{XXX}(\bar X) +\dots \,, 
\end{align}
which is suppressed by at least two powers of $  H \Delta t $. As long as we can neglect the expansion of the universe for some time $  \Delta t \ll H^{-1} $, we can also neglect these additional terms.

Since $  P_{XXX} $ sets both the scale for the time evolution of $  1-c_{s}^{2} $ and the strength of the interaction we re-write
\begin{align}
1-c_{s}^{2}&=-\frac{36}{\sqrt{2}}\frac{H\Delta t\sqrt{XP_{X}}}{\Lambda^{2}}+\mathcal{O}\left(  \left(  X-\bar X\right)^{2}\right)\\
&=-\frac{36}{\sqrt{2}}\left(  \frac{E^{2}}{\Lambda^{2}} \right)\left( \frac{\sqrt{-\dot H \Mpl^{2}}}{E^{2}} \right)\left( H\Delta t \right)+\mathcal{O}\left(  \left(  X-\bar X\right)^{2}\right)\,,
\end{align}
where we introduced the dummy factor $  E $ to represent the energy scale of the scattering process. For the scattering to happen effectively in flat space we need $  E^{2}\gg H^{2},|\dot H| $. To resolve energies of order $  E $ while being able to neglect the expansion of the universe during the experiment, we need the experiment to last a time $ H^{-1} \gg \Delta t \gg E^{-1} $. Finally, perturbativity requires $  E \ll \Lambda $. Then
\begin{align}
1-c_{s}^{2}&\gg -\frac{36}{\sqrt{2}}\left(  \frac{E}{\Lambda} \right)^{2}\left( \frac{\sqrt{-\dot H }}{E} \right)\left( \frac{H}{E}\right)\left(  \frac{\Mpl}{E}\right)+\mathcal{O}\left(  \left(  X-\bar X\right)^{2}\right)\,.
\end{align}
The first three factors must be much smaller than one while $  \Mpl/E $ must be much larger than one. Summarizing, we want the hierarchy of scales 
\begin{align}
H,\sqrt{-\dot H} \ll E\ll  \Lambda \ll \Mpl\,,
\end{align}
while keeping $  1-c_{s}^{2} $ arbitrary small. This is always possible to achieve for any desired $  E/\Lambda $, which parameterizes the strength of the cubic interaction), and $  \Lambda /\Mpl $ simply by taking $  H,\sqrt{-\dot{H}} $ sufficiently small.\\

The upshot of this discussion is that we can find solutions for which a scattering experiment in a small lab in an FLRW spacetime gives Lorentz-breaking amplitudes for massless particles that all move at the same speed to arbitrary but finite precision. Given the assumptions we have made about four-particle amplitudes, our results have shown that if this happened in Minkowski spacetime, there would be a violation of unitarity and/or locality for the amplitudes. But in FLRW those configurations cannot be reached while still neglecting corrections due to the expansion of the universe.

\section{Discussion and conclusion}\label{sec:Discussion}


In this paper we studied scattering amplitudes for massless, luminal, relativistic particles of any spin without demanding  Lorentz invariance of the interactions. This is relevant for many systems that break Lorentz boosts spontaneously, as in cosmology or condensed matter physics. We focussed exclusively on on-shell particles and discussed (analytically continued) amplitudes without reference to unphysical structures such as gauge invariance or off-shell particles. The on-shell approach considerably simplifies the treatment of spinning particles, and our conclusions are independent of perturbative field redefinitions. \\

We systematically derived all possible massless three-particle amplitudes consistent with spacetime translations and rotations and constrained them using unitarity and causality via the requirement that four-particle amplitudes consistently factorize on simple poles into the product of two three-particle amplitudes, a.k.a. the four-particle test \cite{Benincasa:2007xk}. We found that a large number of three-particle amplitudes fail the test and therefore cannot arise in any local, unitary perturbative theory around Minkowski spacetime. One result that stands out is that the existence of an interacting graviton, namely a massless spin-2 particle, enforces all cubic interactions involving particles coupled to it to be Lorentz invariant, including those interactions that do not involve the graviton. This is quite remarkable because, in the absence of a graviton, there could be infinitely many Lorentz-breaking interactions. As a concrete and simple example, consider the theory of a single scalar, 
for which we can write down infinitely many local interactions of the form $  (\partial_{t}^{n_{1}}\phi)( \partial_{t}^{n_{2}}\phi)( \partial_{t}^{n_{3}}\phi) $ for any positive integers $  n_{1,2,3} $. These interactions are not equivalent on-shell, generically giving different amplitudes, yet they are all allowed by the four-particle test. Our results show that in Minkowski, none of these Lorentz-breaking interactions can be consistently coupled to gravity! \\

\noindent Although the form of the three-particle amplitudes that we have derived are completely general, in order to make progress we assumed that the helicity scaling of four-particle amplitudes are fixed by angle and square brackets rather than round ones. As we explained in Section \ref{sec:4p}, this amounts to assuming that the underlying Lagrangian is a function of Lorentz covariant fields with the breaking of boosts driven solely by time derivatives. It would be very interesting to work with a more general ansatz for the four-particle amplitudes such that we can constrain theories constructed out of $SO(3)$ covariant fields.\\

Finally, we have discussed the relation of our analysis to cosmological models, in which spacetime can be approximated as flat only locally, but is never flat asymptotically. We found that, contrary to what happens in Minkowski, one can find models of a massless luminal scalar coupled to dynamical gravity in which sub-Hubble scattering is boost-breaking while no violations of unitarity and locality arise in the IR within the validity of the required approximations. We interpreted this as the observation that \textit{the four-particle test is IR-sensitive} and the expansion of the universe provides an IR modification of the on-shell conditions. This finding mirrors the analogous findings for Lorentz invariant theories, where the four-particle test is not applicable if one deviates ever so slightly from asymptotically flat space \cite{Bekaert:2010hw}. \\

One of our main motivations for studying boostless amplitudes was to use the results to constrain and perhaps fully bootstrap cosmological correlators when de Sitter boosts are not a symmetry of the theory. Our findings shows yet another reason why several clarifications need to be added to the simplistic slogan that the residue of the $  k_{T} $ pole of cosmological correlators is the Minkowski amplitude. In particular, we have shown that consistent factorization (Theorem 2.1) imposes severe constraints on Minkowski amplitudes, but these constraints don't necessary apply to the residue of the total-energy pole of correlators in \eqref{BtoA}. This issue will be discussed in detail elsewhere. \\

There are several ways in which our results could be extended.
\begin{itemize}
\item We used the consistent factorization of four-particle amplitudes to constrain three-particle amplitudes. It would be desirable to extend our analysis to higher $  n $-particle amplitudes. For example, we expect that the coupling to a massless graviton will enforce \textit{all} interactions to be Lorentz invariant. While the pedestrian methods we used in this paper are probably ill-suited to prove this more general result, one would probably want to harvest the power of on-shell recursion relations.
\item It would be interesting to study how unitarity and locality constrains scattering experiments in the sub-Hubble limit of FLRW spacetime. This requires modification of the standard on-shell methods and an analysis will appear elsewhere.
\item It would be interesting to extend our analysis to more general on-shell conditions where different particles can have different speeds, and to allow for a more general form of the four-particle amplitudes such that we capture the type of theories derived in \cite{Zoology}. 
\end{itemize}

\section*{Acknowledgements}
We would like to thank Brando Bellazzini, Paolo Benincasa, Tanguy Grall, Sadra Jazayeri, Scott Melville, and Dong-Gang Wang for useful discussions and comments on a draft. E.P. and D.S. have been supported in part by the research program VIDI with Project No. 680-47-535, which is (partly) financed by the Netherlands Organisation for Scientific Research (NWO). J.S. has been supported by a grant from STFC.

\appendix

\section{Spinor variables and discrete transformations} \label{AppendixSigns}

In this appendix we prove two important results for spinor representations of lightlike momenta, namely their transformation law under spatial reflection and the prescription for transforming the spinors so as to flip the sign of the exchanged particle's energy and momentum, which is necessary to compute the residues correctly.

\subsection*{Spatial reflection}

Under the spatial reflection with respect to the origin, lightlike momentum $p^{\mu}$ tranforms as
\begin{equation}
(E, \bvec{p}) \mapsto (E, -\bvec{p}).
\end{equation}

To the original momentum $p^{\mu}$ we associate a pair of spinors $(\lambda^{\alpha}, \tilde{\lambda}^{\dot{\alpha}})$. One choice is
\begin{equation}
\lambda = \left( \sqrt{p^0 + p^3} , \frac{p^1 + i p^2}{\sqrt{p^0 + p^3}} \right)^T, \ \tilde{\lambda} = \left( \sqrt{p^0 + p^3} , \frac{p^1 - i p^2}{\sqrt{p^0 + p^3}} \right) .
\end{equation}
Spinor helicity variables corresponding to the new momentum must be of the form
\begin{equation} \label{eq:ParityWitht}
\lambda'_{\alpha} = a \epsilon_{\alpha}^{\ \dot{\beta}} \tilde{\lambda}_{\dot{\beta}}, \ \tilde{\lambda}'_{\dot{\alpha}} = a^{-1} \epsilon_{\dot{\alpha}}^{\ \beta},\lambda_{\beta}
\end{equation}
i.e.
\begin{equation}
\lambda' = a (\tilde{\lambda}_2, - \tilde{\lambda}_1)^T, \ \tilde{\lambda}' = a^{-1} (\lambda_2, - \lambda_1).
\end{equation}
It is easy to check that these new variables do indeed give $p'^{\mu} = (E, -\bvec{p})$. Now we must fix the coefficient $a$. To do this, we have to take a look at polarization tensors.\\ 

Consider an exchange diagram with an exchanged particle of spin-$1$. Suppose at the left-hand side vertex, there is an outgoing particle of helicity $+1$ (equivalent to an incoming antiparticle of helicity $-1$). Then the same particle (with helicity $+1$) is incoming at the right-hand side vertex. The $+1$ polarization vector $\xi^+$ of the exchanged particle is mapped to $P \xi^+$ under spatial reflection $P$. But we also require, for consistency, that it be mapped to the $-1$ polarization vector of the particle with reversed momentum. The spatial reflection of $\xi^+$ is, in terms of spinor variables,
\begin{equation}
P \xi_{\alpha \dot{\alpha}}^+(\bvec{p}) = \frac{\epsilon_{\alpha}^{\ \dot{\beta}}\epsilon_{\dot{\alpha}}^{\ \beta} \mu_{\beta} \tilde{\lambda}_{\dot{\beta}}}{\la \mu, \lambda \ra},
\end{equation}
where we used (\ref{eq:ParityWitht}), and $\mu$ is a reference spinor. Now, the $-1$ polarization vector relative to $- \bvec{p}$ momentum is
\begin{equation}
 \xi_{\alpha \dot{\alpha}}^-(-\bvec{p}) =  \frac{\lambda'_{\alpha} \tilde{\zeta}'_{\dot{\alpha}}}{[ \tilde{\lambda}', \tilde{\zeta}' ]}   = -a^2
\frac{\epsilon_{\alpha}^{\ \dot{\beta}} \epsilon_{\dot{\alpha}}^{\ \beta} \zeta_{\beta} \tilde{\lambda}_{\dot{\beta}}}{\la \zeta, \lambda \ra}.
\end{equation}
Setting $\zeta = \mu$ and comparing the two expressions, we conclude that $a^2 = -1$, i.e. $a = \pm i$. Thus, the prescription for mapping $(E, \bvec{p}) \mapsto (E, -\bvec{p})$ is (for example),
\begin{equation}
\lambda' = (-i \tilde{\lambda}_2, i \tilde{\lambda}_1)^T, \quad \tilde{\lambda}' = (i\lambda_2, - i\lambda_1).
\end{equation}
Under spatial reflection, the two inner products then transform as, e.g,
\begin{eqnarray}
[1 2] & \mapsto & [1' 2'] = \la 2 1 \ra = - \la 1 2 \ra, \\
\la 1 2 \ra & \mapsto & \la 1' 2' \ra = [ 2 1 ] = - [ 1 2 ].
\end{eqnarray}
This transformation law leads to consistent results for various 3p amplitudes - see, for example, Appendix \ref{AppendixB}.

\subsection*{The $p_I \mapsto - p_I$ prescription}
Consider again a diagram in which a particle with helicity $+1$ is being exchanged. Let's transform this diagram under $TP$. Then the polarization 4-vector of the intermediate particle flips its sign: $\xi^{\mu} \mapsto - \xi^{\mu}$. On the other hand, this new 4-vector must be precisely the $+1$ polarization vector relative to $-p_I$ (helicity of the exchanged particle doesn't change under $TP$). Schematically, the $\pm 1$ polarization vector is proportional to $\left( \tilde{\lambda}/\lambda \right)^{\pm 1}$. Thus, if $p_I \leftrightarrow (\lambda, \tilde{\lambda})$, then we must have $- p_I \leftrightarrow  (\lambda, -\tilde{\lambda}) $ (or $(-\lambda, \tilde{\lambda})$) to give consistent polarization vectors. We extrapolate this conclusion to spins other than $1$. This convention produces the correct relative signs in the amplitudes - see, for example,  the discussion in Section \ref{sec:GravitonCouplings}. 
\section{Formulas for the framid amplitude} \label{AppFramid}
Here we list the functions we used in \eqref{eq:Framid4pt} to write down the framid exchange four-particle amplitude $\mathcal{A}_{4}(1^0, 2^+, 3^0, 4^-)$:
\begin{eqnarray}
F_{(1,a)}(E_1, E_2, E_3, E_4; s, t)& =  &
-4 e_4 E_{12}^2 - 2 s E_{12} E_{23} f +  s^2 (E_1 - E_2)(E_3 - E_4) - \frac{s^2}{t} g, \\ 
F_{(1,b)}(E_1, E_2, E_3, E_4; s, t) & = & 
12 e_4 E_{12} + \frac{2s}{t}(E_2 - E_4) g + 3s E_{24} f ,
 \\ 
F_{(1,c)}(E_1, E_2, E_3, E_4; s, t)& = & -9 e_4 + \frac{4 E_2 E_4}{t} g, \\ \nn
F_{(2,a)}(E_1, E_2, E_3, E_4; s, t)& = & 
4e_4 (E_1^2 + E_1 E_3 + E_3^2) + t^2 f - s t (E_1 - E_3) (E_2 - E_4) \\ 
& \quad &  + s E_1 E_3 (E_1 - E_3)(E_2 - E_4) \\ \nn
& \quad & - t E_1 E_3 \left( - E_{13}^2 + E_2 E_3 \left( 1 + \frac{2 E_4}{E_1}  \right) + E_1 E_4 \left( 1 + \frac{2E_2}{E_3} \right)  \right),   \\
F_{(2,b)}(E_1, E_2, E_3, E_4; s, t) & = &  2 (E_1 - E_3) (E_2^2 + E_2 E_4 + E_4^2) (t - E_1 E_3 ),
\end{eqnarray}
where we used
\begin{eqnarray}
f & = & E_1 E_4 + E_2 E_3, \\ 
g & = & 4e_4 + \frac{1}{2}E_1 E_3 (2 E_1 + E_2)(2E_3 + E_4) + s f, \\
E_{ij} & = & E_i + E_j,  \\
e_4 & = & E_1 E_2 E_3 E_4.
\end{eqnarray}
For completeness, we also list all on-shell, three-particle amplitudes for the framid, in the case of equal speeds $c_L= c_T$. We find
\begin{eqnarray}
A_3(1^+ 2^+ 3^+) & = & 0, \\
A_3(1^+ 2^+ 3^-) & = & \sqrt{2} g \left( E_1 - E_2 \right) \frac{[12]^3}{[23][31]}, \\
A_3(1^+ 2^+ 3^0) & = & g  [12]^2, \\
A_3(1^+ 2^- 3^0) & = & \frac{1}{2} g (21)^2, \\
A_3(1^+ 2^0 3^0) & = & - \frac{1}{\sqrt{2}} g \left( E_1 + 2 E_2 \right) \frac{[12][31]}{[23]}, \\
A_3(1^0 2^0 3^0) & = & 2 g \left( E_1 E_2 + E_2 E_3 + E_3 E_1 \right).
\end{eqnarray}
where
\begin{equation}
g =  \frac{c_{L}^{2}-1}{c_L^2 M_1}.
\end{equation}

\section{Solutions to constraints on $F(E_i)$} \label{AppendixA}

In this appendix we provide proofs that the only rational functions of the form
\begin{equation}
F(x,y) = \frac{f(x,y)}{x^n y^m (x+y)^k}
\label{eq:RationalForm}
\end{equation}
that solve \eqref{eq:SingleVector2} and \eqref{eq:SingleGraviton} are $F=0$ and $F=\text{const}$ respectively.

\subsection*{Photon constraint}

We begin with the constraint \eqref{eq:SingleVector2}. We allow $F$ to take the form (\ref{eq:RationalForm}) and we have already shown that the antisymmetry in the first two arguments of $F$ requires $n=m$. Thus $F(x,y) = (x y)^{-m} (x+y)^{-k} f(x,y)$, where the function $f$ must be alternating in its two variables. We therefore write $f(x,y) = (x-y)P[x+y,xy]$ where $P$ is another polynomial. Our factorisation constraint \eqref{eq:SingleVector2} is then
\begin{eqnarray} \nonumber
(-1)^k \frac{(E_1 - E_3) (E_2 - E_4)}{E_1^m E_2^m E_4^m (E_1 + E_3)^{2k} } P[E_1 + E_3, E_1 E_3] P[E_2 + E_4, E_2 E_4] \\ \nonumber
+ (-1)^m \frac{(E_1 + 2E_2)(2E_3 + E_4)}{E_1^k E_2^m E_4^k (E_1 + E_2)^{2m} } P[-E_1, -E_2(E_1 + E_2)]  P[-E_4, - E_3 (E_3 + E_4)] \\
- (-1)^m \frac{(E_1 + 2E_4) (E_2 + 2E_3)}{E_1^k E_2^k E_4^m (E_1 + E_4)^{2m}} P[-E_1, -E_4(E_1 + E_4)] P[-E_2, - E_3 (E_2 + E_3)] = 0.
\label{eq:SingleVectorNew1}
\end{eqnarray}
First, we are going to assume that $P$ is non-zero and, by examining the singularities, deduce that $m = k = 0$. By assumption, $P[x,y]$ is not divisible by $x$ or $y$, so the first term in (\ref{eq:SingleVectorNew1}) is singular at $E_1 + E_3 = 0$ for $k > 0$ while neither the second nor the third term are singular there. Thus, we must have $k = 0$. By a similar argument, we also have $m = 0$. Our main equation thus simplifies to
\begin{eqnarray} \nonumber
(E_1 - E_3) (E_2 - E_4) P[E_1 + E_3, E_1 E_3] P[E_2 + E_4, E_2 E_4] \\ \nonumber
+ (E_1 + 2E_2)(2E_3 + E_4) P[-E_1, -E_2(E_1 + E_2)]  P[-E_4, - E_3 (E_3 + E_4)] \\
- (E_1 + 2E_4) (E_2 + 2E_3) P[-E_1, -E_4(E_1 + E_4)] P[-E_2, - E_3 (E_2 + E_3)] = 0,
\label{eq:SingleVectorNew}
\end{eqnarray}
and this equation must be satisfied for all energies subject to $E_{1}+E_{2}+E_{3}+E_{4} = 0$. Now we will aim to show
\begin{equation}
\left( P \left[ x, \frac{(3 \cdot 2^{n+1} - 2)}{(3 \cdot 2^{n+1} - 1)^2} x^2 \right] =0 \quad \forall x \quad \textrm{OR} \quad 
P \left[ x,  3 \cdot 2^{n} \cdot \frac{3 \cdot 2^{n} - 1}{(3 \cdot 2^{n+1} - 1)^2} x^2 \right] = 0 \quad \forall x \right) \forall n \in \mathbb{Z}_{\geq 0},\label{eq:Pconclusion}
\end{equation}
which entails $P \equiv 0$. The reason for this is that $P$ would have to satisfy infinitely many distinct constraints of the form $P[x, a_k x^2] = 0 \quad \forall x$ (it is easy to check that $a_k$ are indeed distinct) and thus we would need $(a_k x^2 - y) \mid P[x,y]$ for all the $a_k$, which is impossible if $P$ is a nonzero polynomial. \\

To prove (\ref{eq:Pconclusion}), let
\begin{eqnarray} \nonumber
E^{(n)}_1 & = & (3 \cdot 2^{n+1} - 2)x, \\ \nonumber
E^{(n)}_2 & = & -(3 \cdot 2^{n})x, \\ \nonumber
E^{(n)}_3 & = & x, \\ \label{eq:Induction}
E^{(n)}_4 & = & -(3 \cdot 2^{n}-1)x,
\end{eqnarray}
for $n = 0,1,2,\ldots$. Note that $E_1 = -2E_4$ for any $n$, in which case the third term in \eqref{eq:SingleVectorNew} vanishes and the main equation becomes
\begin{align} \nonumber
(E_3 + 2E_4) P[E_3 - 2E_4, -2 E_3 E_4] P[E_2 + E_4, E_2 E_4]  \\
= 2(2E_3 + E_4) P[2 E_4, -E_2(E_2 - 2E_4)]  P[-E_4, - E_3 (E_3 + E_4)].
\label{eq:MainPEquation1}
\end{align}
Taking $n = 0$, we get
\begin{equation}
- 3 x P[5x, 4x^2] P[-5x, 6x^2] = 0,
\end{equation}
so
\begin{equation}
P[5x, 4x^2] =0 \quad \forall x \quad \textrm{OR} \quad P[-5x, 6x^2] = 0 \quad \forall x ,
\end{equation}
or equivalently,
\begin{equation}
P[x, \frac{4}{25}x^2] =0 \quad \forall x \quad \textrm{OR} \quad P[-x, \frac{6}{25}x^2] = 0 \quad \forall x ,
\end{equation}
which is precisely the condition from (\ref{eq:Pconclusion}) for $n = 0$. Now we will prove (\ref{eq:Pconclusion}) for any $n > 0$ by induction. Suppose (\ref{eq:Pconclusion}) is true for some $n-1$. Then set $E_i$ to the values specified in (\ref{eq:Induction}). We get
\begin{align} \nonumber
(3 - 3 \cdot 2^{n+1}) x P[(3 \cdot 2^{n+1} - 1) x , (3 \cdot 2^{n+1} - 2)x^2] 
P[-(3 \cdot 2^{n+1} - 1) x, 3 \cdot 2^n (3 \cdot 2^n - 1) x^2] = \\
= 
2 (3 - 3 \cdot 2^{n}) x 
P[(3 \cdot 2^{n} - 1) x, (3 \cdot 2^{n} - 2) x^2]
P[-(3 \cdot 2^{n+1} - 2) x, 3 \cdot 2^{n} \cdot (3 \cdot 2^{n} - 2) x^2].
\end{align}
The right hand side is zero by virtue of the previous induction step. Thus, the left hand side is also zero, which entails
\begin{equation}
P[ x, \frac{(3 \cdot 2^{n+1} - 2)}{(3 \cdot 2^{n+1} - 1)^2} x^2] =0 \quad \forall x \quad \textrm{OR} \quad 
P[x,  3 \cdot 2^{n} \cdot \frac{3 \cdot 2^{n} - 1}{(3 \cdot 2^{n+1} - 1)^2} x^2] = 0 \quad \forall x,
\end{equation}
thereby completing the proof. This proves that there are no consistent $(+1,-1\pm1)$ amplitudes under the assumption made in (\ref{eq:SemiCovLagr}) and discussed in that section. \\

\subsection*{Graviton constraint}
We now show that the only solution to the system of equations\footnote{In fact, we need only 2 equations - those relating the second, third and fifth expression in (\ref{eq:SingleGraviton}) - and we can drop the condition that the residue must be the same regardless of how the pole is approached.} (\ref{eq:SingleGraviton}) is $F = \text{const}$ thereby reducing the $(+2,-2,\pm2)$ amplitudes to their boost-invariant limits. \\

Here $F$ must be of the form
\begin{equation}
F(x,y) = \frac{f(x,y)}{x^m y^m (x+y)^k}
\label{eq:RationalForm2}
\end{equation}
where $f$ is a symmetric polynomial, so $f(x,y) = P[x+y, xy]$ for some polynomial $P$. Thus (\ref{eq:SingleGraviton}) takes the form
\begin{eqnarray} \nonumber
& \ & \frac{(-1)^{k+m}}{E_2^m E_3^m E_4^m (E_1 + E_3)^{2k}} P[E_1 + E_3, E_1 E_3] P[E_2 + E_4, E_2 E_4]  \\ \nonumber
& = & \frac{1}{E_2^k E_3^k E_4^m (E_1 + E_2)^{2m}} P[-E_2, -E_1(E_1 + E_2)]  P[-E_3, - E_4 (E_3 + E_4)] \\
& = & \frac{1}{E_2^m E_3^k E_4^k (E_1 + E_4)^{2m}} P[-E_4, -E_1(E_1 + E_4)] P[-E_3, - E_2 (E_2 + E_3)] .
\end{eqnarray}
As in the case of the photon, we see that singularities generally don't match. If $k > 0$, then the first line contains a singularity at $E_1+E_3 = 0$ which does not appear in the other two expressions. If $m > 0$, then the second line has a singularity at $E_1 + E_2 = 0$ which does not correspond to the behaviour of the other two functions. Thus, we must have $m = k = 0$ and the equations become
\begin{eqnarray} \nonumber
& \ & P[E_1 + E_3, E_1 E_3] P[E_2 + E_4, E_2 E_4]  \\ \nonumber
& = &  P[-E_2, -E_1(E_1 + E_2)]  P[-E_3, - E_4 (E_3 + E_4)] \\
& = &  P[-E_4, -E_1(E_1 + E_4)] P[-E_3, - E_2 (E_2 + E_3)] .
\end{eqnarray}
This must hold for any $E_i$ that satisfy $\sum_i E_i = 0$. Now if we let $E_1 = E_2 = 0$, $E_3 = -E_4 = E$, our constraint becomes
\begin{equation}
P[E, 0] P[-E, 0] = P[0, 0]  P[-E, 0] = P[E, 0] P[-E, 0],
\end{equation}
and so $P[-E, 0] (P[E,0] - P[0,0]) = 0$. This implies that $P[-E,0] = 0$ for all $E$ or $P[E,0] = P[0,0]$ for all $E$. But the first alternative entails the latter, so we can just assume
\begin{equation}
P[E, 0] = P[0,0] := P_0 \quad \forall E.
\end{equation}
Now let $E_1 + E_2 = E_3 + E_4 = 0$. Our factorisation constraint is then
\begin{align}
&P[E_1 + E_3, E_1 E_3] P[- (E_1 + E_3), E_1 E_3] \\
=&P[E_1, 0]  P[-E_3, 0]  \\
=&P[E_3, -E_1(E_1 - E_3)] P[-E_3,  - E_1 (E_1 - E_3)].
\end{align}
Because $E_1$ and $E_3$ are effectively independent variables, we can write $x = E_1 + E_3$, $y = E_1 E_3$ and find that the following equation must hold for all $x,y$:
\begin{equation}
P[x, y] P[- x, y] = P_0^2.
\end{equation}
It is then easy to show (e.g. by observing that any zero of $P[x,y]$ would correspond to a singularity of $P[-x,y]$, which a polynomial cannot have) that the only polynomial solution to this equation is $P[x,y] = P_0$.

\section{Tree level three-point amplitudes for broken Maxwell theory} \label{AppendixB}

Maxwell theory of electromagnetism is a Lorentz invariant theory of a massless spin-$1$ particle, with just two degrees of freedom corresponding to the two helicities $\pm 1$ of the photon. The quadratic Lagrangian is
\begin{equation}
\mathcal{L}_2 =  \frac{1}{4} F_{\mu \nu} F^{\mu \nu},
\label{eq:Maxwell}
\end{equation}
where $F_{\mu \nu} = \partial_{\mu} A_{\nu} - \partial_{\nu}A_{\mu}$. By counting first class and second class constraints, one can show that the free theory indeed has two degrees of freedom. This is because $A_0$ is non-dynamical and we also have a one-dimensional gauge freedom. In the boost-invariant theory, there are no cubic interactions, as we have shown in Section \ref{sec:3p}. Interactions can only start at quartic order in the fields. \\

As for the boost-breaking amplitudes in a theory of a single photon, we have shown that they are allowed: they are the $(\pm1,\pm1,\pm1)$ amplitudes with at least three powers of energy as dictated by Bose symmetry. The simplest such amplitudes are 
\begin{align}  
\mathcal{A}_3(1^{-1} 2^{-1} 3^{-1}) &= g \la12 \ra \la 2 3\ra \la 3 1\ra (E_{1}-E_{2})(E_{2}-E_{3})(E_{1}-E_{3}), \label{AppPhotonAmplitudes1} \\
\mathcal{A}_3(1^{+1} 2^{+1} 3^{+1}) &= \pm g [12][23][31] (E_{1}-E_{2})(E_{2}-E_{3})(E_{1}-E_{3}),\label{AppPhotonAmplitudes2}
\end{align}
and in Section \ref{sec:photon} we suggested that such amplitudes arise from 
\begin{align} \label{VectorConsistentVertexApp}
\ddot{F}^{\mu}{}_{\nu}\dot{F}^{\nu}{}_{\rho}F^{\rho}{}_{\mu}, \qquad \epsilon^{\mu\nu\rho\sigma}\ddot{F}_{\mu\nu}\dot{F}_{\rho \kappa}F_{\sigma}{}^{\kappa},
\end{align} 
operators in the Lagrangian. In this Appendix we consider the second of these operators showing that it does indeed give rise to the parity-odd form of the above amplitudes. Extending the following to the first of these operators is straightforward and yields the parity-even form of the above amplitudes. \\

We will use the following, elegant identity:
\begin{equation} \label{eq:EpsContraction}
\epsilon^{\mu \nu \rho \sigma} p^1_{\mu}  p^2_{\nu}  p^3_{\rho}  p^4_{\sigma} = -4 i \left( 
\la 1 2 \ra [2 3] \la 3 4 \ra [41] - [1 2] \la 2 3 \ra [3 4] \la 4 1 \ra \right),
\end{equation}
which is valid for any four, null $4$-momenta (not necessarily conserved). The identity can be proven efficiently using symbolic manipulation in Mathematica. The tree-level, $(+1,+1,+1)$, S-matrix element $S^+_{3 \to 0}$ due to $\epsilon^{\mu\nu\rho\sigma}\ddot{F}_{\mu\nu}\dot{F}_{\rho \kappa}F_{\sigma}{}^{\kappa}$ is 
\begin{align}
S^+_{3 \to 0} & = \la 0 |(- i) \int d^3 x dt H_{int}(x,t) \left[ \prod\limits_{i=1}^3 \sqrt{2E_i} a^{+ \dag}_{\bvec{p}_i} \right] | 0 \ra \nn  \\
& = i g' \int d^3 q_1 d^3 q_2 d^3 q_3 \delta^{(4)} \left( \sum q^{\mu}_i \right) \nn \\
& \quad \times \sum\limits_{\Lambda_{1,2,3}} \epsilon_{\mu \nu \rho \sigma} E_{q1}^2 \left( q^{\mu}_{1} \xi_1^{\Lambda_1, \nu} - q^{\nu}_{1} \xi_1^{\Lambda_1, \mu} \right) E_{q2}  \left( q^{\rho}_{2} \xi_2^{\Lambda_2, \alpha} - q^{\alpha}_{2} \xi_2^{\Lambda_2, \rho} \right) \left( q^{\sigma}_{3} \xi^{\Lambda_3}_{3,\alpha} - q_{3,\alpha} \xi_3^{\Lambda_3, \sigma} \right) \nn \\
& \quad \times \sum\limits_{\sigma \in S_3} \left( 
\delta(p_{\sigma(1)} - q_1) \delta(p_{\sigma(2)} - q_2) \delta(p_{\sigma(3)} - q_3) \delta_{+, \Lambda_1}  \delta_{+, \Lambda_2}  \delta_{+, \Lambda_3}
\right) \nn \\
& = i g' \delta^{(4)}\left( \sum p^{\mu}_i \right) \epsilon_{\mu \nu \rho \sigma} E_1^2 E_2 \nn \\
& \quad \times \left( p^{\mu}_{1} \xi_1^{+, \nu} - p^{\nu}_{1} \xi_1^{+, \mu} \right) \left( p^{\rho}_{2} \xi_2^{+, \alpha} - p^{\alpha}_{2} \xi_2^{+, \rho} \right) \left( p^{\sigma}_{3} \xi^{+}_{3,\alpha} - p_{3,\alpha} \xi_3^{+, \sigma} \right) + 5 \ \textrm{perms} \nn \\
& = 2 i g' \delta^{(4)}\left( \sum p^{\mu}_i \right)  \epsilon_{\mu \nu \rho \sigma} E_1^2 E_2 p^{\mu}_{1} \xi_1^{+, \nu}   \left( p^{\rho}_{2} \xi_2^{+, \alpha} - p^{\alpha}_{2} \xi_2^{+, \rho} \right) \left( p^{\sigma}_{3} \xi^{+}_{3,\alpha} - p_{3,\alpha} \xi_3^{+, \sigma} \right) + 5 \ \textrm{perms} \nn .
\end{align}

Once we expand the product of two brackets into a sum, each permutation seems to include four terms, but one of these trivially vanishes as it involves a factor  $p_2 \cdot p_3 = 0$. We therefore have
\begin{align}
S^+_{3 \to 0} & =  2 i g' \delta^{(4)}\left( \sum p^{\mu}_i \right)  \epsilon_{\mu \nu \rho \sigma} E_1^2 E_2 p^{\mu}_{1} \xi_1^{+, \nu} \nn   \\
& \quad \times
\left( p^{\rho}_{2} p^{\sigma}_{3} (\xi^+_2 \cdot \xi^{+}_3)
-\xi_2^{+, \rho} p^{\sigma}_{3} (p_2 \cdot \xi^{+}_{3})
- p^{\rho}_{2} \xi_3^{+, \sigma} (p_3 \cdot \xi^+_2)  \right)
+ 5 \ \textrm{perms} .
\end{align}

Using (\ref{eq:EpsContraction}), we get 
\begin{align}
S^+_{3 \to 0} & =  2 i g' \delta^{(4)}\left( \sum p^{\mu}_i \right)  (-4i) E_1^2 E_2  
\big\{ \left( \la 1 \xi_1 \ra [\xi_1 2] \la 2 3 \ra [3 1] - [1 \xi_1] \la \xi_1 2 \ra [2 3] \la 3 1 \ra \right) 
(\xi^+_2 \cdot \xi^{+}_3) \nn \\
& \quad \quad \quad \quad - \left( \la 1 \xi_1 \ra [\xi_1 \xi_2] \la \xi_2 3 \ra [3 1] - [1 \xi_1] \la \xi_1 \xi_2 \ra [\xi_2 3] \la 3 1 \ra
 \right) 
(p_2 \cdot \xi^{+}_{3}) \nn \\ 
& \quad \quad \quad \quad  - \left(  \la 1 \xi_1 \ra [\xi_1 2] \la 2 \xi_3 \ra [ \xi_3 1] - [1 \xi_1] \la \xi_1 2 \ra [ 2 \xi_3 ] \la \xi_3  1 \ra \right) 
(p_3 \cdot \xi^+_2)  \big\}
+ 5 \ \textrm{perms} .
\end{align}

\noindent
(Spinors constructed from the momenta are written as numbers $1,2,3$; spinors constructed from the polarization vectors are written as $\xi_i$.) Recall that for three-particle, on-shell interactions, we have $\la ij \ra = 0$ for all $i,j$ or $[ij] = 0$ for all $i,j$; so the first line vanishes. We also have $[1 \xi_1] = 0$, so all terms involving this factor vanish as well. Thus,
\begin{align} \label{eq:XYZ}
S_{3 \to 0} =  - 8 g' \delta^{(4)}\left( \sum p^{\mu}_i \right)   E_1^2 E_2  
\big\{ 
& \left( \la 1 \xi_1 \ra [\xi_1 \xi_2] \la \xi_2 3 \ra [3 1]
 \right) 
(p_2 \cdot \xi^{+}_{3}) \nn \\
+ & \left(  \la 1 \xi_1 \ra [\xi_1 2] \la 2 \xi_3 \ra [ \xi_3 1] \right) 
(p_3 \cdot \xi^+_2)  \big\}
+ 5 \ \textrm{perms} .
\end{align}

To make further progress, we have to choose a concrete spinor representation of the polarization vectors $\xi_i$. Recall that
\begin{equation*}
\xi^{+}_{a\dot{a}}(\bvec{p}) = \frac{\eta_a \tilde{\lambda}_{\dot{a}}}{\la \eta, \lambda \ra}, 
\end{equation*}
with an almost arbitrary reference spinor $\eta$. At this point, we are free to make a choice that breaks the Lorentz symmetry and we do so such that 
\begin{equation}
\xi_i^{+} =  \frac{(\epsilon . \tilde{\lambda}_i^T) \tilde{\lambda}_i}{(ii)} .
\end{equation}
So $\eta_{i,1} = \tilde{\lambda}_{i,2}$ and $\eta_{i,2} = -\tilde{\lambda}_{i,1}$. Then, we have the following identities:
\begin{eqnarray}
\la i \xi^+_j \ra & = & -(i j), \\
{} [ i \xi^+_j ]  & = & \frac{[ij]}{(jj)}, \\
{} [\xi^+_i , \xi^+_j] & = & \frac{[ij]}{(ii)(jj)}, \\
p_i \cdot \xi^+_j & = & \frac{1}{2} \la i \xi_j \ra [i \xi_j ] = - \frac{1}{2} \frac{(i j) [i j]}{(jj)} .
\end{eqnarray}
Now we can simplify (\ref{eq:XYZ}). The first line (dropping the prefactor $- 8 g' \delta $) gives: \\

$
\begin{aligned}
& \sum\limits_{\text{perms}} E_1^2 \left( E_2 \la 1 \xi_1 \ra [\xi_1 \xi_2] \la \xi_2 3 \ra [3 1]  (p_2 \cdot \xi^{+}_{3})
\right)  \\
& =
\frac{1}{8 E_1 E_2 E_3} \sum\limits_{\text{perms}}  E_1^2 \left(  E_2 (-(11))  [12] (32) [3 1] \left( - \frac{1}{2}  (23)[23] \right)
\right) \\
& =
\frac{1}{8 E_1 E_2 E_3} \sum\limits_{\text{perms}}  E_1^3  E_2  [12] (32) [3 1]  (23)[23] \\
& = \frac{1}{2} \sum\limits_{\text{perms}}  E_1^2  E_2  [12][23][31]  = \frac{1}{2} [12]  [23]  [3 1] \sum\limits_{\text{cyc}}   E_1^2 \left(E_2 - E_3 \right).
\end{aligned}
$\\

\noindent
Meanwhile, the second line of (\ref{eq:XYZ}) (again dropping the prefactor $-8 g' \delta$) gives: \\

$
\begin{aligned}
& \sum\limits_{\text{perms}} E_1^2 \left( E_2 \la 1 \xi_1 \ra [\xi_1 2] \la 2 \xi_3 \ra [\xi_3 1]  (p_3 \cdot \xi^{+}_{2})
\right)  \\
& =
\frac{1}{2} \frac{1}{8E_1E_2 E_3} \sum\limits_{\text{perms}}  E_1^2 \left(  E_2 (-(11)) [12] (-(23)) [31] (-(32))[32]
\right) \\
& =  \frac{1}{8E_1E_2 E_3} \sum\limits_{\text{perms}}  E_1^3 E_2 [12] (23) [31] (32) [23] \\
& = 
 \frac{1}{2} \sum\limits_{\text{perms}}  E_1^2 E_2 [12][23][31] = 
 \frac{1}{2}  [12][23][31] \sum\limits_{\text{cyc}}  E_1^2 \left(E_2 - E_3 \right). \\
\end{aligned}
$ \\

\noindent
We see that the two contributions are exactly the same. In conclusion, we get

\begin{align} \nonumber
S^+_{3 \to 0} & =  -8 g' \delta^{(4)}\left( \sum p^{\mu}_i \right)   [12][23][31]  \sum\limits_{cyc}  E_1^2 \left(E_2  -
E_3
\right) \\
& = 8 g' \delta^{(4)}\left( \sum p^{\mu}_i \right)   [12][23][31] (E_1 - E_2) (E_2 - E_3) (E_3 - E_1) .
\label{eq:+++}
\end{align}

\noindent
The analogue of (\ref{eq:XYZ}) for all-minus helicities is
\begin{align} \label{eq:XYZ-}
S^-_{3 \to 0} =  8 g' \delta^{(4)}\left( \sum p^{\mu}_i \right)  E_1^2 E_2  
\big\{ 
& \left( [ 1 \xi_1 ] \la \xi_1 \xi_2 \ra [ \xi_2 3 ] \la 3 1 \ra
 \right) 
(p_2 \cdot \xi^-_3) \\ \nonumber
+ & \left(  [ 1 \xi_1 ] \la \xi_1 2 \ra [ 2 \xi_3 ] \la \xi_3 1 \ra \right) 
(p_3 \cdot \xi^-_2)  \big\}
+ 5 \ \textrm{perm-s} .
\end{align}

\noindent
We choose reference spinors similarly as before,
\begin{equation}
\xi_i^{-}  =  \frac{\lambda_i (\epsilon . \lambda_i^T)}{(ii)} .
\end{equation}
Then
\begin{eqnarray}
\la i \xi^{-}_j \ra & = & \la i j \ra, \\
{} [ i \xi^{-}_j ] & = & -\frac{(j i )}{(jj)}, \\
\la \xi^{-}_i , \xi^{-}_j \ra & = & \la i j \ra, \\
p_i \cdot \xi^{-}_j & = & \frac{1}{2} \la i \xi_j \ra [i \xi_j ] = - \frac{1}{2} \frac{\la i j \ra (j i)}{(jj)} .
\end{eqnarray}

The first line of (\ref{eq:XYZ-}), after dropping the prefactor $8 g' \delta$, gives \\

$
\begin{aligned}
& \sum\limits_{\text{perms}} E_1^2 \left( E_2  [ 1 \xi_1 ] \la \xi_1 \xi_2 \ra [ \xi_2 3 ] \la 3 1 \ra (p_2 \cdot \xi^-_3)  \right) \\
& = \frac{1}{2}\sum\limits_{\text{perms}} E_1^2 \left( E_2 \cdot (-1) \cdot \la 1 2 \ra \left( -\frac{(23)}{(22)} \right) \la 3 1 \ra \frac{-\la 2 3 \ra (32)}{(33)} \right) \\
& = \frac{1}{2}\sum\limits_{\text{perms}} \frac{E_1^2 E_2}{4 E_2 E_3} \la 1 2 \ra \la 2 3 \ra \la 3 1 \ra \left( -(23)\right)  (32)  = - \frac{1}{2} \la 1 2 \ra \la 2 3 \ra \la 3 1 \ra \sum\limits_{cyc} E_1^2 (E_2 - E_3).
\end{aligned}
$ \\

The second line of (\ref{eq:XYZ-}) yields \\

$
\begin{aligned}
&  \sum\limits_{\text{perms}} E_1^2 \left( E_2  [ 1 \xi_1 ] \la \xi_1 2 \ra [ 2 \xi_3 ] \la \xi_3 1 \ra  (p_3 \cdot \xi^-_2)  
\right)  = \frac{1}{2} \sum\limits_{\text{perms}} E_1^2 \left( E_2 \cdot (-1) \cdot \la 1 2 \ra  \frac{(32)}{(33)}\la 3 1 \ra \frac{-\la 3 2 \ra (23)}{(22)} 
\right) \\
& = -\frac{1}{2} \sum\limits_{\text{perms}} \frac{E_1^2 E_2}{4 E_2 E_3} \la 1 2 \ra  \la 2 3 \ra \la 3 1 \ra (23) (32)  = -\frac{1}{2} \la 1 2 \ra  \la 2 3 \ra \la 3 1 \ra   \sum\limits_{\text{cyc}} E_1^2 (E_2-E_3).
\end{aligned}
$ \\

So
\begin{align}  \nonumber
S^-_{3 \to 0} & =  - 8 g' \delta^{(4)}\left( \sum p^{\mu}_i \right)   \la 1 2 \ra \la 2 3 \ra \la 3 1 \ra \sum\limits_{cyc}  E_1^2 
(E_2  - E_3) \\
& = 8 g' \delta^{(4)}\left( \sum p^{\mu}_i \right)   \la 1 2 \ra \la 2 3 \ra \la 3 1 \ra (E_1 - E_2) (E_2 - E_3) (E_3 - E_1) .
\label{eq:---}
\end{align}

Comparing (\ref{eq:+++}) and (\ref{eq:---}) with (\ref{eq:ParityTransformation}), we see that the amplitude due to $\epsilon^{\mu \nu \rho \sigma} \ddot{F}_{\mu \nu} \dot{F}_{\rho \alpha} F_{\sigma}^{\ \alpha}$ is parity-odd, as expected from the presence of the $\epsilon$ tensor.

\section{Boost-breaking massless QED} \label{AppendixC}

In this Appendix we provide Lagrangians for the boost-breaking versions of massless QED we derived using the four-particle test in Section \ref{sec:PhotonCouplings}. In the boost-invariant limit massless scalar QED is described by the Lagrangian 
\begin{align} \label{MasslessQED}
\mathcal{L} =  \frac{1}{4}F_{\mu\nu}^{2} +\frac{1}{2}D^{\mu}\phi D_{\mu} \phi^{*} 
\end{align}
where the covariant derivative is as usual $D_{\mu} \phi = \partial_{\mu} \phi - i e \phi A_{\mu}$. This gives rise to the standard kinetic terms plus cubic and quartic vertices. The Lagrangian is invariant under the gauge symmetry
\begin{align}
\phi \rightarrow e^{ie \alpha(x)} \phi, \qquad A_{\mu} \rightarrow A_{\mu} + \partial_{\mu}\alpha(x).
\end{align}
By choosing the basis $\phi = \phi^{1} + i \phi^{2}$ the anti-symmetric nature of the cubic vertices is manifest and the three-particle amplitude has $F_{ab} = \epsilon_{ab}$ in \eqref{Compton1} and \eqref{Compton2}. Now to realise the function of energy in the amplitude we need to add time derivatives to \eqref{MasslessQED}. We saw that in the boost-breaking case we have $F_{ab} = \epsilon_{ab}F(E_{1}+E_{2})$ and since $E_{1}+E_{2} = -E_{3}$ we can add time derivatives to the vector only, and we find that the correct Lagrangian is given by
\begin{align} \label{MasslessQEDBV}
\mathcal{L} =  \frac{1}{4}F_{\mu\nu}^{2} +\frac{1}{2}\hat{D}^{\mu}\phi \hat{D}_{\mu} \phi^{*} 
\end{align}
where we have defined the new boost-breaking covariant derivative
\begin{align}
\hat{D}_{\mu} \phi = \partial_{\mu} \phi - i e \phi \hat{\partial}_{t}A_{\mu}, 
\end{align}
in terms of the derivative operator
\begin{align}
\hat{\partial}_{t} = a_{1}\partial_{t} + a_{2} \partial^{2}_{t} + a_{3}\partial^{3}_{t} + \ldots.
\end{align}
In comparison to the boost-invariant theory, this theory also has a gauge symmetry given by 
\begin{align}
\phi \rightarrow e^{ie \hat{\partial}_{t}\beta(x)} \phi, \qquad A_{\mu} \rightarrow A_{\mu} + \partial_{\mu}\beta(x).
\end{align}
If we again write $\phi = \phi^{1}+i \phi^{2}$ we see that 
\begin{align}
\mathcal{L} \supset i e \epsilon_{ab} \phi^{a}\partial^{\mu}\phi^{b} \hat{\partial}_{t}A_{\mu},
\end{align} 
and these cubic vertices give rise to our three-particle amplitudes. We therefore have a consistent boost-breaking theory of massless scalar QED. \\

For $S=1/2$ the story is a simple generalisation of the above discussion. In the boost-invariant limit, massless fermionic QED is described by the Lagrangian
\begin{align} \label{FermionicQED}
\mathcal{L} =\frac{1}{4}F_{\mu\nu}F^{\mu\nu} + i \bar{\psi} \gamma^{\mu}D_{\mu} \psi,
\end{align}
where $\psi$ is a four-component Dirac spinor\footnote{Recall that a Dirac spinor is not a irreducible respresentation of the Lorentz group. It is really comprised of two 2-component spinors reflecting the fact that we need two $S=1/2$ particles each with $\pm 1/2$ helicities.}, $\gamma^{\mu}$ are the gamma matrices and $D_{\mu} = \partial_{\mu} + i e A_{\mu}$. This Lagrangian is invariant under the $U(1)$ gauge symmetry
\begin{align}
\psi \rightarrow e^{-ie \alpha(x)} \psi, \qquad A_{\mu} \rightarrow A_{\mu} + \partial_{\mu} \alpha(x).
\end{align}
Guided by the scalar case, we can instead define a new covariant derivative as
\begin{align}
\hat{D}_{\mu} = \partial_{\mu} + i e \hat{\partial}_{t}A_{\mu},
\end{align}
and if we replace $D_{\mu}$ by $\hat{D}_{\mu}$ in \eqref{FermionicQED} then we find a consistent boost-breaking theory of massless fermionic QED invariant under the gauge symmetry
\begin{align}
\psi \rightarrow e^{-ie \hat{\partial}_{t}\beta(x)} \psi, \qquad A_{\mu} \rightarrow A_{\mu} + \partial_{\mu} \beta(x).
\end{align}
Again this theory gives rise to our boost-breaking amplitudes derived in Section \ref{sec:PhotonCouplings}.

\section{More details on the inconsistency of $\dot{\phi}^3$ coupled to gravity}  \label{AmplitudeCalculation}

In this appendix we consider a self-interacting scalar minimally coupled to $h_{\mu \nu}$ in Minkowski space and directly compute the $\mathcal{A}_{4}(1^{0},2^{0},3^{0},4^{+2})$ amplitude due to scalar exchange, showing that the final result is gauge invariant only in the absence of Lorentz-violating interactions. Thus the aim is to provide further clarity on why an interaction of the form $\dot{\phi}^3$ is inconsistent. We take the graviton self-interactions and the minimal coupling between the scalar and the graviton to be Poincar\'{e} invariant and consider a Lagrangian of the form
\begin{align}
\mathcal{L} = \mathcal{L}_{EH} + \frac{1}{2}(\partial \phi)^2 - \frac{1}{\sqrt{2} M_{\text{pl}}}h^{\mu\nu}\partial_{\mu} \phi \partial_{\nu} \phi + \mathcal{L}_{\phi},
\end{align}
where $\mathcal{L}_{EH}$ contains the quadratic and cubic terms in the canonically normalised graviton fluctuation $h_{\mu\nu}$ arising from expanding $\sqrt{-g}R$ around Minkowski space and $\mathcal{L}_{\phi}$ contains cubic self-interactions for the scalar with an unspecified number of time derivatives (all Lorentzian derivatives can be removed by field redefinitions). The results of this appendix will therefore capture $\dot{\phi}^3$ but also a more general class of self-interactions where the on-shell three-scalar amplitude is $\mathcal{A}_{3}(1^0,2^0,3^0) = F(E_{1},E_{2},E_{3})$ where $F$ is a symmetric polynomial. \\

First consider the $s$-channel of the $\mathcal{A}_{4}(1^{0},2^{0},3^{0},4^{+2})$ amplitude. Up to unimportant $\mathcal{O}(1)$ factors and inverse powers of $M_{\text{pl}}$, we have 
\begin{align}
\mathcal{A}_{4}^{s}(1^{0},2^{0},3^{0},4^{+2}) = \frac{F(E_{1},E_{2})}{s} \epsilon^{+}_{\mu\nu}(p_{4})p_{3}^{\mu} (p_{3}^{\nu} + p_{4}^{\nu}) = \frac{F(E_{1},E_{2})}{s} \epsilon^{+}_{\mu\nu}(p_{4})p_{3}^{\mu}p_{3}^{\nu},
\end{align}
where we have used the fact that the graviton's on-shell polarisation tensor is transverse and have used energy conservation to eliminate the energy of the exchanged scalar particle. The $t$ and $u$ channel expressions are 
\begin{align}
\mathcal{A}_{4}^{s}(1^{0},2^{0},3^{0},4^{+2}) = \frac{F(E_{1},E_{3})}{t} \epsilon^{+}_{\mu\nu}(p_{4})p_{2}^{\mu}p_{2}^{\nu}, \\
\mathcal{A}_{4}^{s}(1^{0},2^{0},3^{0},4^{+2}) = \frac{F(E_{2},E_{3})}{u} \epsilon^{+}_{\mu\nu}(p_{4})p_{1}^{\mu}p_{1}^{\nu}.
\end{align}
Now we can write these expressions in the spinor helicity formalism using 
\begin{align}
4 \epsilon_{\mu\nu}^{+}(p_{4}) p_{i}^{\mu}p_{i}^{\mu} = e^{+}_{\alpha \dot{\alpha}}(p_{4})e^{+}_{\beta \dot{\beta}}(p_{4})\lambda^{\alpha}_{i} \tilde{\lambda}^{\dot{\alpha}}_{i}\lambda^{\beta}_{i} \tilde{\alpha}^{\dot{\beta}}_{i} = \left(\frac{\langle \eta i \rangle [4i]}{\langle \eta 4 \rangle} \right)^2.
\end{align}
Now we have infinitely many choices for the reference spinor $\eta$, but it is sufficient to consider only three options, $\eta = 1,2,3$, so that $\eta$ corresponds to a spinor of one of the particles other than the graviton. The three choices for each channel yield (again dropping unimportant common factors)
\begin{align}
A_{4}^{s}(1^0,2^0,3^0,4^{+2})&=F_{12}\left(\langle 12 \rangle [14][24]\right)^{2} \times \left\{ \begin{array}{ll} \frac{1}{su^{2}} &\eta =1 \\[.2cm]\frac{1}{st^{2}} & \eta=2 \\[.2cm] 0 & \eta =3 \end{array} \right. \\
A_{4}^{t}(1^0,2^0,3^0,4^{+2})&=F_{13}\left(\langle 12 \rangle [14][24]\right)^{2} \times \left\{ \begin{array}{ll} \frac{1}{tu^{2}} & \eta=1 \\[.2cm] 0& \eta =2 \\[.2cm] \frac{1}{ts^{2}} &\eta =3 \end{array} \right. \\
A_{4}^{u}(1^0,2^0,3^0,4^{+2})&=F_{23}\left(\langle 12 \rangle [14][24]\right)^{2}  \times \left\{ \begin{array}{ll} 0 & \eta =1\\[.2cm] \frac{1}{ut^{2}} & \eta=2 \\[.2cm] \frac{1}{us^{2}} &\eta =3 \end{array} \right.
\end{align}
where we have introduced the shorthand $F(E_{i},E_{j}) = F_{ij}$. Using $s+t+u = 0$, we can therefore write the full amplitude as 
\begin{align}
A_{4}(1^0,2^0,3^0,4^{+2})=- \left(\langle 12 \rangle [14][24]\right)^{2} \times \left\{ \begin{array}{ll}  \frac{F_{12}}{stu} + \frac{F_{12}-F_{13}}{tu^{2}}&\eta =1 \\[.2cm]
\frac{F_{23}}{stu}+\frac{F_{23}-F_{12}}{st^{2}} & \eta =2\\[.2cm] 
\frac{F_{13}}{stu}+\frac{F_{13}-F_{23}}{us^{2}} & \eta =3. \end{array} \right.
\end{align}
For general boost-breaking scalar self-interactions, $  F_{12}\neq F_{13} $ and so on. Hence we see that the above amplitude could change as different choices for the unphysical reference spinor are made. This certainly indicates an inconsistency. Demanding that the amplitude is the same for each choice of reference spinor leads to the constraints
\begin{align}
F_{12} = F_{13} = F_{23}\,.
\end{align}
This is only solved by $F = constant$ for generic energies, and so the three-particle amplitude for a scalar coupled to gravity must be Poincar\'{e} invariant.

\bibliographystyle{utphys}
\bibliography{refs}

\end{document}